\PassOptionsToPackage{table}{xcolor}
\documentclass[numsec,webpdf,modern,medium,namedate,dvipsnames]{oup-authoring-template}

\onecolumn

\graphicspath{{Fig/}}

\theoremstyle{thmstyleone}%
\newtheorem{theorem}{Theorem}
\newtheorem{corollary}{Corollary}
\newtheorem{proposition}{Proposition}
\newtheorem{lemma}{Lemma}
\newtheorem{illustration}{Illustration}
\theoremstyle{thmstyletwo}%

\theoremstyle{thmstylethree}%

\makeatletter
\@ifundefined{description}{%

}{}
\makeatother

\usepackage{bbm}
\usepackage{bm}
\usepackage{booktabs}
\usepackage{enumitem}
\usepackage[caption=false]{subfig}
\usepackage[normalem]{ulem}
\usepackage{CJKutf8}
\usepackage[doublespacing]{setspace}
\usepackage[fontsize=12pt]{fontsize}
\usepackage{tikz}
\usepackage[T1]{fontenc}
\shortcites{ghcnd_data,Menne2012}
\allowdisplaybreaks

\newcommand{\tikzcircle}[2][black,fill=black]{\tikz[baseline=-0.5ex]\draw[#1,radius=#2] (0,0) circle ;}%
\newcommand{\bs}{\boldsymbol{s}}

\newcommand{\pr}{\text{Pr}}
\newcommand{\trans}{^\text{T}}
\newcommand{\iid}{\stackrel{\text{iid}}{\sim}}
\newcommand{\indep}{\perp \!\!\! \perp}

\newcommand{\bX}{\boldsymbol{X}}
\newcommand{\bx}{\boldsymbol{x}}
\newcommand{\bY}{\boldsymbol{Y}}

\newcommand{\bZ}{\boldsymbol{Z}}
\newcommand{\bS}{\boldsymbol{S}}

\newcommand{\bsigma}{\boldsymbol{\sigma}}
\newcommand{\bxi}{\boldsymbol{\xi}}
\newcommand{\bphi}{\boldsymbol{\phi}}
\newcommand{\brho}{\boldsymbol{\rho}}
\newcommand{\bgamma}{\boldsymbol{\gamma}}
\newcommand{\bSigma}{\boldsymbol{\Sigma}}
\newcommand{\bbeta}{\boldsymbol{\beta}}
\newcommand{\bc}{\boldsymbol{c}}
\newcommand{\bd}{\boldsymbol{d}}
\newcommand{\bI}{\boldsymbol{I}}
\usepackage{varioref}
\labelformat{equation}{(#1)}

\makeatletter
\def\equationautorefname{\@gobble}
\makeatother

\newcommand{\BAS}[1]{\color{black}{#1}\color{black}}

\newcommand{\MS}[1]{\color{black}{#1}\color{black}}
\newcommand{\LZ}[1]{\color{black}{#1}\color{black}}

\makeatletter
\def\section{%
  \@startsection{section}{1}{\z@}
  {-1\p@ plus -0\p@}{0.01\p@}
  {\reset@font\raggedright\secsize}}
\def\subsection{%
  \@startsection{subsection}{2}{\z@}
  {-1\p@ plus -0\p@}{0.01\p@}
  {\reset@font\raggedright\subsecsize}}
\def\subsubsection{%
  \@startsection{subsubsection}{3}{\z@}
  {-1\p@ plus -0\p@}{0.01\p@}
  {\reset@font\raggedright\subsecsize}}
\makeatother

\AtBeginDocument{
  \setlength{\abovedisplayskip}{2pt plus 1pt minus 1pt}
  \setlength{\belowdisplayskip}{2pt plus 1pt minus 1pt}
  \setlength{\abovedisplayshortskip}{2pt plus 1pt minus 1pt}
  \setlength{\belowdisplayshortskip}{2pt plus 1pt minus 1pt}
}

\setlist[enumerate]{topsep=2pt, partopsep=0pt, parsep=0pt}

\makeatletter
\renewcommand{\maketitle}{%
  \begingroup
    \renewcommand\thefootnote{\@fnsymbol\c@footnote}%
    \thispagestyle{plain}%
    {\sffamilyfontbold\fontsize{18}{23}\selectfont\raggedright \@title \par}%
    \vspace{8pt}%
    {\sffamilyfontbold\boldmath\fontsize{12}{15}\selectfont\raggedright \@author \par}%
    \vspace{6pt}%
    {\sffamilyfontcn\fontsize{10}{13}\selectfont\raggedright \@address \par}%
    \vspace{4pt}%
    {\sffamilyfontcn\fontsize{8}{11.5}\selectfont\raggedright \@corresp \par}%
    \vspace{14pt}%
    {\sffamilyfont\fontsize{10}{12}\bfseries\selectfont Abstract\par}%
    \vspace{5pt}%
    {\sffamilyfont\fontsize{9}{12}\selectfont\noindent\@abstract\par}%
    \@thanks
  \endgroup
  \setcounter{footnote}{0}%
}
\makeatother

\begin{document}
\pagestyle{plain}


\title[Log-Laplace Nuggets for Threshold Exceedances]{Log-Laplace Nuggets for Fully Bayesian Fitting of Spatial Extremes Models to Threshold Exceedances}

\author[1,$\ast$]{Muyang Shi}
\author[2]{Likun Zhang}
\author[1]{Benjamin A. Shaby}

\authormark{Shi et al.}

\address[1]{\orgdiv{Department of Statistics}, \orgname{Colorado State University}, \orgaddress{\state{Colorado}, \country{USA}}}
\address[2]{\orgdiv{Department of Statistics}, \orgname{University of Missouri}, \orgaddress{\state{Missouri}, \country{USA}}}

\corresp[$\ast$]{Muyang Shi, Colorado State University, Fort Collins, Colorado, 80524, U.S.A. \href{muyang.shi@colostate.edu}{muyang.shi@colostate.edu}}


\abstract{Flexible \LZ{random scale-mixture models } provide a framework for capturing a broad range of extremal dependence structures. However, likelihood-based inference under the peaks-over-threshold setting is often computationally infeasible, due to the censored likelihood requiring repeated evaluation of high-dimensional Gaussian distribution functions. We propose a multiplicative log-Laplace nugget that yields conditional independence in the censored likelihood, resulting in a joint likelihood function that is the product of univariate densities which are available in closed form.  This eliminates multivariate Gaussian distribution function evaluations and thereby enables inference for threshold exceedances in high dimensions,
\MS{which represents a major shift for spatial extremes modelling as the total computational cost is now primarily driven by standard spatial statistics operations. }
We further show that a broad class of scale-mixture processes augmented with the proposed nugget preserves the extremal dependence structure of the underlying smooth process. 
The proposed methodology is illustrated through simulation studies and an application to precipitation extremes. 
}
\keywords{Censored likelihood, Computation, Peaks-over-Threshold, Tail dependence}
\maketitle

\section{Introduction}\label{sec:Introduction}

Flexible spatial extreme-value models based on \LZ{random scale-mixture models } provide a powerful framework for capturing a wide range of tail-dependence structures, including smooth transitions between different tail dependence regimes and spatially heterogeneous extremal behaviour. However, despite their modelling appeal, these constructions are difficult to fit under the peaks-over-threshold (POT) framework due to severe computational bottlenecks.  In this work, we slightly modify the models in a way that sidesteps the main computational challenge yet retains all of the desirable tail dependence characteristics.

Spatial extreme events such as intense precipitation, prolonged heat waves, and severe windstorms can affect broad geographic regions and trigger cascading impacts on infrastructure and interconnected services 
\citep{Forzieri_et_al_2018}.
Quantifying how such extremes co-occur across space is therefore fundamental for risk assessment, mitigation planning, and climate-resilient design 
\citep{milly2008stationarity}.

In response, a large literature has emerged to address the challenge of accurately modelling dependence in the tails of spatial processes.  A prominent example is the class of \LZ{random scale-mixture models built on latent Gaussian processes}. The preferred tool of inference for these models when applied to POT data is the censored likelihood, which \MS{retains partial information from observations below the threshold while avoiding the need to model their exact sub-threshold behaviour, thereby improving efficiency without unduly increasing bias from bulk misspecification}. However, evaluation of the censored likelihood requires repeated computation of high-dimensional Gaussian distribution functions, rendering inference infeasible for even moderately large numbers of locations \citep{zhang2021hierarchical}.
In this paper, we introduce a straightforward modification to this class of models by incorporating an independent multiplicative log-Laplace nugget at each spatial location. This construction yields conditional independence in the censored likelihood, \BAS{resulting in a joint likelihood function that is the product of univariate densities which are available in closed form.  This eliminates } the need for multivariate Gaussian distribution function evaluations and enables scalable Bayesian inference for high-dimensional threshold exceedance data.
We show that, under mild regular variation conditions, the spatial process that includes the proposed nugget preserves the extremal dependence structure of the underlying smooth process, with residual tail dependence coefficients unchanged and upper tail dependence coefficients modified only by multiplicative constants.
We demonstrate that the proposed approach applies broadly across several modern \LZ{scale-mixture } models, including stationary and nonstationary constructions with spatially varying extremal dependence structures.

\subsection{Tail Dependence for Spatial Extremes}


Let $\{Y(\bs): \bs \in \mathcal{S} \subseteq \mathbb{R}^2\}$ denote a \BAS{spatial } stochastic process of interest.
Throughout, for any function $A(\bs)$ evaluated at spatial locations $\bs_1,\ldots,\bs_D$, we use the shorthand $A_j := A(\bs_j)$. For example, $Y_j := Y(\bs_j)$, $j = 1, \ldots, D$. Let $F_{Y_j}$ denote the marginal distribution of $Y_j$. In spatial extremes modelling, it is common to separate marginal behaviour from spatial dependence using a copula representation. Specifically, applying the probability integral transform yields $U_j = F_{Y_j}(Y_j)$, so that the resulting copula captures the dependence structure independently of the marginals. Interest then centres on characterize dependence in the joint upper tail of the copula.

Extremal dependence between two locations \MS{$\bs_i, \bs_j$ is commonly summarized by the upper tail dependence coefficient
\begin{equation}\label{eqn:chi_measure}
\chi_{ij} = \lim_{u \to 1} \chi_{ij}(u), \qquad
\chi_{ij}(u) = \Pr(U_i > u \mid U_j > u).
\end{equation}}
The coefficient $\chi_{ij}$ quantifies how often extreme events co-occur at two sites as the quantile level $u$ approaches one. A pair of variables exhibits asymptotic dependence (AD) if $\chi_{ij} > 0$, indicating a non-vanishing probability of joint extremes, and asymptotic independence (AI) if $\chi_{ij} = 0$, indicating that joint exceedances become increasingly rare in the limit.
In the case of asymptotic independence, the coefficient $\chi_{ij}$ alone does not capture the rate at which joint exceedance probabilities decay. Additional information is provided by the residual tail dependence coefficient $\eta_{ij}$ \citep{ledford1996statistics}, defined through
\MS{\begin{equation}\label{eqn:eta_measure}
\Pr(U_i > u \mid U_j > u)
= \mathcal{L}_{ij}(1-u)(1-u)^{-(1 - 1/\eta_{ij})},
\end{equation}
where $\mathcal{L}_{ij}$ is a slowly varying function at zero, that is, $\lim_{t\rightarrow 0}\mathcal{L}(tx)/\mathcal{L}(t) = 1$ for any $x > 0$. The parameter $\eta_{ij}$ } \BAS{summarises } the strength of extremal dependence under asymptotic independence. \MS{For a stationary isotropic process, these pairwise coefficients depend only on $h_{ij} = \lVert \bs_i - \bs_j \rVert$, in which case we write $\chi_u(h_{ij})$, $\chi(h_{ij})$, and $\eta(h_{ij})$.}

Distinguishing asymptotic dependence (AD) and asymptotic independence (AI) is a central challenge in spatial extremes modelling, especially for extrapolation beyond observed data. Incorrectly specifying the extremal dependence class can result in substantial underestimation or overestimation of the joint risk of concurrent extreme events \citep{Huser_Opitz_Wadsworth_2025}. 
However, in practice, the tail region typically contains few observations; empirical tail diagnostics are often highly uncertain and difficult to make a definitive AD/AI classification. This motivates the need for flexible model classes that can represent both dependence regimes and allow the data to inform the dependence class.

Empirical evidence from environmental data for spatial extremes also often suggest that the conditional exceedance probability \MS{$\chi_{ij}(u)$ } typically decreases as spatial separation \MS{$h_{ij}$ } increases, and also with higher quantile levels $u$ \citep{shi2026}. 
These observations motivate the development of models that can accommodate \BAS{a variety of dependence features, including ``scale awareness'', wherein a process may exhibit strong } short-range extremal (in)dependence 
\BAS{and } long-range asymptotic independence.


\subsection{Flexible Spatial Extreme Models}\label{sec:flexible_models}

In response to these modelling needs, a growing body of work has developed flexible spatial extremes models that can bridge asymptotic dependence and asymptotic independence within a single framework, represent decreasing tail dependence with increasing spatial separation, and accommodate spatially heterogeneous dependence across large domains.
An especially important and widely used class of such models is based on \LZ{random scale-mixtures}, which provide a unified structure for several modern constructions.

A general random scale-mixture model can be written as
\begin{equation}\label{eqn:base_model}
X^*(\bs) = R\BAS{(\bs)}^{\phi\BAS{(\bs)}} \cdot g_{\phi}\{Z(\bs)\}
\end{equation}
where $Z(\bs)$ is a latent spatial process that is asymptotically independent at any two locations, typically a Gaussian process with covariance $\boldsymbol{\Sigma}_{\boldsymbol{\rho}}$; $R\BAS{(\bs)}$ is a random scaling variable \BAS{which can vary over space}; $\phi\BAS{(\bs)}$ serves as an extremal dependence parameter that indexes how the random scaling interacts with the latent spatial process\BAS{ and can also vary over space}; and $g_{\phi}(\cdot)$ is a \BAS{univariate } link function.

\BAS{A key assumption we make is that the combination of the link function $g_{\phi}(\cdot)$ and scaling variable $R(\bs)$ are chosen such that the resultant process $X^*(\bs)$ has regularly varying marginal tails.  That is, we assume that at every $\bs$, for some slowly varying function $\mathcal{L}_{\bs}$, 
$\Pr\{X^*(\bs)>x\}=\mathcal{L}_{\bs}(x)\,x^{-\alpha^*(\bs)}$ as $x\to\infty$  for some $\alpha^*(\bs) > 0$, where $\alpha^*(\bs)$ is called the tail index.}

Here we consider four representative constructions within the \BAS{transformed } Gaussian scale-mixture framework \MS{of \autoref{eqn:base_model}}. These models differ primarily in whether the scaling mechanism is global or spatially varying and whether tail parameters are allowed to vary across space, which in turn determines the degree of scale awareness and spatial heterogeneity in extremal dependence. Together, these models illustrate the main types of extremal dependence behaviour achieved by modern spatial extreme models \citep{huser2019modeling,majumder2024modeling,hazra2021realistic,shi2026}. We briefly describe how the scale-mixture components are specified in each of the four models:

\begin{enumerate}[label=(M\arabic*), labelsep=0.3cm]
  \item \label{Model-HW}
  The model of \citet{huser2019modeling} can be written as
  \[
    X^*(\bs) = R^\phi\, g\{Z(\bs)\}^{1-\phi},
  \]
  where $R$ is a global scaling variable with standard Pareto distribution, $Z(\bs)$ is a stationary Gaussian process, and $g(\cdot)$  transforms $Z(\bs)$ to have standard Pareto margins. This construction yields a smooth transition between asymptotic dependence and asymptotic independence through the parameter $\phi$, but it imposes the same extremal dependence structure \BAS{over all spatial scales and } across the \BAS{entire } spatial domain.

  \item \label{Model-Reetam}
  The model of \citet{majumder2024modeling} introduces a spatially varying random scale $R(\bs)$ derived from a Brown-Resnick max-stable process. After a monotone marginal transformation, it can be written in the \BAS{transformed } Gaussian scale-mixture form
  \[
    X^*(\bs) = R(\bs)^\phi\, g\{Z(\bs)\}^{1-\phi},
  \]
  where $Z(\bs)$ is a Gaussian process and both $R(\bs)$ and $g\{Z(\bs)\}$ have standard Pareto margins. This construction yields short-range asymptotic dependence with long-range asymptotic independence.

  \item \label{Model-Hazra}
  The model of \citet{hazra2021realistic} also specifies spatially varying random scale through
  \[
    X^*(\bs) = R(\bs)\, g\{Z(\bs)\},
    \qquad
    R(\bs)=\sum_{k=1}^K B_k(\bs;l)^{1/\gamma} R_k^*,
  \]
  where $\{B_k(\cdot;l)\}$ are compactly supported basis functions with radius $l$, $k = 1, \ldots, K$, $\{R_k^*\}$ are independent $\mathrm{Pareto}(\gamma)$ variables \citep[corresponding to $\beta=0$ in][]{huser2017bridging}, and $g(\cdot)$ is the identity link. This construction yields short-range asymptotic dependence with long-range asymptotic independence.

  \item \label{Model-Shi}
  The model of \citet{shi2026} can be written as
  \[
    X^*(\bs) = R(\bs)^{\phi(\bs)}\, g\{Z(\bs)\}, \qquad R(\bs) = \sum_{k=1}^K B_k(\bs;l) S_k,
  \]
  where $\{B_k(\cdot;l)\}$ are compactly supported basis functions with radius $l$, $k = 1, \ldots, K$; $\{S_k\}$ are independent $\mathrm{Stable}(\alpha, \beta, \gamma_k, \delta)$ variables; $\phi(\bs)$ is a spatially varying tail parameter surface; \BAS{and } $Z(\bs)$ is a Gaussian process, transformed by $g(\cdot)$ to standard Pareto margins. 
  This construction allows both the AD and AI local dependence class\BAS{es } to vary across space while retaining long-range AI.  \BAS{\citet{shi2026} set $\alpha=1/2, \beta=\MS{1}, \delta=1$ so that each $S_k \sim \text{L\'evy}(0, \gamma_k)$, yielding $R(\bs_j) \sim \text{L\'evy}(0, \bar\gamma_j)$ with $\bar\gamma_j=\left[\sum_{k=1}^K \sqrt{B_k(\bs_j;l)\,\gamma_k} \, \right]^2$.} 
  

\end{enumerate}

In summary, Model~\ref{Model-HW} achieves AD/AI flexibility through a global scaling mechanism but imposes a single stationary dependence class. Models~\ref{Model-Reetam} and \ref{Model-Hazra} introduce spatially varying scaling surfaces, enabling short-range AD with long-range AI. Model~\ref{Model-Shi} further generalises this framework by allowing the local dependence class \BAS{to be \textit{either} AD \textit{or} AI and } to vary across space through $\phi(\bs)$.
\autoref{tab:tail_summary_models} summarises the marginal tail index $\alpha^*$ and describes the joint tail behaviour through \BAS{the indices } $\chi$ and $\eta$ of Models \ref{Model-HW}--\ref{Model-Shi}.

\begin{table}[!t]
\begin{adjustwidth}{-0.5cm}{-0.5cm}
\centering
\caption{Summary of marginal and bivariate tail behaviour of $(\bs_i, \bs_j)$ for  \LZ{random scale-mixture models } \ref{Model-HW}--\ref{Model-Shi}. 
Let $W_i := g\{Z_i\}$, $\phi_i \in (0,\BAS{\infty})$ be the dependence parameter, and $\LZ{\eta_{ij}^Z}\in (1/2,1)$ be the residual tail dependence coefficient induced by the latent $\{Z(\bs)\}$ process. For \ref{Model-Hazra}, $\gamma>0$ is \BAS{the Pareto tail index for $R(\bs)$}. For \ref{Model-Shi}, $\alpha$ is the Stable-index. For \ref{Model-Hazra} and \ref{Model-Shi}, $\mathcal{K}_i := \{k: B_k(\bs_i; l) > 0\}$, the set of knots that are active for location $\bs_i$. See \autoref{sec:flexible_models} for model construction details.}
\label{tab:tail_summary_models}
\begingroup
\setlength{\tabcolsep}{3pt}
\renewcommand{\arraystretch}{1.0}
\small
\begin{tabular}{@{}p{0.06\linewidth} p{0.21\linewidth} p{0.7\linewidth}@{}}
\toprule
Model & $\overline{F}_{X^*} \in \mathrm{RV}_{-\alpha^*}$ & Joint Tail \BAS{Indices} \\
\midrule

\ref{Model-HW}
& $\alpha^*=\min\{\frac{1}{\phi},\,\frac{1}{1-\phi}\}$
& $\chi_{ij} = \begin{cases}
\frac{2\phi-1}{\phi}\,\mathbb{E}\!\left[\BAS{\min\{W_i, W_j\}}^{(1-\phi)/\phi}\right] & \text{if } \delta > \frac{1}{2}, \\
0 & \text{if } \delta \le \frac{1}{2}.
\end{cases}$ \par
$\eta_{ij} = \begin{cases}
1, & \text{if } \delta \ge \tfrac{1}{2},\\
\max\left\{\delta/(1-\delta),\LZ{\eta_{ij}^Z}\right\}, & \text{if }\delta < \tfrac{1}{2}.
\end{cases}$ \\

\cmidrule(lr){2-3}

\ref{Model-Reetam}
& $\alpha^*=\min\{\frac{1}{\phi},\,\frac{1}{1-\phi}\}$
& Same as \ref{Model-HW} (\textit{novel} analytical results in \autoref{prop:pmm_joint_tail}) \\

\cmidrule(lr){2-3}

\ref{Model-Hazra}
& $\alpha^*=\gamma$
& $\chi_{ij} = \displaystyle\sum_{k\in\mathcal{K}_i\cap \mathcal{K}_j}\Big[ B_k(\bs_i;l)\,\overline{T}_{\gamma+1}\{a_{ij}^{(k)}\}+B_k(\bs_j;l)\,\overline{T}_{\gamma+1}\{a_{ji}^{(k)}\}\Big]$\par
\quad where $a_{ij}^{(k)}=\sqrt{\gamma+1}\,\frac{B_k(\bs_i;l)^{1/\gamma}B_k(\bs_j;l)^{-1/\gamma}-\rho(\bs_i,\bs_j)}{\sqrt{1-\rho(\bs_i,\bs_j)^2}}$, and
$\overline{T}_{\text{df}}$ denotes\par
\quad Student's $t$ survival function with df degrees of freedom\par
$\eta_{ij} = \begin{cases}
1, & \text{if } \mathcal{K}_i \cap \mathcal{K}_j \not=\emptyset,\\
1/2, & \text{if } \mathcal{K}_i \cap \mathcal{K}_j =\emptyset.
\end{cases}$
\\

\cmidrule(lr){2-3}

\ref{Model-Shi}
& $\alpha^*_j=\min\{\frac{\alpha}{\phi_j},\,1\}$
& $\chi_{ij} = \begin{cases}
\mathbb{E}\!\left[
\min\!\left\{
\frac{W_i^{\alpha/\phi_i}}{\mathbb{E}\!\left[W_i^{\alpha/\phi_i}\right]},
\frac{W_j^{\alpha/\phi_j}}{\mathbb{E}\!\left[W_j^{\alpha/\phi_j}\right]}
\right\}\right]
\displaystyle\sum_{k \in \mathcal{K}_i \cap \mathcal{K}_j} v_{k,\wedge},
& \text{if } \alpha <\phi_i<\phi_j, \\
0 & \text{otherwise},
\end{cases}$
\par\quad  where $v_{k,\land} = \min\{v_{ki},v_{kj}\}$ with $v_{ki}$ and $v_{kj}$ defined in~\autoref{eqn:v_ki}.
\par
If $\mathcal{K}_i\cap\mathcal{K}_j\neq\emptyset$, \par
\quad $\begin{cases}
\eta_{ij} = 1 & \text{if } \alpha<\phi_i<\phi_j, \\
\eta_{ij} \in[\max(\LZ{\eta_{ij}^Z},\frac{\phi_i}{\alpha}),\max(\LZ{\eta_{ij}^Z},\frac{\phi_j}{\alpha})]\;
&\text{if } \phi_i<\phi_j<\alpha, \\
1/\eta_{ij}\in\Big[\frac{\min\{\phi_i+\phi_j,2\alpha\LZ{\eta_{ij}^Z}\}-\phi_j+\alpha}{2\alpha\LZ{\eta_{ij}^Z}},\;2-\frac{\phi_i}{\alpha}\Big] & \text{if } \phi_i<\alpha<\phi_j.
\end{cases}$
\par
If $\mathcal{K}_i\cap\mathcal{K}_j=\emptyset$, \par
\quad $\begin{cases}
\eta_{ij} \in \Big[\max(\frac{1}{2}, \frac{\alpha\LZ{\eta_{ij}^Z}}{\phi_j}),\max(\frac{1}{2}, \frac{\alpha}{\phi_i})\Big] & \text{if } \alpha < \phi_i<\phi_j,\\
\eta_{ij} \in \Big[\LZ{\eta_{ij}^Z}, \max(\LZ{\eta_{ij}^Z}, \frac{\phi_j}{\alpha})\Big] & \text{if } \phi_i<\phi_j<\alpha, \\
1/\eta_{ij}\in\big[\frac{\min\{\phi_j,2\alpha\LZ{\eta_{ij}^Z}\}+\alpha}{2\alpha\LZ{\eta_{ij}^Z}},\,2\big] & \text{if } \phi_i<\alpha<\phi_j.
\end{cases}$
\\
\bottomrule
\end{tabular}
\endgroup
\end{adjustwidth}
\end{table}

Despite their modelling appeal, these \MS{transformed } Gaussian scale-mixture constructions pose substantial computational challenges under the peaks-over-threshold framework.
Likelihood-based inference for multivariate threshold exceedances is typically based on a censored likelihood, and for \BAS{transformed } Gaussian scale-mixture models 
\BAS{including} ~\ref{Model-HW}, \ref{Model-Hazra}, and \ref{Model-Shi}, this likelihood requires repeated evaluation of high-dimensional Gaussian distribution functions. This quickly becomes infeasible once the number of spatial locations is just moderately large. \BAS{For Model~\ref{Model-Reetam}, the joint likelihood is not available in closed form even in low dimensions, and \citet{majumder2024modeling} therefore relied on a combination of Vecchia-type approximations and neural-network emulators for approximate Bayesian inference. }
These computational barriers motivate the scalable approach developed in the next sections. We first review the censored likelihood formulation and identify the key bottleneck in \autoref{sec:censored_likelihood}.

\subsection{The Censored Likelihood}\label{sec:censored_likelihood}


In multivariate POT, inference is driven by exceedances above a high threshold, but the data also contain many sub-threshold observations. \BAS{Treating observations below the threshold as censored provides an effective compromise between, on one hand, treating them as fully observed and thereby allowing them to bias estimates of properties specific to the tail, and on the other hand, removing them entirely and ignoring any information they provide \citep{zhang2021hierarchical,huser2016non}. } 
\BAS{Specifically, consider } observations at $D$ locations $\bs_1,\ldots,\bs_D$, the joint distribution of $\boldsymbol{X^*}$ defined in \autoref{eqn:base_model} can be obtained by conditioning on $R$ as
\begin{equation}\label{eqn:joint_dist_base_model}
F_{\bX^*}(\bx^*) =\int_{\mathcal{R}}\Phi_D\!\left\{g_{\phi}^{-1}\!\left(\bx^*/r^{\phi}\right);\,\bSigma_{\rho}\right\} f_R(r) \mathrm{d}r,
\end{equation}
where $\Phi_D$ denotes the $D$-variate Gaussian CDF with mean $\mathbf{0}$ and covariance  $\boldsymbol{\Sigma}_{\boldsymbol{\rho}}$.

Let $\mathcal{E}$ index exceedances and $\mathcal{C}$ censored components, and partition $\bSigma_{\boldsymbol{\rho}}$ accordingly. The joint censored likelihood, obtained by differentiating \autoref{eqn:joint_dist_base_model} with respect to $\bx^*_{\mathcal{E}}$, is
\begin{equation}\label{eqn:joint_lik_base_model}
\begin{split}
L(\bx^*)=\int_{\mathcal{R}}
&\Phi_{|\mathcal{C}|}\!\left(\boldsymbol{z}_{\mathcal{C}}-\bSigma_{\mathcal{CE}}\bSigma_{\mathcal{EE}}^{-1}\boldsymbol{z}_{\mathcal{E}};\;\bSigma_{\mathcal{C}\mid\mathcal{E}}\right)\,
\phi_{|\mathcal{E}|}\!\left(\boldsymbol{z}_{\mathcal{E}};\bSigma_{\mathcal{EE}}\right) \\
&\times\left\{\prod_{i\in\mathcal{E}}\left(g_{\phi}^{-1}\right)'\!\left(x_i^*/r^{\phi}\right)\right\}
\, r^{-\phi|\mathcal{E}|}\, f_R(r)\,\mathrm{d}r,
\end{split}
\end{equation}
where $\boldsymbol{z}_{\mathcal{A}}=g_{\phi}^{-1}(\bx^*_{\mathcal{A}}/r^{\phi})$
and $\bSigma_{\mathcal{C}\mid\mathcal{E}}=\bSigma_{\mathcal{CC}}-\bSigma_{\mathcal{CE}}\bSigma_{\mathcal{EE}}^{-1}\bSigma_{\mathcal{EC}}$.

The key difficulty in evaluating \autoref{eqn:joint_lik_base_model} is the term $\Phi_{|\mathcal{C}|}(\cdot)$, a $|\mathcal{C}|$-dimensional Gaussian distribution function. As a result, for each time replicate, the likelihood evaluation requires repeated computation of Gaussian distribution functions in dimensions that quickly become prohibitive once $|\mathcal{C}|$ reaches even the low tens \citep{Huser_Opitz_Wadsworth_2025, zhang2021hierarchical}.
The burden is further exacerbated for models with spatially varying scaling mechanisms (e.g., $R(\bs)$ and/or $\phi(\bs)$ in Models~\ref{Model-Reetam}, \ref{Model-Hazra} and \ref{Model-Shi}), which introduce additional latent structure that must be repeatedly integrated or sampled within each likelihood evaluation.

This Gaussian CDF bottleneck is a major obstacle that motivates more efficient fitting of flexible Gaussian scale-mixture models to high-dimensional threshold exceedance data.  \BAS{\citet{zhang2021hierarchical} circumvent this problem by supplementing $X^*(\bs)$ in \autoref{eqn:base_model} with an additive Gaussian nugget, as $X(\bs) = X^*(\bs) + \varepsilon(\bs_i)$, with $\varepsilon(\bs_i) \iid \text{N}(0, \sigma^2)$ for $\bs_1, \ldots, \bs_D$. Then, while the observations themselves are  left-censored below the threshold, the (now-latent) smooth process $X^*(\bs)$ is not.  The resulting likelihood treats the highly-multivariate $\{X^*(\bs)\}_{i=1}^D$ as uncensored while, conditional on $\{X^*(\bs)\}_{i=1}^D$, the univariate nugget terms $\{\varepsilon(\bs)\}_{i=1}^D$ are censored. Consequently, the full likelihood requires multivariate Gaussian densities but only univariate CDFs. This eliminates the intractable CDF calculation but the addition operation introduces a \textit{convolution} which is unavailable in closed form and must be computed numerically.  For models like \ref{Model-Hazra}--\ref{Model-Shi} with spatially varying parameters, the numerical integral must be computed at every observation location for each MCMC iteration.  This renders model-fitting infeasible for more complicated models, even though the intractable high-dimensional Gaussian CDF is bypassed.}

In the following sections, we introduce a \emph{multiplicative} log-Laplace nugget that \BAS{similarly } yields conditional independence in the censored likelihood. This \BAS{also } eliminates multivariate Gaussian distribution function evaluations\BAS{, but unlike the additive nugget of \citet{zhang2021hierarchical}, it gives closed-form marginal density and distribution functions, eliminating the need for any numerical integration. This } enables scalable Bayesian inference for high-dimensional threshold exceedance data while\BAS{, as we show below, } preserving the extremal dependence \BAS{properties } of the underlying smooth process.

\section{Model} \label{sec:Model}

\subsection{Construction} \label{sec:Construction}

Let $X^*(\bs)$ denote the latent smooth process that captures spatial extremal dependence. We define the \BAS{modified model as } 
\begin{equation}\label{eqn:noisy_model}
X(\bs_i)=\epsilon(\bs_i)\,X^*(\bs_i) \, \BAS{ = \epsilon(\bs_i)\cdot R(\bs_i)^{\phi(\bs_i)}  g_{\phi}\{Z(\bs_i)\}},\qquad i=1,\ldots,D,
\end{equation}
where the nugget terms $\epsilon(\bs_i)$ are independent and identically distributed $\text{log-Laplace}(0,1/\alpha_0)$ \BAS{random variables }across sites, concentrated around 1.


\subsection{Computational Implications} \label{sec:Computational_Implications}

\subsubsection{Conditional Independence}

Under the nuggeted model \autoref{eqn:noisy_model}, the components of $\bX=(X_1,\ldots,X_D)^{\trans}$ are independent  across sites\BAS{, conditional on the latent smooth process values $\bX^*=(X_1^*,\ldots,X_D^*)^{\trans}$, } because $X_i=\epsilon_i X_i^*$ with independent nugget variables $\{\epsilon_i\}_{i=1}^D$. This conditional independence \BAS{dramatically simplifies the censored likelihood. } Specifically, for each component $i$,
\begin{equation*}
\Pr\{X_i\le x_{0i}\mid X_i^*\} = F_{\epsilon}\!\left(\frac{x_{0i}}{X_i^*}\right),
\qquad
f_{X\mid X^*}(x_i\mid X_i^*) = f_{\epsilon}\!\left(\frac{x_i}{X_i^*}\right)\frac{1}{X_i^*},
\end{equation*}
where $x_{0i}$ denotes the censoring threshold on the $X$-scale. 
Conditional on $\bX^*$ and other model parameters, the joint censored likelihood factorises into a product of one-dimensional terms,
\begin{equation*}
\prod_{i\in\mathcal{C}}F_{\epsilon}\!\left(\frac{x_{0i}}{X_i^*}\right)
\;\times\;
\prod_{i\in\mathcal{E}} f_{\epsilon}\!\left(\frac{x_i}{X_i^*}\right)\frac{1}{X_i^*}.
\end{equation*}

As a result, the nuggeted construction eliminates the need for repeated evaluation of the multivariate Gaussian distribution function $\Phi_{|\mathcal{C}|}(\cdot)$ in censored likelihood computation.
The remaining computational cost is the marginal evaluations of the chosen scale-mixture construction under the nugget multiplication, which we address next.

\subsubsection{Marginal Tractability} \label{sec:marginal_tractability}

While conditional independence removes the high-dimensional Gaussian CDF bottleneck, it does not by itself guarantee that the resulting likelihood is numerically efficient. In particular, a nugget can eliminate multivariate dependence at the likelihood level while simultaneously making marginal evaluation more expensive\BAS{, as in \citet{zhang2021hierarchical}. This burden is especially severe in settings where marginal distributions vary across space, rendering computations required for model fitting infeasible. }

In contrast, the multiplicative log-Laplace nugget in \autoref{eqn:noisy_model} avoids introducing an additional \BAS{numerical integration } layer. For a broad subclass of \LZ{random scale-mixture } models, the nugget can be absorbed into the latent scale variable, so marginal evaluation retains the same functional form as the base model and often remains available in closed form. We summarize this property for Models~\ref{Model-HW}--\ref{Model-Shi} below.

\begin{enumerate}[label={}, labelsep=0.3cm]
\item[\ref{Model-HW}]
The base model $X^*(\bs)$ admits a closed-form marginal distribution,
\begin{equation*}\label{eqn:HW-CDF}
F_{X^*}(x) = 1 - \frac{\phi}{2\phi-1}x^{-1/\phi} + \frac{1-\phi}{2\phi-1}x^{-1/(1-\phi)}.
\end{equation*}
With the log-Laplace nugget, $X(\bs)$ also admits a closed-form marginal distribution,
\begin{align*}\label{eqn:HW-CDF-nugget}
F_X(x) &= 1 - \frac{\phi}{2\phi-1}\,x^{-1/\phi}\,A_{1/\phi}(x) + \frac{1-\phi}{2\phi-1}\,x^{-1/(1-\phi)}\,A_{1/(1-\phi)}(x) - \frac{1}{2}x^{-\alpha_0},
\end{align*}
where
\begin{equation*}
A_q(x) = \frac{\alpha_0}{2}\left(\frac{1}{q+\alpha_0} + \frac{x^{q-\alpha_0}-1}{q-\alpha_0}\right)\mathbbm{1}(\alpha_0\neq q) + \left(\frac{1}{4}+\frac{\alpha_0}{2}\log x\right)\mathbbm{1}(\alpha_0=q).
\end{equation*}

\item[\ref{Model-Reetam}]
After a monotone marginal transformation, the model of \citet{majumder2024modeling} can be written in the \BAS{marginally } equivalent scale-mixture form as Model~\ref{Model-HW}. Consequently, the nuggeted \BAS{version of Model \ref{Model-Reetam} }inherits the same closed-form marginal tractability \BAS{as Model~\ref{Model-HW}}.

\item[\ref{Model-Hazra}]

The model of \citet{hazra2021realistic} extends from the stationary construction of \citet{huser2017bridging}, in which the marginal distribution is \BAS{known only up to }a one-dimensional integral,
\begin{equation}\label{eqn:Hazra-base-CDF}
F_{X^*}(x)=\int_{0}^{\infty}\Phi(x/r)\,f_R(r)\,dr,
\end{equation}
where $\Phi(\cdot)$ denotes the standard Gaussian CDF.
Under the $\mathrm{Pareto}(\gamma)$ specification of $R$ used for analysis by \citet{hazra2021realistic}, incorporating the log-Laplace nugget retains the same one-dimensional integral complexity \BAS{required for } the marginal evaluation with no additional numerical integration layer introduced:
\begin{equation}\label{eqn:Hazra-base-CDF-nugget}
F_{X}(x)=\int_{0}^{\infty}\Phi(x/s)\,f_{\tilde{R}}(s)\,ds,
\end{equation}
where \MS{$\tilde{R} = \epsilon R$}, and \MS{$f_{\tilde{R}}$ } is available in closed form as
{\setstretch{1}
\setlength{\abovedisplayskip}{2pt}
\setlength{\belowdisplayskip}{2pt}
\setlength{\abovedisplayshortskip}{2pt}
\setlength{\belowdisplayshortskip}{2pt}
\small
\begin{equation*}
f_{\tilde{R}}(s) = \frac{\alpha_0 \gamma}{2}\, s^{-(\alpha_0 + 1)} \left[\frac{s^{\alpha_0 - \gamma}- 1}{\alpha_0 - \gamma}\,\mathbbm{1}(\alpha_0 \neq \gamma) + \log(s)\,\mathbbm{1}(\alpha_0 = \gamma) \right] + \frac{\alpha_0 \gamma}{2(\gamma + \alpha_0)}\, s^{-(\gamma + 1)}.
\end{equation*}
}
The model of \citet{hazra2021realistic} replaces the global $\{R(\bs)\}$ with a low-rank representation through deterministic weighted sum of finitely many latent scalars $R_k^*$ (see \autoref{sec:flexible_models}), so marginal evaluation at each site $\bs$ retains the one-dimensional \BAS{integral } form and the nugget does not introduce any additional numerical integration.

\item[\ref{Model-Shi}]
For the construction of \citet{shi2026}, using the standard-Pareto link $g(\cdot)$ as in \autoref{sec:flexible_models}, the smooth process $X^*(\bs)$ admits a closed-form marginal distribution expressed in terms of incomplete gamma functions:
\begin{equation*}
F_{X^*_j}(x)
= 1 - \sqrt{\frac{1}{\pi}}\,
\gamma\left(
\frac{1}{2},
\frac{\bar{\gamma}_{j}}{2x^{1/\phi_j}}
\right)
- x^{-1}\sqrt{\frac{1}{\pi}}
\left(\frac{\bar{\gamma}_{j}}{2}\right)^{\phi_j}
\Gamma\left(
\frac{1}{2} - \phi_j,
\frac{\bar{\gamma}_{j}}{2x^{1/\phi_j}}
\right),
\end{equation*}
where $\gamma(\cdot,\cdot)$ and $\Gamma(\cdot,\cdot)$ denote the lower and upper incomplete gamma functions.
Under the log-Laplace nugget, the marginal survival function of $X_j$ remains available in closed form, with marginal distribution function
\begin{align*}
F_{X_j}(x)
&= 1-\sqrt{\dfrac{1}{\pi}}\Bigg[\gamma\left(\tfrac{1}{2}, \,\lambda_j(x)\right)
    \;+\; \frac{\alpha_0^{2}}{\alpha_0^{2}-1}
        \left(\frac{\bar\gamma_j}{2}\right)^{\phi_j}
        \frac{1}{x}\,
        \Gamma\left(\frac{1}{2}-\phi_j, \,\lambda_j(x)\right) \\
&\qquad \;-\;\frac{1}{2(\alpha_0+1)}
        \,x^{\alpha_0}
        \left(\frac{\bar\gamma_j}{2}\right)^{-\phi_j\alpha_0}
        \gamma\left(\frac{1}{2}+\phi_j\alpha_0, \,\lambda_j(x)\right) \\
&\qquad \;-\;\frac{1}{2(\alpha_0-1)}
        \,x^{-\alpha_0}
        \left(\frac{\bar\gamma_j}{2}\right)^{\phi_j\alpha_0}
        \Gamma\left(\frac{1}{2}-\phi_j\alpha_0, \,\lambda_j(x)\right)\Bigg],
\end{align*}
where $\lambda_j(x)=\bar\gamma_j/(2x^{1/\phi_j})$.

\end{enumerate}

\subsection{Tail Properties} \label{sec:tail_properties}

\subsubsection{Marginal tail equivalence}

We first show that the nugget can be chosen to preserve the marginal tail of the latent smooth process. Specifically, it will have negligible impact on the tail as long as it is lighter-tailed than $X^*(\bs)$, in the sense that it possesses sufficiently high moments. \BAS{Because $X^*(\bs)$ will ultimately be re-scaled to have GPD margins during inference, the marginal tail properties that we derive here are interesting only insofar as they allow us to deduce the tail dependence properties that we describe in Section \ref{sec:joint_tail_dependence}.}

\begin{theorem}[Marginal Tail Equivalence]\label{thm:marg_tail_equiv}\upshape
Assume $X^*(\bs)$ has a regularly varying upper tail with index $\alpha^*(\bs)$; that is, for slowly varying $\mathcal{L}_{\bs}$,
\begin{equation}\label{eqn:rv_marg}
\Pr\{X^*(\bs)>x\}=\mathcal{L}_{\bs}(x)\,x^{-\alpha^*(\bs)},\qquad x\to\infty.
\end{equation}
If $\alpha_0>\sup_{\bs\in\mathcal{S}}\alpha^*(\bs)$, then $X(\bs)=\epsilon(\bs)X^*(\bs)$ is marginally tail equivalent to $X^*(\bs)$:
\[
\Pr\{X(\bs)>x\}\sim \mathbb{E}\{\epsilon(\bs)^{\alpha^*(\bs)}\}\,\Pr\{X^*(\bs)>x\},\qquad x\to\infty.
\]
\end{theorem}

\begin{corollary}[Marginal tail equivalence for models~\ref{Model-HW}--\ref{Model-Shi}, and sufficient conditions on $\alpha_0$]
\label{cor:marg_tail_equiv}\upshape
For each model in \autoref{sec:flexible_models}, the smooth process $X^*(\bs)$ has a regularly varying upper tail with index $\alpha^*(\bs)$ as in \autoref{eqn:rv_marg}. Therefore, \autoref{thm:marg_tail_equiv} yields marginal tail equivalence whenever $\alpha_0>\sup_{\bs\in\mathcal{S}}\alpha^*(\bs)$.
In particular:
\begin{enumerate}[label=(M\arabic*), labelsep=0.3cm]

\item[\ref{Model-HW}]
By the marginal Pareto construction, $X^*(\bs)$ has a regularly varying tail with index
\begin{equation*}
\alpha^* = \min\{1/\phi,1/(1-\phi)\},
\end{equation*}
Thus, it suffices to take $\alpha_0>2$.

\item[\ref{Model-Reetam}]
After a monotone marginal transformation, Model \ref{Model-Reetam} can be written in the equivalent form as \ref{Model-HW}, and therefore the same condition applies and it suffices to take $\alpha_0>2$.

\item[\ref{Model-Hazra}]
Under the $\mathrm{Pareto}(\gamma)$ specification of \citet{huser2017bridging}, \BAS{$X^*(\bs)$ } has a regularly varying upper tail with index $\alpha^*=\gamma$, \LZ{since } $\Pr\{RW>x\}=\tfrac12\Pr\{R|W|>x\}$ \LZ{given the } symmetry of $W$ and \LZ{\citet[Lemma]{breiman1965some}}. In \citet{hazra2021realistic}, \BAS{a } finite nonnegative weighted sum of the independent $\mathrm{RV}_\gamma$ variables \BAS{is } used to construct the spatially varying $R(\bs)$, and hence $X^*(\bs)$ retains tail index $\gamma$. Thus, it suffices to take $\alpha_0>\gamma$.

\item[\ref{Model-Shi}]
By construction and Proposition 1 of \citet{shi2026}, $X^*(\bs)$ has regularly varying upper tail with tail index
\[
\alpha^*(\bs)=\min\left[\frac{\alpha}{\phi(\bs)},1\right].
\]
Thus, it suffices to take $\alpha_0>1$.
\end{enumerate}
\end{corollary}

\subsubsection{Joint tail dependence}
\label{sec:joint_tail_dependence}

We next show that the multiplicative log-Laplace nugget preserves the extremal dependence \BAS{properties } of the underlying smooth process. Specifically, for any two sites $(\bs_i, \bs_j)$, we compare the joint upper-tail decay of $(X(\bs_i),X(\bs_j))$ and $(X^*(\bs_i),X^*(\bs_j))$ through the coefficients \MS{$(\chi_{ij}, \eta_{ij})$ and $(\chi_{ij}^*, \eta_{ij}^*)$, respectively, defined as in } \autoref{eqn:chi_measure} and \autoref{eqn:eta_measure}. Under mild regular variation conditions on the bivariate tail of the smooth process \MS{and the corresponding moment condition on the nugget}, the nugget leaves \MS{$\eta_{ij}$ } unchanged and modifies \MS{$\chi_{ij}$ } only up to multiplicative constants. We formalise this in \autoref{thm:joint_tail_equiv}, and then \LZ{validate } its conditions for Models~\ref{Model-HW}--\ref{Model-Shi} in a corollary, using \LZ{both } existing tail dependence results and a new proposition establishing the joint tail decay for Model~\ref{Model-Reetam}, which was previously only assessed numerically \LZ{in \citet{majumder2024modeling}}.

\begin{theorem}[Joint Tail Equivalence]\label{thm:joint_tail_equiv}\upshape
\MS{Consider a random scale-mixture model and suppose its underlying $\{Z(\bs)\}$ process is asymptotically independent but positively correlated (i.e., $\eta_{ij}^Z \in (1/2, 1)$). }
If \MS{$X_i^*$, $X^*_j$, and }$\min(X_i^*,X_j^*)$ all have regularly varying tails and satisfy \MS{$\alpha_0 \ge2\sup_{\bs\in\mathcal{S}}\alpha^*(\bs)$ } for any two sites  $\bs_i$ and $\bs_j$, then
\[
\eta_{ij}=\eta_{ij}^*
\quad \text{and} \quad
\chi_{ij}\in \big[c_{ij}\,\chi_{ij}^*,\,C_{ij}\,\chi_{ij}^{*}\big],
\]
where
\begin{equation}\label{eqn:c_ij}
    c_{ij} = \mathbb{E}\left[\min\left(\dfrac{\epsilon_i^{\alpha_i^*}}{\mathbb{E}[\epsilon_i^{\alpha_i^*}]}, \dfrac{\epsilon_j^{\alpha_j^*}}{\mathbb{E}[\epsilon_j^{\alpha_j^*}] }\right)\right]
\quad \text{and} \quad
C_{ij} = \mathbb{E}\left[\max\left(\dfrac{\epsilon_i^{\alpha_i^*}}{\mathbb{E}[\epsilon_i^{\alpha_i^*}]}, \dfrac{\epsilon_j^{\alpha_j^*}}{\mathbb{E}[\epsilon_j^{\alpha_j^*}] }\right)\right].
\end{equation}
Explicit expressions of the expectations in $c_{ij}$ and $C_{ij}$ are provided in Appendix \ref{sec:appendix_tail_equiv}.
\end{theorem}

\begin{proposition}[Joint Distribution of \ref{Model-Reetam}]\label{prop:pmm_joint_tail}\upshape

Consider the \MS{random } scale-mixture representation \ref{Model-Reetam}. 
For any two sites $\bs_i,\bs_j$, the joint exceedance probability is regularly varying in $x$; as $x\to\infty$,
\begingroup
\renewcommand{\arraystretch}{0.9}
\[
\Pr(X_i > x,\, X_j > x)=
\left\{
\begin{array}{@{}l@{\quad}l@{}}
L(x)\,x^{-1/\{(1-\phi)\eta_{ij}^Z\}},
& \eta_{ij}^Z \ge \dfrac{\phi}{1-\phi},\\
\mathbb{E}\!\bigl[\min(W_i,W_j)^{(1-\phi)/\phi}\bigr]\,
x^{-1/\phi}\{1+o(1)\},
& \eta_{ij}^Z < \dfrac{\phi}{1-\phi}.
\end{array}
\right.
\]
\endgroup
where $\LZ{\eta_{ij}^Z}$ is the residual tail dependence coefficient induced by the Gaussian copula \citep{ledford1996statistics}.
\end{proposition}

\begin{corollary}[Joint tail equivalence for models (M1)--(M4)]
\label{cor:joint_tail_equiv}\upshape
For each \MS{random } scale-mixture model in Section \ref{sec:flexible_models}\BAS{, assume the underlying  \MS{$Z(\bs)$ is asymptotically independent but positively correlated}.  Then \autoref{thm:joint_tail_equiv} applies. In particular:}


\begin{enumerate}[label=(M\arabic*), labelsep=0.3cm]
\item[\ref{Model-HW}]
The regularly varying tail of \BAS{$\min(X^*_i, X^*_j)$  } and the corresponding joint tail behavior is given in Proposition 1 and Corollary 1 of \citet{huser2019modeling}. 
\MS{Since $\sup_{\bs\in\mathcal S}\alpha^*(\bs)=2$, it suffices to take \MS{$\alpha_0 \ge 4$}. }
Thus the dependence classes and tail coefficients characterized in \citet{huser2019modeling} are preserved under the multiplicative log-Laplace nugget.

\item[\ref{Model-Reetam}]
Proposition~\ref{prop:pmm_joint_tail} of this paper establishes the regularly varying tail of \MS{$\min(X^*_i, X^*_j)$ } for Model~\ref{Model-Reetam}.
\MS{Since $\sup_{\bs\in\mathcal S}\alpha^*(\bs)=2$, it suffices to take \MS{$\alpha_0 \ge 4$}. } 

\item[\ref{Model-Hazra}]
\citet{hazra2021realistic} established the joint tail behavior in Proposition 3 of their main paper and the Appendix, showing that \BAS{$(X_i, X_j)$ } is $\mathrm{MRV}(\gamma, a(\cdot), \nu_Z(\cdot))$ with $a(t) = t^{1/\gamma}$ and $\nu_Z(\mathcal{A}_r := [r_1, \infty) \times [r_2, \infty)) = \sum_{k=1}^K\mathbb{E}\left[\min\left\{\frac{B_k(\bs_i; l)Z_+^\gamma(\bs_i)}{r_1^\gamma}, \frac{B_k(\bs_j; l)Z_+^\gamma(\bs_j)}{r_2^\gamma}\right\}\right]$. \MS{Since $\sup_{\bs\in\mathcal S}\alpha^*(\bs)=\gamma$, it suffices to take \MS{$\alpha_0 \ge 2\gamma$}. } Then, \autoref{thm:joint_tail_equiv} gives joint tail equivalence under the log-Laplace nugget multiplication.

\item[\ref{Model-Shi}]
\citet{shi2026} established the regular variation of \MS{$\min(X^*_i, X^*_j)$ } in Theorem 3 of their paper and Section A.4 of the Appendix. \MS{Since $\sup_{\bs\in\mathcal S}\alpha^*(\bs)=1$, it suffices to take \MS{$\alpha_0 \ge 2$}. } Hence, \autoref{thm:joint_tail_equiv} leads to the same conclusion of joint tail equivalence.
\end{enumerate}
\end{corollary}

\subsection{Numerical Illustration}

\BAS{W}e illustrate the theoretical results of \autoref{thm:joint_tail_equiv} \BAS{in the context of the most flexible available model, \ref{Model-Shi}, by providing } \autoref{cor:Shi_joint_tail} and an empirical example in \autoref{ill:Shi_joint_tail}. 
Theorem~\ref{thm:joint_tail_equiv} and \autoref{cor:joint_tail_equiv} imply that the multiplicative log-Laplace nugget leaves the residual tail dependence coefficient unchanged, i.e., $\eta_{ij}=\eta_{ij}^*$ \BAS{and } modifies $\chi_{ij}$ by multiplicative constants. \BAS{Below, } we summarize the expression for $\chi_{ij}$, \BAS{and } refer to \autoref{tab:tail_summary_models} and Theorem 3 of \citet{shi2026} for the details of $\eta_{ij}$.

\begin{corollary}\label{cor:Shi_joint_tail}
\upshape
Under the model of \citet{shi2026} in \ref{Model-Shi}, consider two locations $\bs_i$ and $\bs_j$.
Let $\mathcal{K}_i:={k:,B_k(\bs_i;l)>0}$ and $\mathcal{K}_j:={k:,B_k(\bs_j;l)>0}$ denote the sets of basis indices whose compactly supported kernels overlap $\bs_i$ and $\bs_j$, respectively. Define
\begin{equation}\label{eqn:v_ki}
v_{ki}=\frac{\{{B_k(\bs_i;l)\gamma_k}\}^\alpha}{\sum_{k'\in \mathcal{K}i}{\{B_{k'}(\bs_i;l)\gamma_{k'}}\}^\alpha},\qquad
\alpha_i^* = \min\left(\frac{\alpha}{\phi_i}, 1\right).
\end{equation}
\BAS{Then }
\begin{enumerate}[label=(\alph*), labelsep=0.3cm]
\item\label{cor:Shi_joint_tail_AD}
If $\mathcal{K}_i\cap \mathcal{K}_j\neq \emptyset$ and $\alpha<\phi_i<\phi_j$, then $(X_i,X_j)^{\trans}$ is asymptotically dependent with $\eta_{ij}=1$ and
\begin{equation*}
\chi_{ij}\in\Big[\LZ{c_{ij}}\,\chi_{ij}^*,\; \LZ{C_{ij}}\,\chi_{ij}^*\Big],
\end{equation*}
where the nugget constants \LZ{$c_{ij}$ } and \LZ{$C_{ij}$ } are \LZ{defined in~\autoref{eqn:c_ij} }
and
\begin{equation*}
\chi_{ij}^*=\mathbb{E}\left[\min\left\{\frac{W_i^{\alpha/\phi_i}}{\mathbb{E}[W_i^{\alpha/\phi_i}]},\frac{W_j^{\alpha/\phi_j}}{\mathbb{E}[W_j^{\alpha/\phi_j}]}\right\}\right]\sum_{k=1}^K \BAS{\min(v_{ki},v_{kj})}.
\end{equation*}
where $W_i=g(Z_i)$.

\item\label{cor:Shi_joint_tail_AI}
Otherwise, $(X_i,X_j)$ is asymptotically independent with $\chi_{ij} = 0$\BAS{, and $\eta_{ij}$ is inhereted from $X^*(\bs)$, as shown in } \autoref{tab:tail_summary_models} and Theorem 3 of \citet{shi2026}.
\end{enumerate}
\end{corollary}

\begin{illustration}[Empirical evaluations of the bounds in \autoref{cor:Shi_joint_tail}]\label{ill:Shi_joint_tail}
\upshape
We simulate $N=300{,}000{,}000$ \BAS{draws from Model \ref{Model-Shi} } on the domain $\mathcal{S}=[0,10]\times [0,10]$ using a $\{\phi(\bs):\;\bs\in\mathcal{S}\}$ surface \BAS{as shown in } Figure \ref{fig:phi_surface_demon}. For each simulation, we generate independent Stable variables with $\alpha = 0.5$ at 9 knots on a uniform grid and \BAS{combine } them using Wendland basis functions centered at the knots \citep{Wendland1995}. We generate independent log-Laplace nugget terms with scale \BAS{parameters } $\alpha_0 = 2.0, 5.0$, and $10.0$.
We \BAS{then } empirically estimate the $\chi_{ij}(u)$ and $\eta_{ij}(u)$ functions 
using the $N$ independent simulations of $(X_i,X_j)$.

\begin{figure}[!h]
    \centering
  \begin{minipage}[c]{0.45\textwidth}
  \centering
    \includegraphics[width=\textwidth]{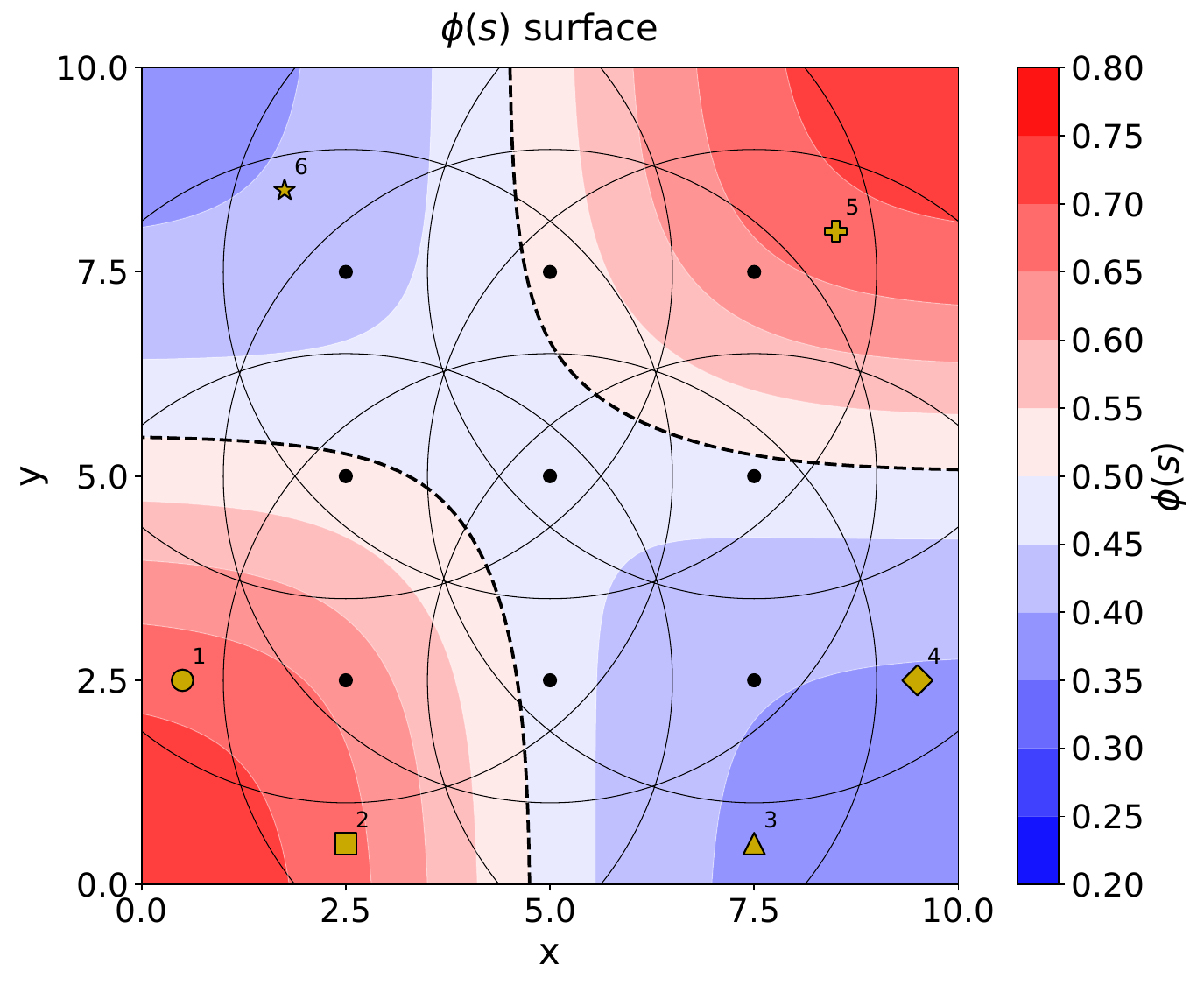}
  \end{minipage}
  \hspace{5mm}
  \begin{minipage}[c]{0.45\textwidth}
    \caption{A $\phi(\bs)$ surface on $[0,10]^2$, in which the dashed line marks the transition between local AI and AD. The points with `$\tikzcircle{1.5pt}$' are centers for the Wendland basis functions. The points with other signs/marker-styles are chosen sample points that we use to illustrate the dependence properties in \autoref{cor:Shi_joint_tail}.}
    \label{fig:phi_surface_demon}
  \end{minipage}
\end{figure}

\autoref{fig:emp_chi_example} shows \BAS{that } empirical estimates of $\chi_{ij}$ and $\eta_{ij}$ fall within the theoretical bounds, for various values of the log-Laplace scale \BAS{parameter } $\alpha_0$.
Figure~\ref{fig:emp_chi_example_a} considers sample points 1 and 2, which share a common Wendland kernel and satisfy $\alpha<\phi_2<\phi_1$, so the pair is asymptotically dependent. Figure~\ref{fig:emp_chi_example_b} considers points 3 and 4, which also share a kernel but satisfy $\phi_3<\phi_4<\alpha$, so the pair is asymptotically independent. Figure~\ref{fig:emp_chi_example_c} considers points 4 and 5, which share a kernel yet satisfy $\phi_4<\alpha<\phi_5$, and therefore remain asymptotically independent. Finally, Figure~\ref{fig:emp_chi_example_d} considers points 1 and 5, which do not share any common Wendland kernel; consistent with the long-range case, the pair is asymptotically independent even though $\phi_5>\phi_1>\alpha$. 

\begin{figure}[htbp]
    \centering
    \subfloat[\autoref{cor:Shi_joint_tail} \ref{cor:Shi_joint_tail_AD} Asymptotic Dependence: $\chi_{12}$ and $\eta_{12}$\label{fig:emp_chi_example_a}]{%
        \parbox{\textwidth}{\centering
            \includegraphics[width=0.485\textwidth]{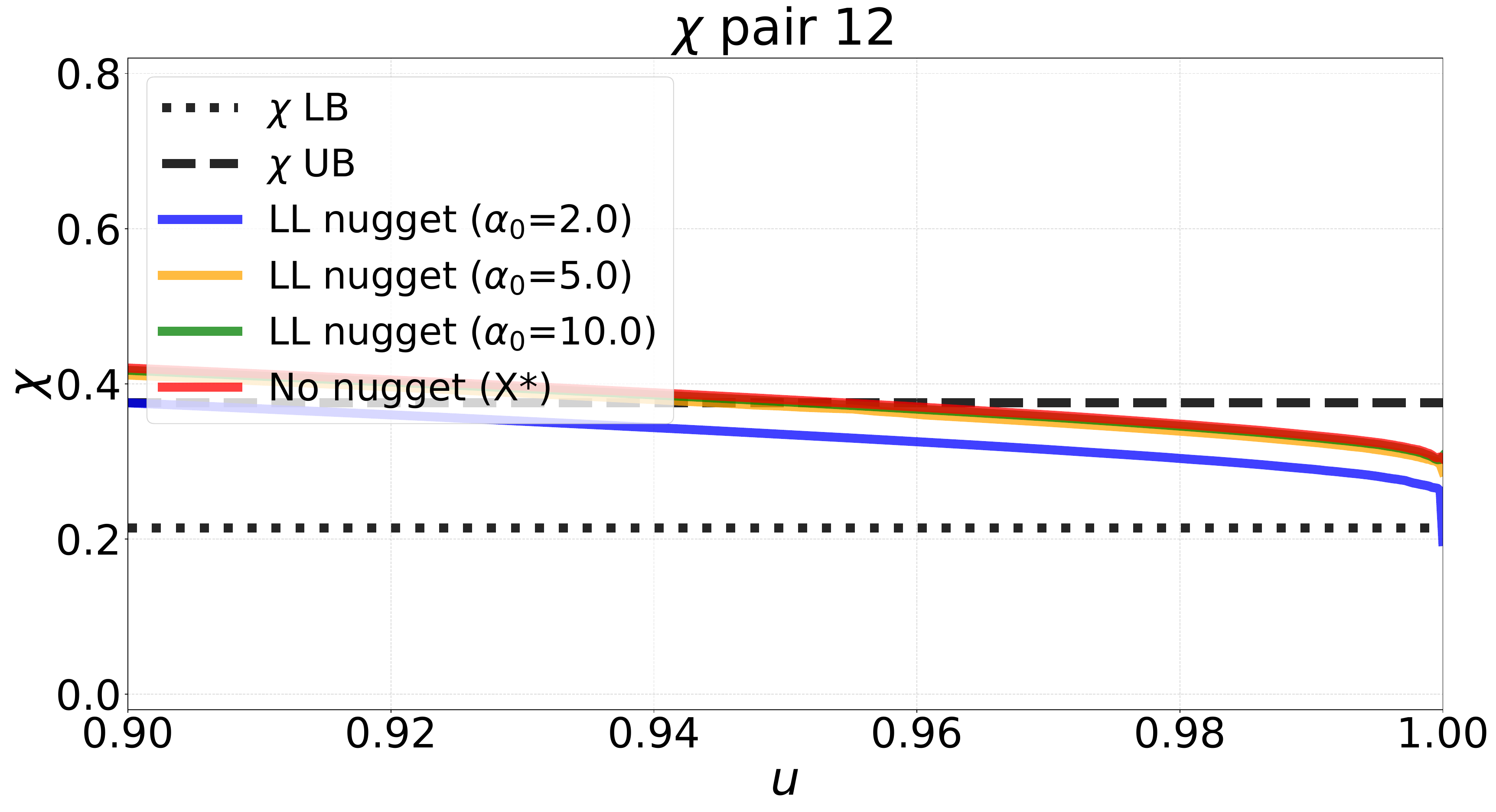}%
            \includegraphics[width=0.485\textwidth]{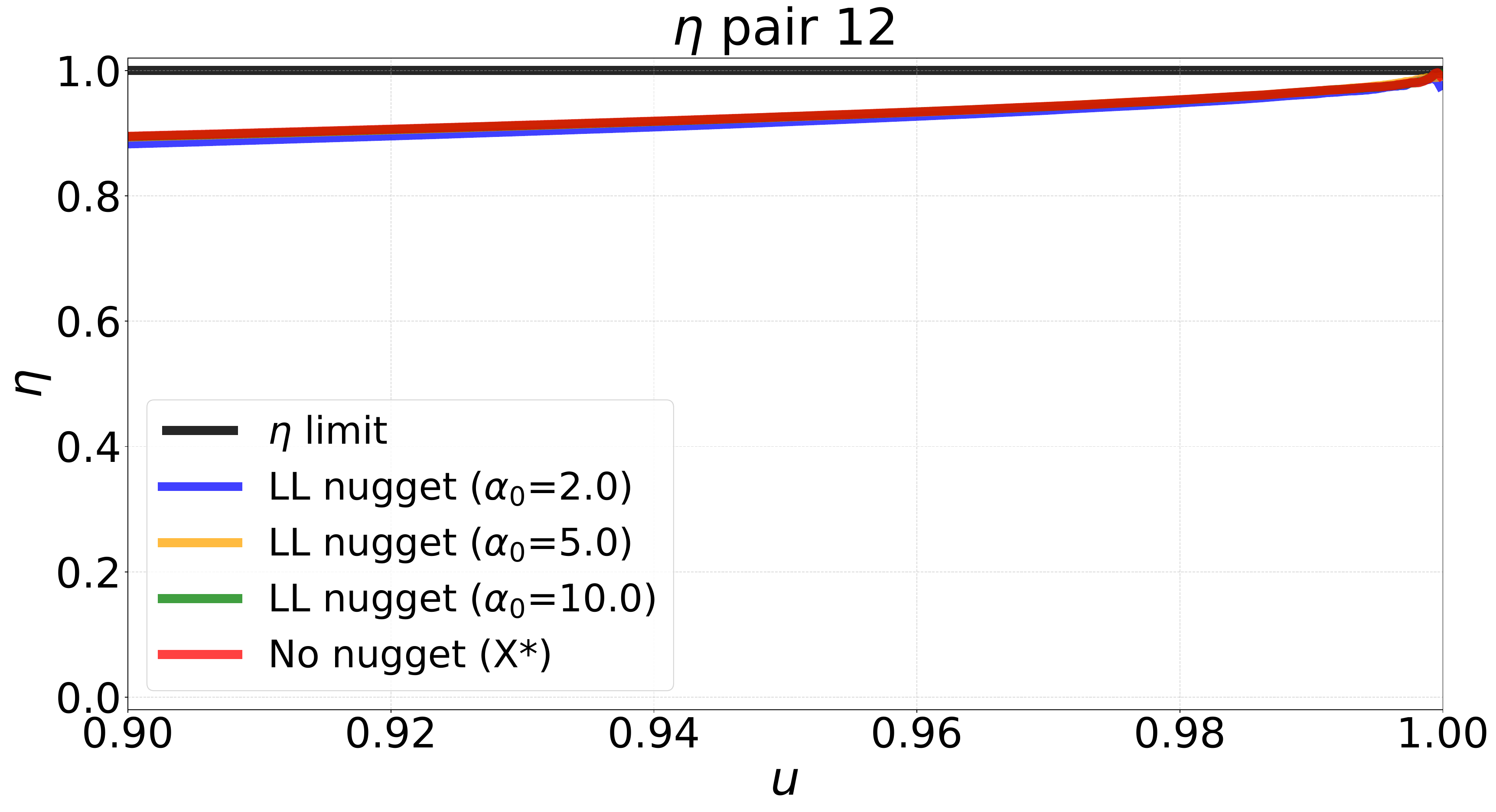}%
        }%
    }\par
    \subfloat[\autoref{cor:Shi_joint_tail} \ref{cor:Shi_joint_tail_AI} Asymptotic independence: $\chi_{34}$ and $\eta_{34}$\label{fig:emp_chi_example_b}]{%
        \parbox{\textwidth}{\centering
            \includegraphics[width=0.485\textwidth]{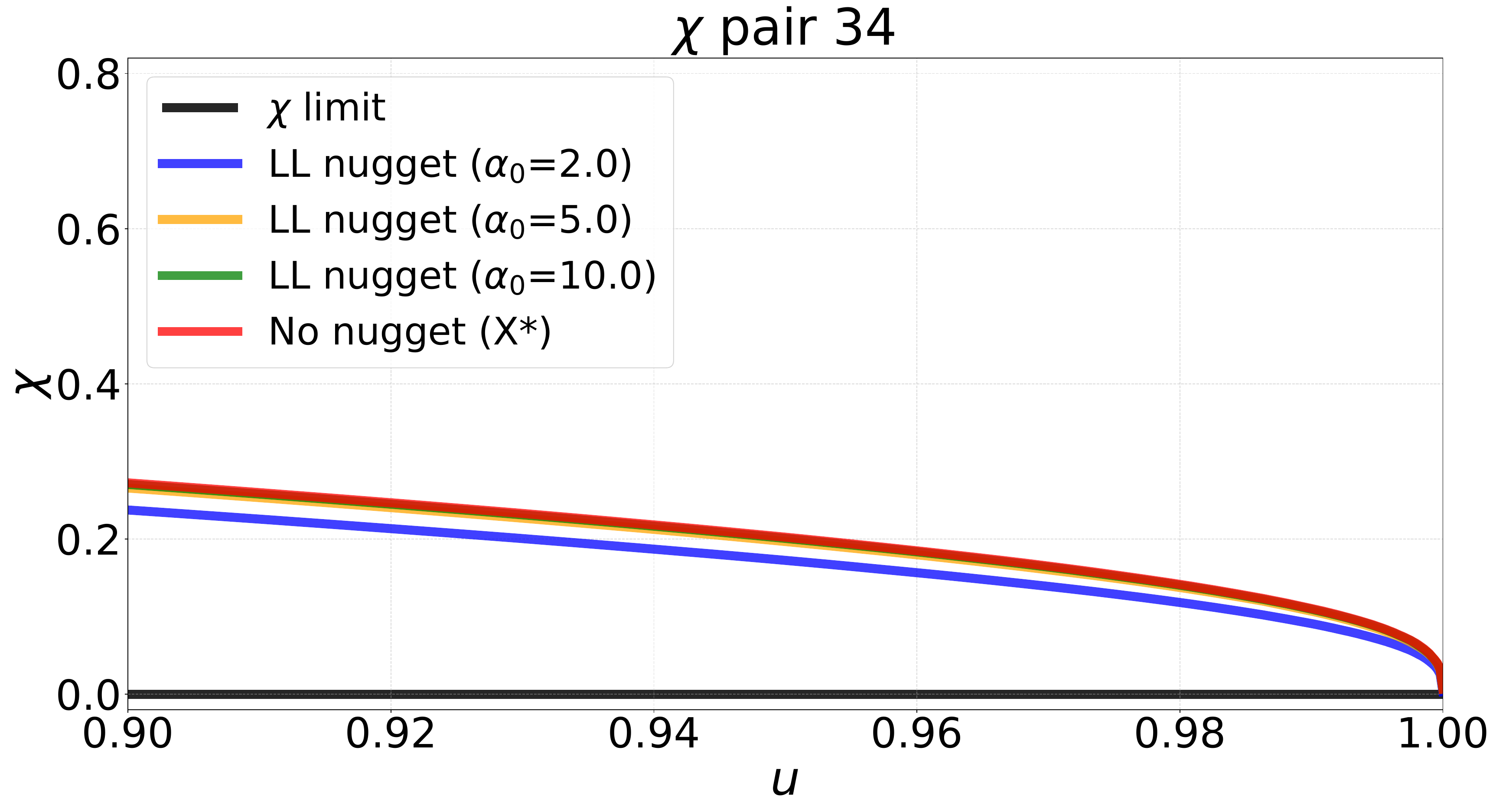}%
            \includegraphics[width=0.485\textwidth]{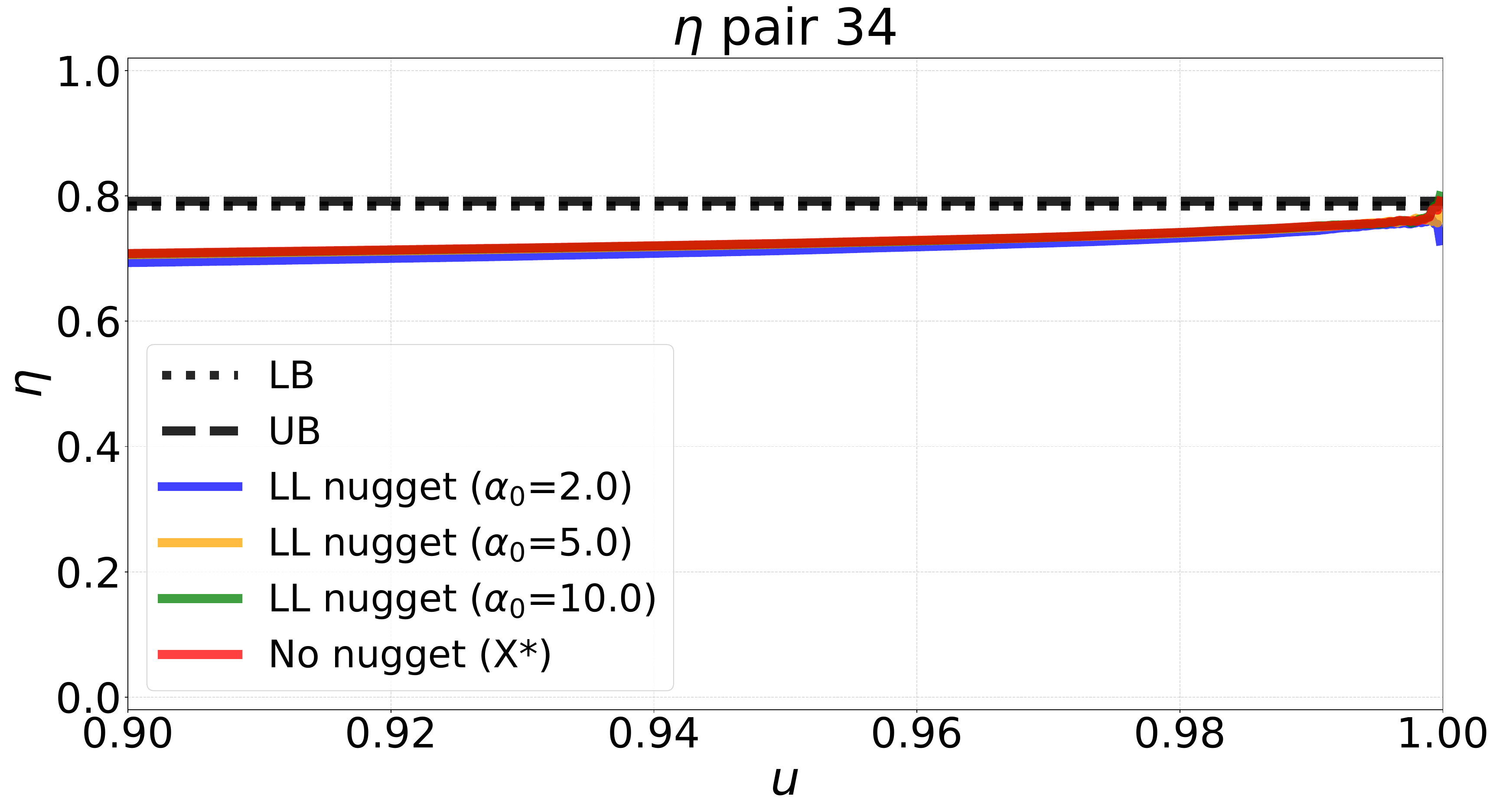}%
        }%
    }\par
    \subfloat[\autoref{cor:Shi_joint_tail} \ref{cor:Shi_joint_tail_AI} Asymptotic independence: $\chi_{45}$ and $\eta_{45}$\label{fig:emp_chi_example_c}]{%
        \parbox{\textwidth}{\centering
            \includegraphics[width=0.485\textwidth]{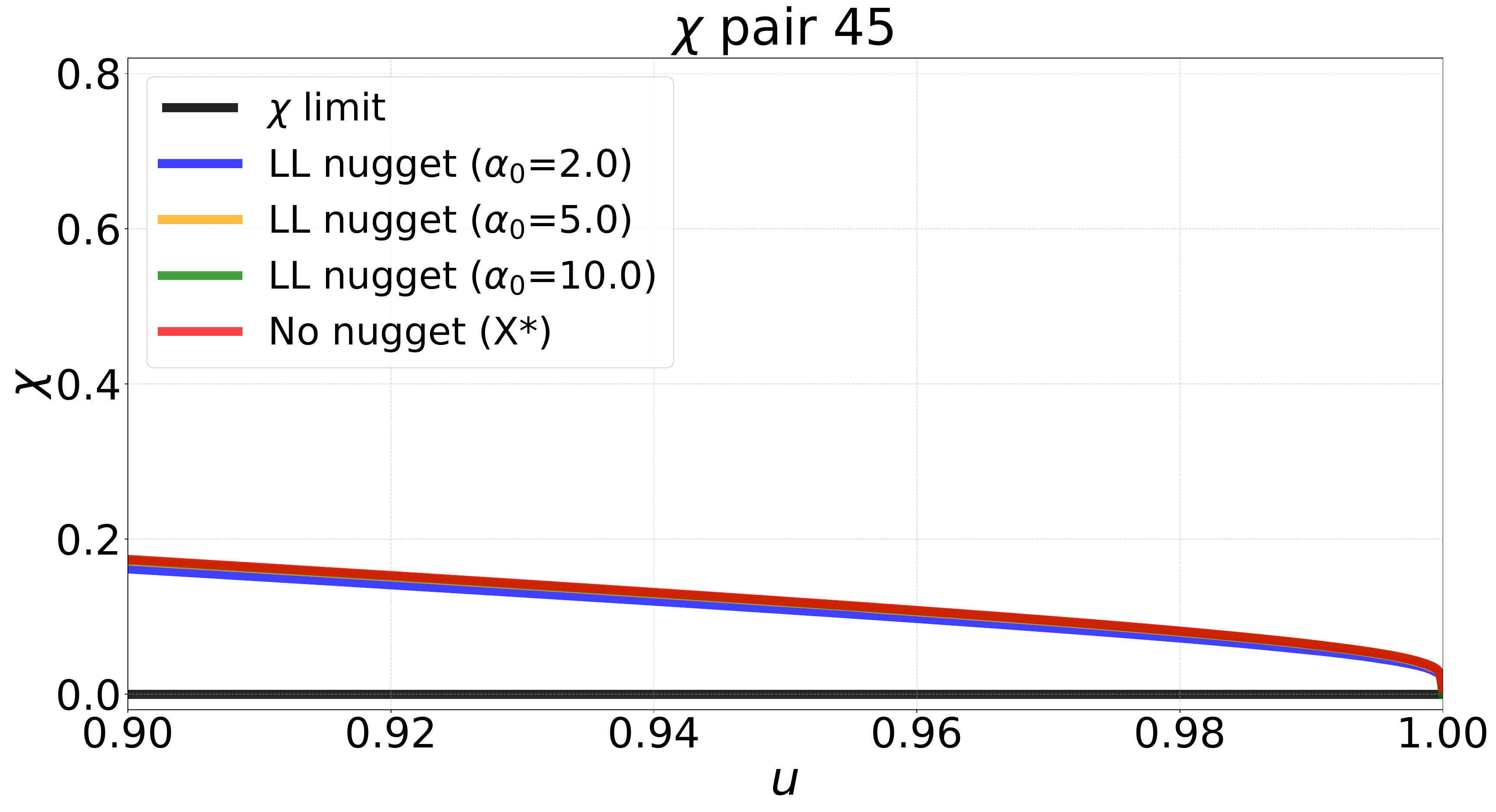}%
            \includegraphics[width=0.485\textwidth]{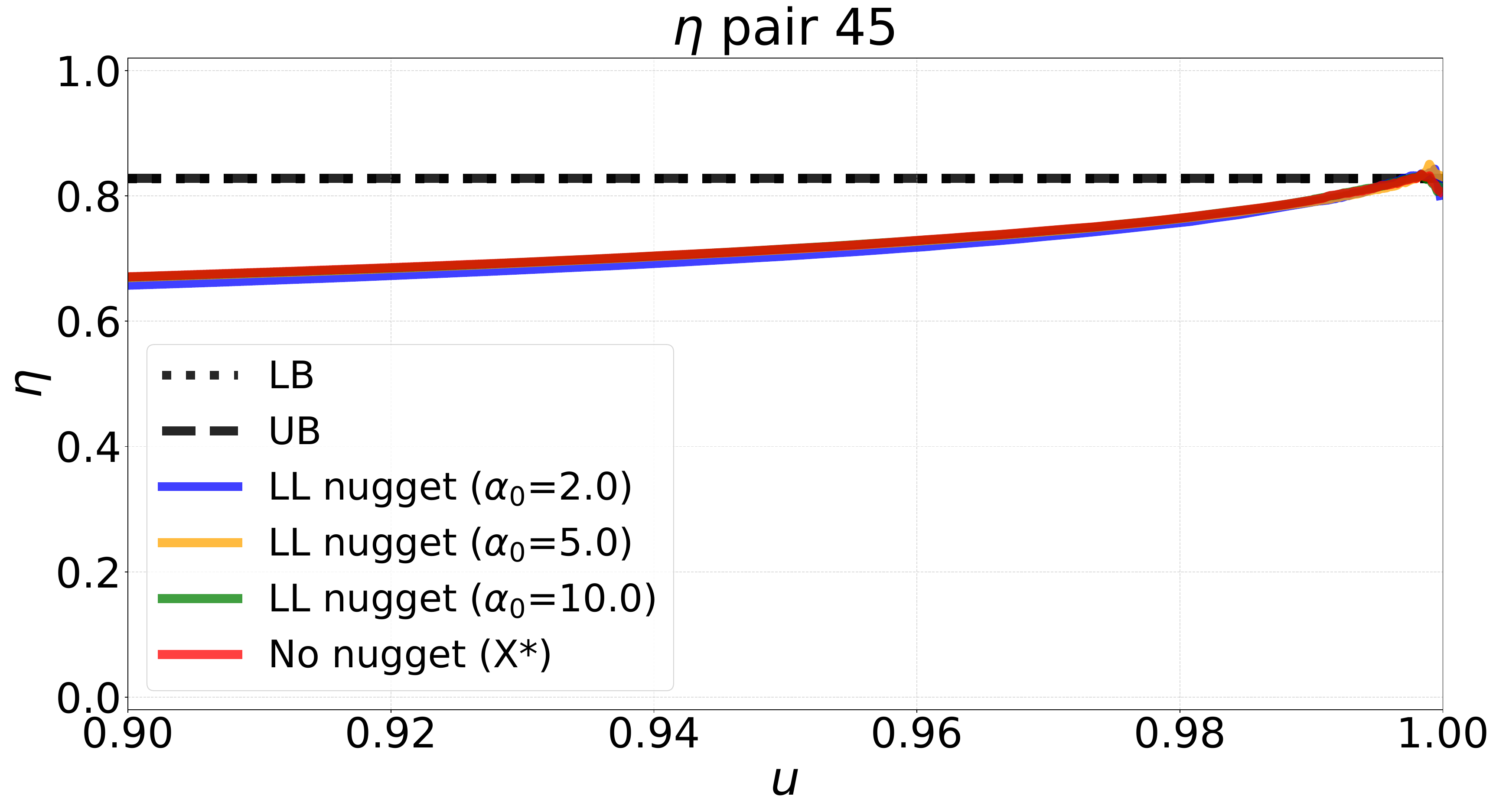}%
        }%
    }\par
    \subfloat[\autoref{cor:Shi_joint_tail} \ref{cor:Shi_joint_tail_AI} Asymptotic independence: $\chi_{15}$ and $\eta_{15}$\label{fig:emp_chi_example_d}]{%
        \parbox{\textwidth}{\centering
            \includegraphics[width=0.485\textwidth]{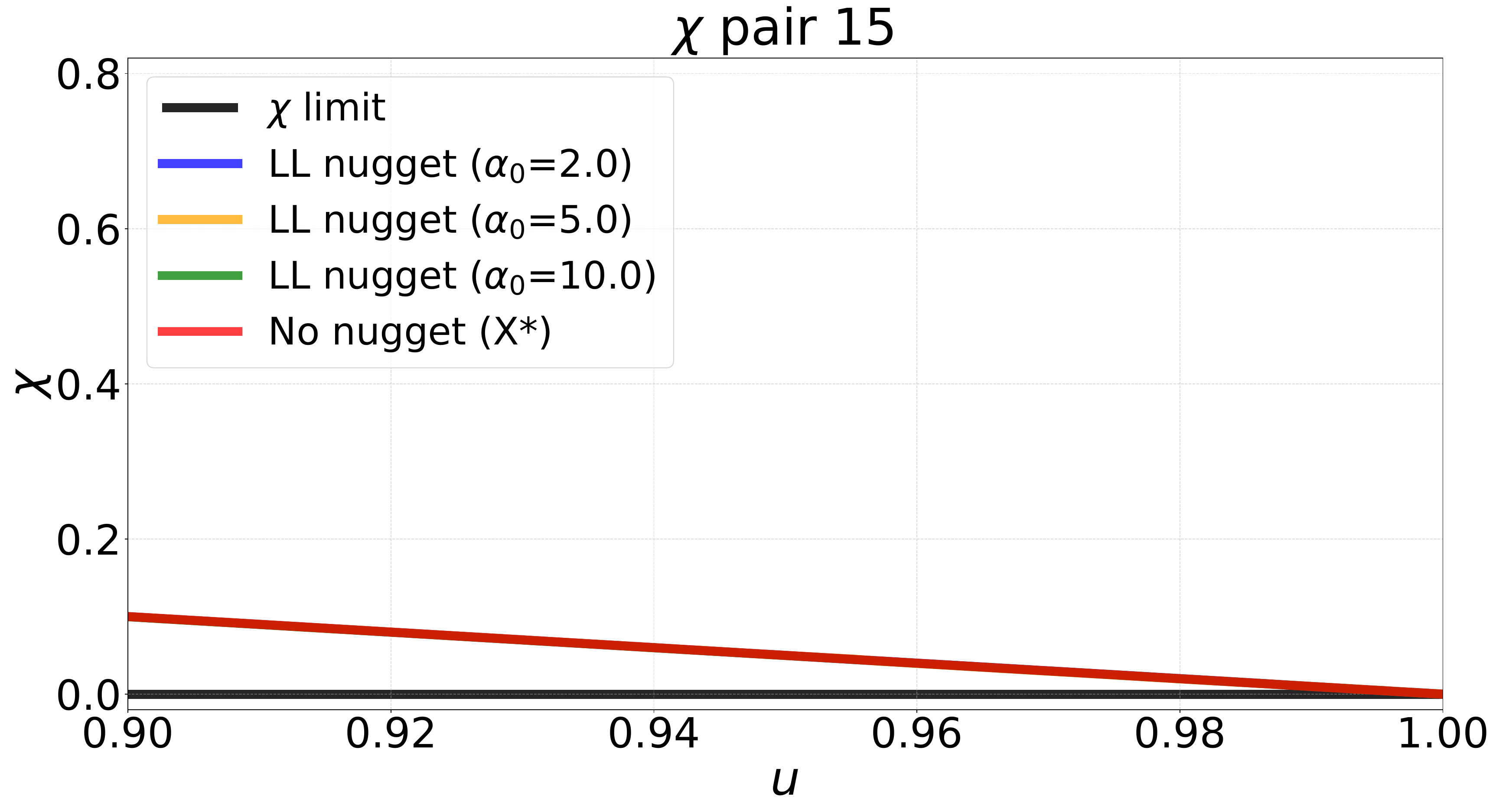}%
            \includegraphics[width=0.485\textwidth]{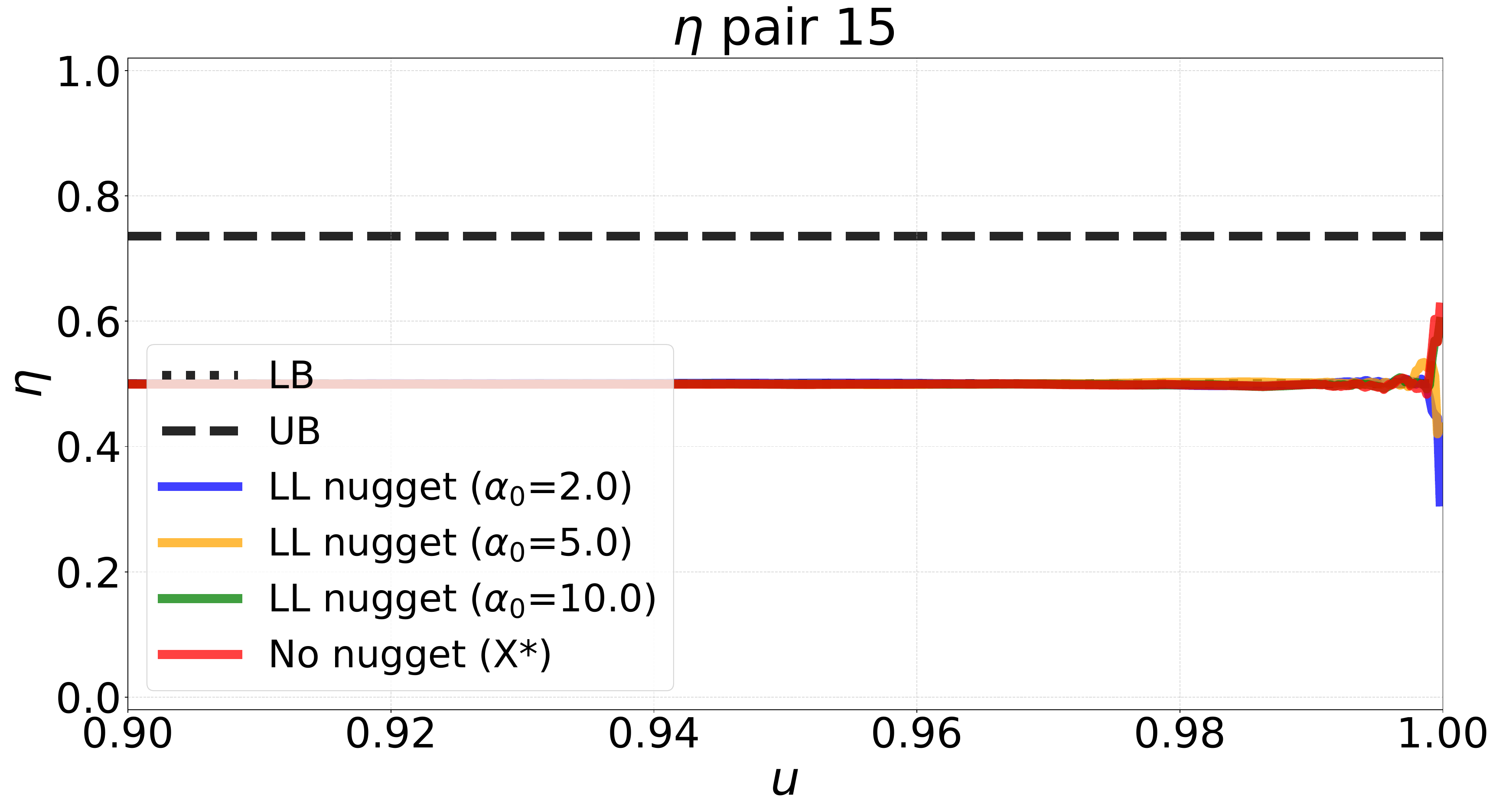}%
        }%
    }
    \caption{Empirical estimates of the dependence coefficients $\chi_{ij}(u)$ and $\eta_{ij}(u)$ \citep{R_mev} between sample points $i$ and $j$ in Figure \ref{fig:phi_surface_demon}. The black solid, dotted, and dashed lines mark the theoretical limit, upper, and lower bounds, respectively. 
    }
    \label{fig:emp_chi_example}
\end{figure}
\end{illustration}

\section{Bayesian Inference}\label{sec:Bayesian Inference}

We define a Bayesian hierarchical model based on Equation \autoref{eqn:noisy_model} and use an MCMC algorithm to fit to the data. The dependence model \autoref{eqn:noisy_model} is displayed again here for convenience, now with each time replicate denoted with a subscript $t = 1, \dots, T$:
\begin{equation*}
    X_t(\bs)=\epsilon_t(\bs) X_t^*(\bs) = \epsilon_t(\bs) \cdot R_t(\bs)^{\phi(\bs)}g(Z_t(\bs)).
\end{equation*}

\subsection{Hierarchical Model and Computation}\label{sec:hier_model}

We connect the dependence model in \autoref{eqn:noisy_model} to a \BAS{GPD } marginal specification through a probability integral transform.
For each replicate $t=1,\ldots,T$ and site $j=1,\ldots,D$, let $Y_{tj}$ denote
the \BAS{observations, which } are independent across $t$.
Fix \BAS{the exceedance probability } $p\in(0,1)$ and let $\MS{y_{0j}}$ denote a high threshold at site $j$. Define
\begingroup
\renewcommand{\arraystretch}{0.9}
\[
T(Y_{tj})=
\left\{
\begin{array}{@{}l@{\quad}l@{}}
\MS{y_{0j}}, & Y_{tj}\le \MS{y_{0j}},\\
Y_{tj}, & Y_{tj}>\MS{y_{0j}}.
\end{array}
\right.
\]
\endgroup
For exceedances, assume
\[
Y_{tj}-\MS{y_{0j}}\mid (Y_{tj}>\MS{y_{0j}})\sim \mathrm{GPD}(\sigma_{j},\xi_{j}),
\]
with tail CDF
\[
H_{\MS{y_{0j}}}(y\mid \sigma_{j},\xi_{j})
=1-\left(1+\xi_{j}\frac{y-\MS{y_{0j}}}{\sigma_{j}}\right)^{-1/\xi_{j}},
\qquad 1+\xi_{j}(y-\MS{y_{0j}})/\sigma_{j}>0.
\]
Hence the censored marginal CDF on the $Y$-scale is
\begingroup
\renewcommand{\arraystretch}{0.9}
\[
G_{j}(y)=
\left\{
\begin{array}{@{}l@{\quad}l@{}}
p, & y\le \MS{y_{0j}},\\
p+(1-p)\,H_{\MS{y_{0j}}}(y\mid \sigma_{j},\xi_{j}), & y>\MS{y_{0j}}.
\end{array}
\right.
\]
\endgroup
Let $x_{0j}:=F_{X_{j}}^{-1}(p)$. We map observations to the latent $X$-scale by
\begingroup
\renewcommand{\arraystretch}{0.9}
\[
X_{tj}=
\left\{
\begin{array}{@{}l@{\quad}l@{}}
x_{0j}, & Y_{tj}\le \MS{y_{0j}},\\
F_{X_{j}}^{-1}\!\left\{G_{j}(Y_{tj})\right\}, & Y_{tj}>\MS{y_{0j}}.
\end{array}
\right.
\]
\endgroup
Define $\mathcal{C}_t=\{j: Y_{tj}\le \MS{y_{0j}}\}$ and $\mathcal{E}_t=\{j: Y_{tj}>\MS{y_{0j}}\}$. Under the multiplicative nugget model, components are conditionally independent across sites given $\bX_t^*$, so the censored likelihood factorises as
\[
L_t(\bY_t\mid \bS_t, \bZ_t, \bphi, \bgamma, \brho, l, \alpha_0, \bsigma, \bxi)
=
\prod_{j\in\mathcal{C}_t}
F_\epsilon\!\left(\frac{x_{0j}}{X^*_{tj}}\right)
\;\times\;
\prod_{j\in\mathcal{E}_t}
f_{X\mid X^*}(x_{tj}\mid X^*_{tj})
\left|\frac{dX_{tj}}{dY_{tj}}\right|
\]
\BAS{for each $t=1, \ldots, T$}.  Under the temporal independence assumption, the \BAS{full joint } likelihood is simply the product of the likelihood for each time replicate.


Following \citet{shi2026}, we \BAS{fix } \(\alpha=1/2\), which simplifies the analytical formulation. \BAS{Likewise, } we represent \(\phi(\bs)\) and \(\rho(\bs)\) using Gaussian kernel basis functions centered at the knots. \BAS{We model } the latent Gaussian process \(Z_t(\bs)\) with a locally isotropic \textit{nonstationary} Mat\'ern covariance \citep{paciorek2006spatial,risser2015regression}. 
We implement an adaptive random-walk Metropolis algorithm with parallel updates of replicate-specific latent blocks across \(t\) \citep{shaby2010exploring}. 
The full hierarch\BAS{ical } specification and MCMC implementation details are given in \autoref{sec:Appendix_MCMC}.

\subsection{Simulation and Coverage Analysis}

\begin{figure}[!t]
    \centering
  \begin{minipage}[c]{0.7\textwidth}
  \centering
    \includegraphics[width=\textwidth, trim=0cm 2.4cm 0cm 2.4cm, clip]{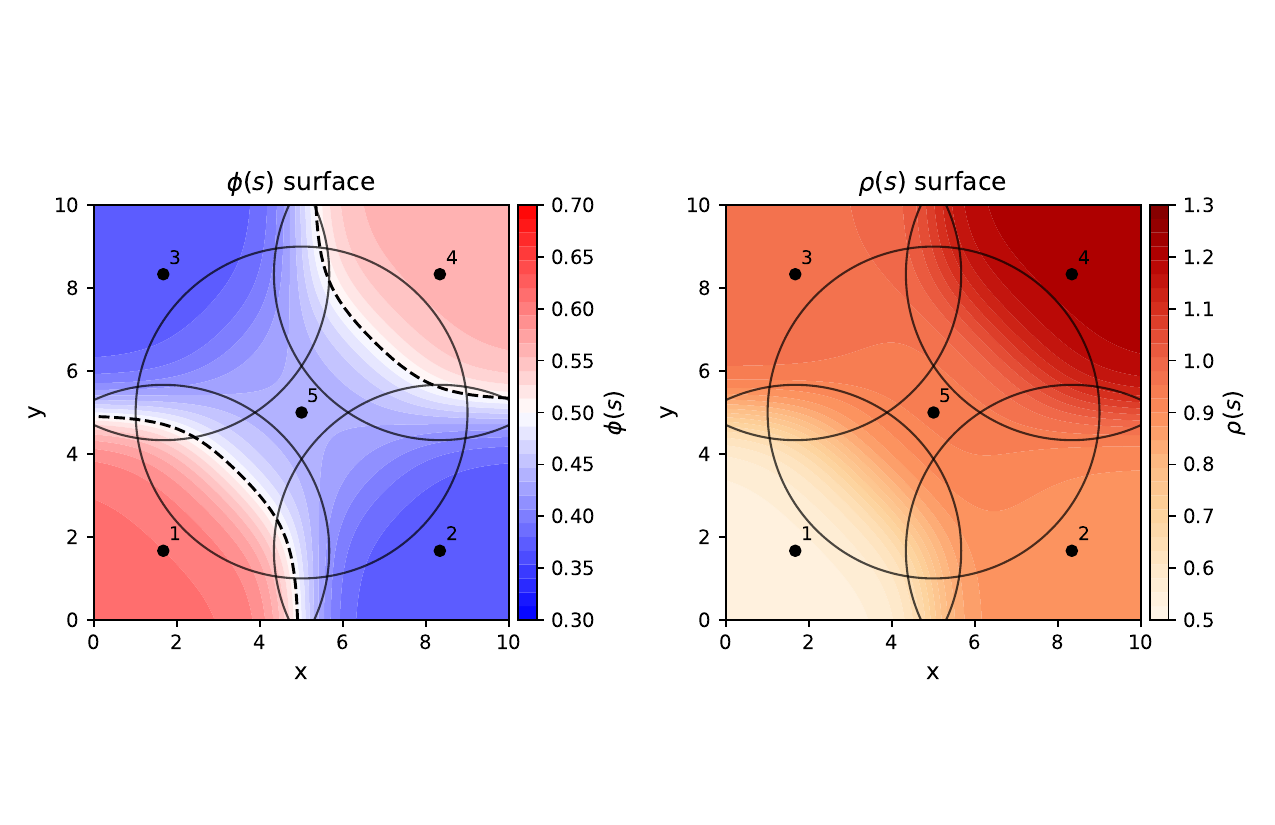}
  \end{minipage}
  \begin{minipage}[c]{0.29\textwidth}
    \caption{Simulation surfaces of \(\phi(\bs)\) and \(\rho(\bs)\). The black dashed line marks the local AD/AI transition. The \(\bullet\) symbols denote knot locations, and circles indicate kernel radii.}
  \label{fig:simulation_scenario}
  \end{minipage}
\end{figure}

\MS{We assessed posterior coverage using 50 independent datasets simulated from the model in \autoref{sec:hier_model}. For each dataset, \(D=100\) sites were sampled uniformly over \(\mathcal{S}=[0,10]^2\), with \(T=64\) replicates. The simulation used a locally isotropic nonstationary Mat\'ern latent field with \(\nu=1\), \(K=5\) knots on a regular grid, Wendland basis radius \(l=4\), Gaussian kernel bandwidth 4 for \(\phi(\bs)\) and \(\rho(\bs)\), L\'evy parameter \(\gamma=1\), and log-Laplace nugget parameter \(\alpha_0=5\); the resulting \(\phi(\bs)\) and \(\rho(\bs)\) surfaces are shown in \autoref{fig:simulation_scenario}. Observations were then mapped to a censored peaks-over-threshold scale with \(p_0=0.95\), threshold \(y_0=60\), and Generalised Pareto parameters \((\sigma,\xi)=(\exp(3),0.15)\). }


\begin{figure}[tp]
  \centering
  \includegraphics[width=0.9\textwidth]{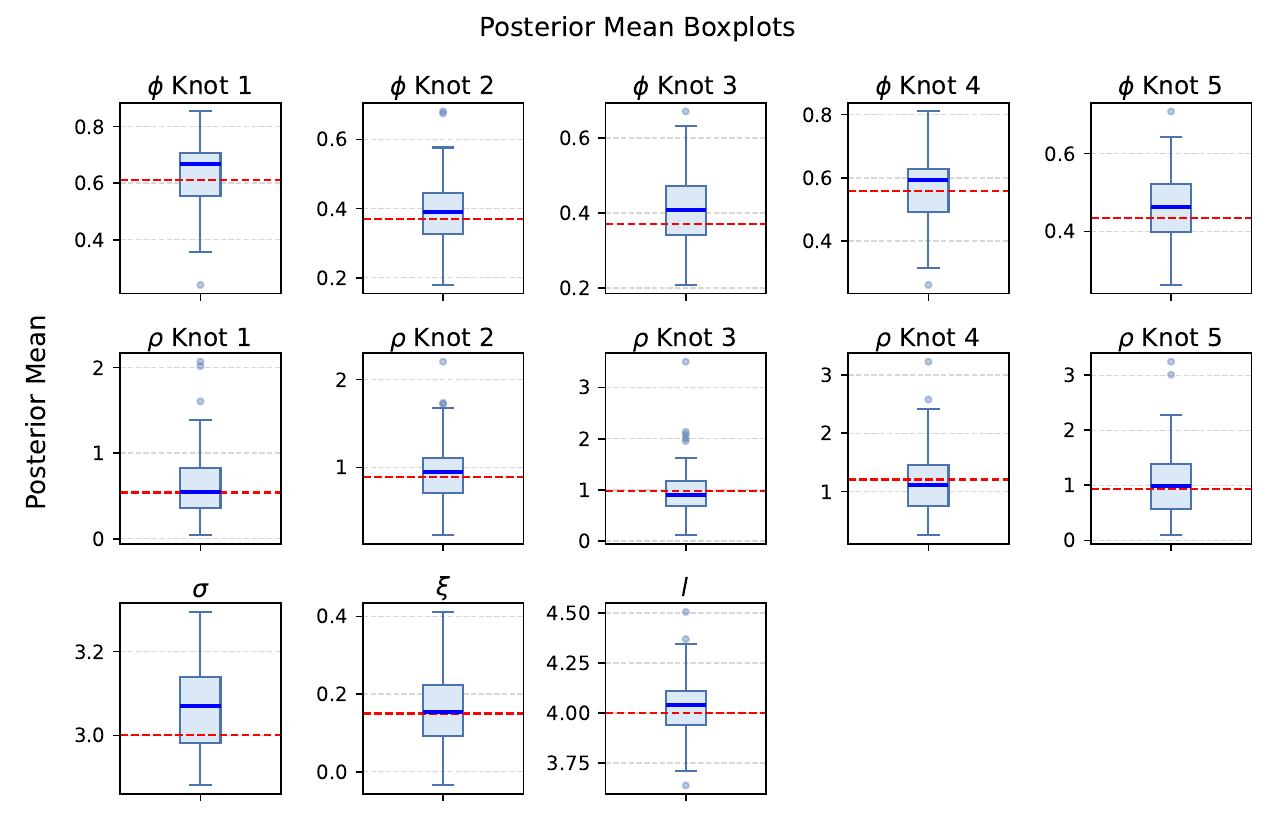}
  \includegraphics[width=0.9\textwidth]{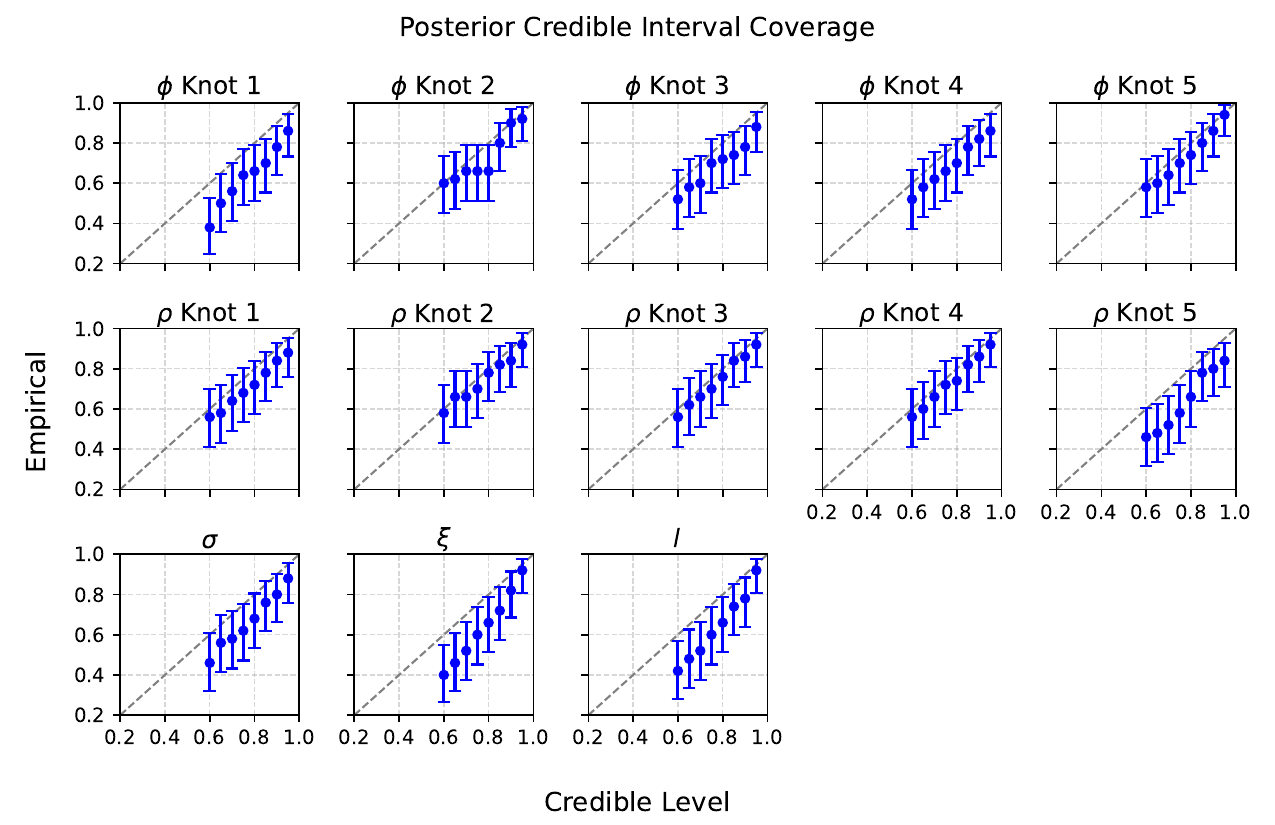}
  \caption{Posterior means (top) and standard binomial empirical coverage confidence intervals (bottom) for dependence ($\phi$, $\rho$), marginal ($\sigma$, $\xi$), and kernel radius ($l$) parameters across 50 simulated datasets. Red dashed lines in boxplots mark the true values.}
  \label{fig:coverage_analysis}
\end{figure}

For each simulated dataset, we fit the model and compute posterior credible intervals for dependence, marginal, and kernel radius parameters. \autoref{fig:coverage_analysis} reports empirical coverage rates at knot locations, with standard binomial confidence intervals around the empirical proportions. The observed coverage is close to nominal, indicating well-calibrated posterior inference.

\section{Extreme of \emph{in situ} Daily Precipitation}

\subsection{Data Analysis}
\BAS{We re-analyse the precipitation dataset from \citet{shi2026}.  That work fit Model \ref{Model-Shi} to the summer seasonal maxima using a GEV response.  Here, we again fit Model \ref{Model-Shi}, but instead analyse daily exceedances over a high threshold using a GPD response.  Revisiting the previous analysis of seasonal maxima  is worthwhile for two reasons.  First, exploratory analysis in \citet{shi2026} found that  a single field of summer maxima often combines extremes from many distinct storms; out of roughly 90 summer days, the station-wise seasonal maxima occur on about 75 different days on average. As a result, seasonal maxima may artificially inflate spatial tail dependence by pooling extremes that occur at different times. Second, the threshold-exceedance dataset is unusually large for spatial extremes, providing a stringent test of the computational feasibility of the proposed censored-likelihood approach in high dimensions.}

\begin{figure}[!t]
  \centering
  \includegraphics[width=0.85\textwidth]{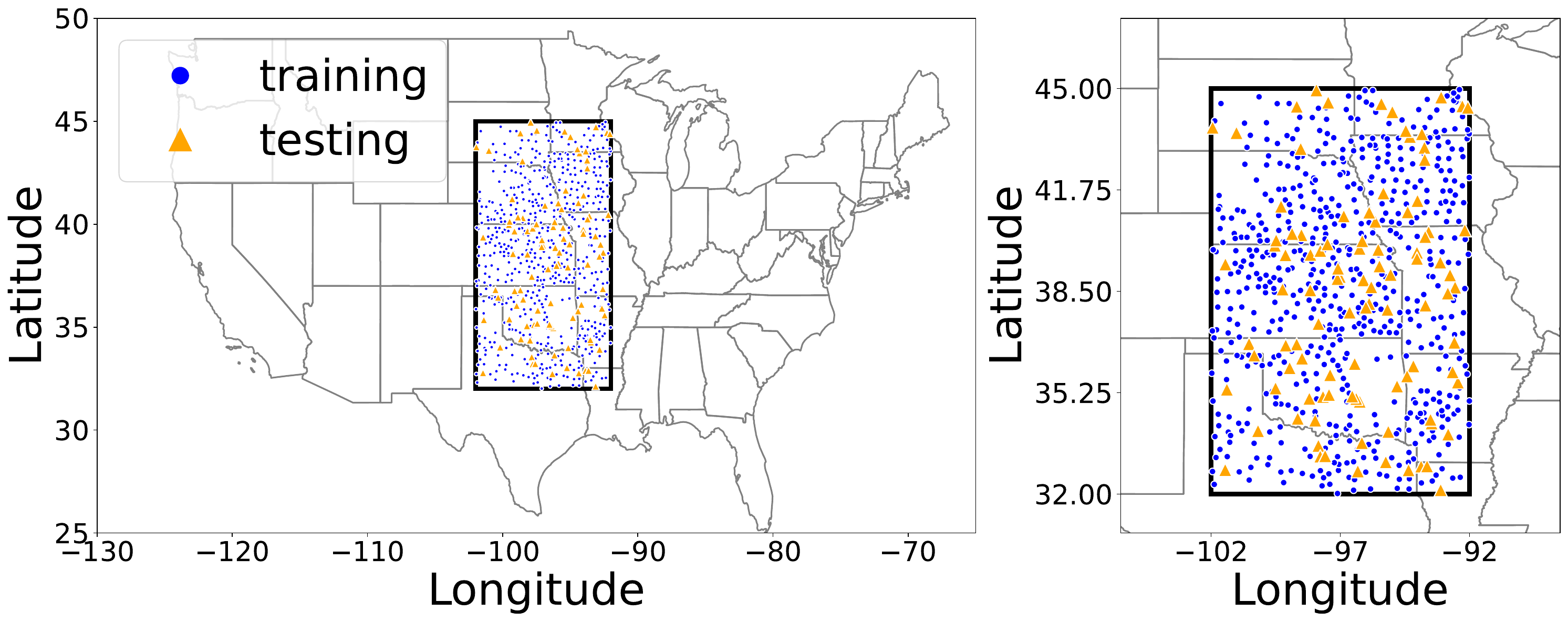}
  \caption{Location of the 590 stations (blue circle) and the 99 out-of-sample testing stations (orange triangle) in the central US used for analysis.}
  \label{fig:us_map}
\end{figure}

To reduce short-range temporal dependence, we divide each summer into nine consecutive 10-day periods and retain the maximum daily precipitation within each period. This yields 675 time points at each of the 590 stations \BAS{(see Figure \ref{fig:us_map})}. We \BAS{fit the exceedances of a } site-specific 0.95 quantile threshold.

\begin{figure}[!t]
  \centering
  \includegraphics[width=\textwidth]{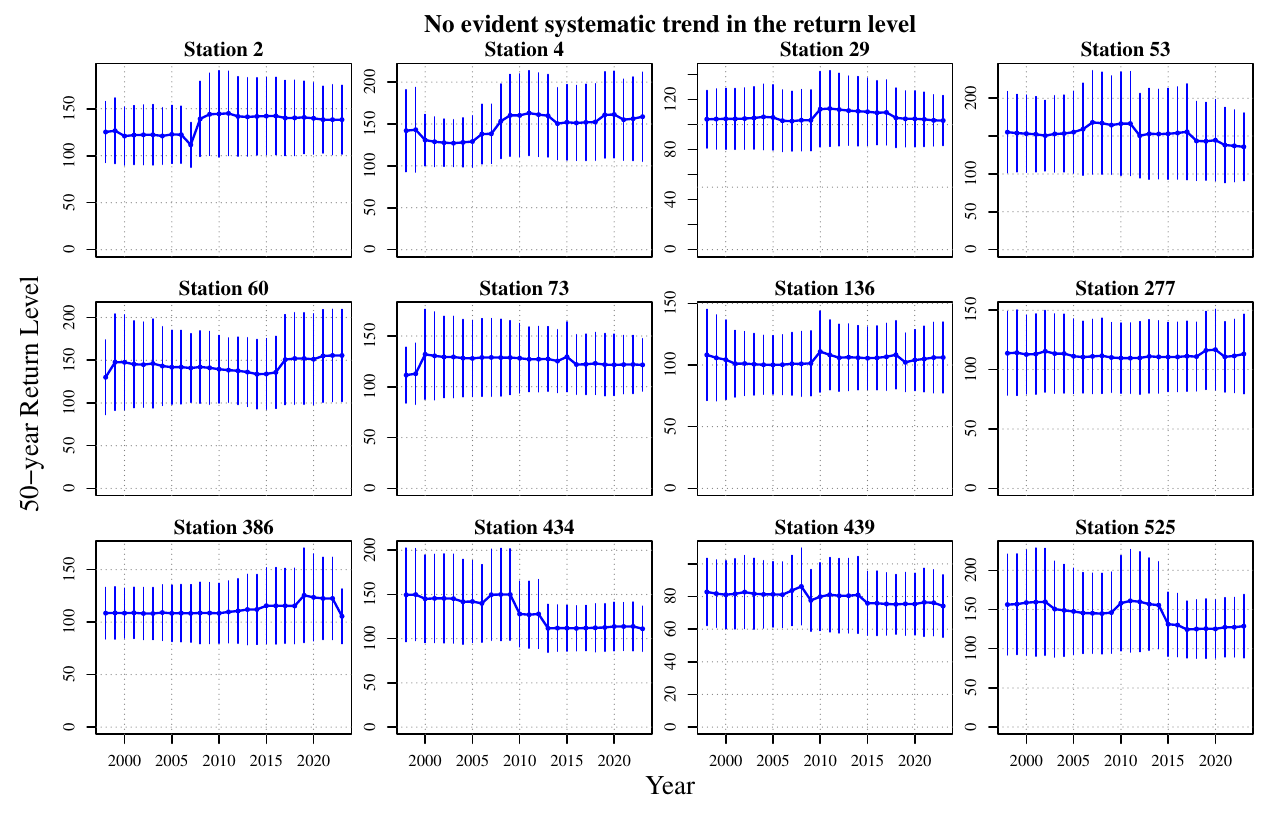}
  \caption{Point estimates and 95\% confidence intervals for 50-year precipitation return levels at 12 randomly selected stations, estimated from annual maxima using 50-year sliding windows. No evident systematic trend is observed, supporting the assumption of temporally constant marginal parameters.}
  \label{fig:temporal_independence}
\end{figure}

Empirical diagnostics indicate that treating the 10-day maxima as approximately temporally independent is reasonable. \autoref{fig:temporal_independence} shows 50-year return levels estimated from annual maxima using 50-year sliding windows at 12 randomly selected stations. The absence of systematic trends supports the assumption of temporally constant marginal parameters. Although such assumptions can be \BAS{unrealistic } for meteorological variables, especially temperature, they appear reasonable for precipitation extremes in this setting. To account for broad terrain effects, we model the marginal GPD parameters as functions of elevation, as 
\[
\log \sigma(\bs)=\beta_{\sigma,0}+\beta_{\sigma,1}\,\mathrm{elev}(\bs),
\qquad
\xi(\bs)=\beta_{\xi,0}+\beta_{\xi,1}\,\mathrm{elev}(\bs).
\]

For the dependence model, we place knots on a regular grid over the spatial domain. Motivated by the analysis of \citet{shi2026}, we consider four candidate configurations that vary in the number of knots and in whether the marginal parameters are \LZ{\textit{fixed} } at their smoothed site-wise estimates or updated \LZ{\textit{jointly} } with the spatial dependence model. Across these models, the Wendland basis radius governing the latent scale surface is treated as unknown and updated within the MCMC. The model specifications are summarized in \autoref{tab:model_config}. 
We run \BAS{each } chain for approximately 250{,}000 iterations, retain every fifth draw, and after discarding burn-in obtain 15{,}000 posterior samples for inference.
\MS{After parallelisation, one chain took about 70 hours on an AMD Milan EPYC CPU, or about 1 second per iteration.}

\begin{table}[!t]
\centering
\caption{Candidate model configurations considered for the precipitation analysis. Model naming convention is as follows: $k$, $b$, and $m$ respectively denote the number of knots, effective range of the Gaussian basis (the distance at which the kernel function drops below 0.05), and restriction indicator on marginal GPD parameters. 
}
\label{tab:model_config}
\begin{tabular}{lccc}
\toprule
\textbf{Model} & \textbf{Number of knots} & \textbf{Gaussian basis effective range} & \textbf{Constraint} \\
\midrule
\rowcolor{gray!8}
\texttt{k25b4}  & 25 & 4 & --- \\
\texttt{k25b4m} & 25 & 4 & Fixed $\sigma$, $\xi$ \\
\rowcolor{gray!8}
\texttt{k41b4}  & 41 & 4 & --- \\
\texttt{k41b4m} & 41 & 4 & Fixed $\sigma$, $\xi$ \\
\bottomrule
\end{tabular}
\end{table}

\subsection{Model Evaluation}

\begin{figure}[!t]
  \centering
  \includegraphics[width=0.75\textwidth]{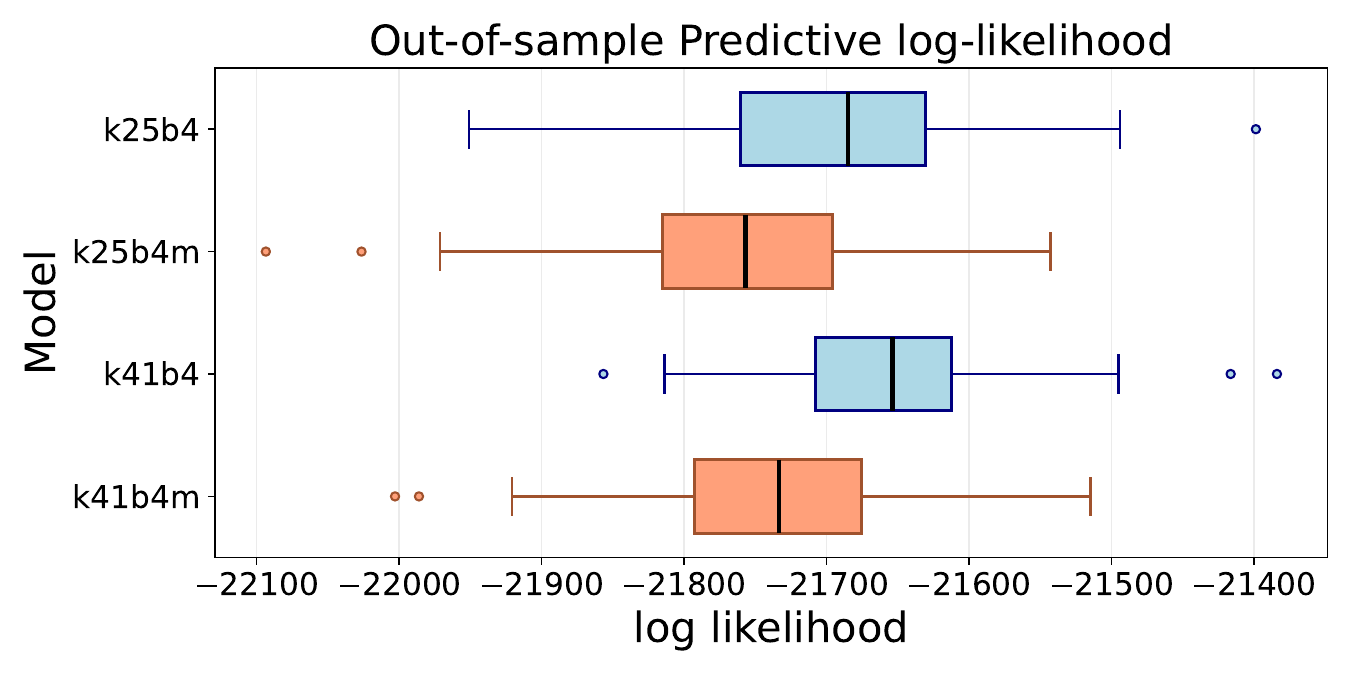}
  \caption{Predictive log scores at holdout sites for the four candidate models.}
  \label{fig:loglik_boxplot}
\end{figure}

To assess model performance, we use an additional set of \BAS{99 } holdout stations from the same spatial domain and time period \BAS{(see Figure \ref{fig:us_map})}. 
We compare the four candidate models using predictive log scores at these holdout sites. 
\autoref{fig:loglik_boxplot} summarises the resulting predictive log scores. Based on this criterion, we select the \texttt{k41b4} model for the remainder of the analysis.

The comparison also shows a systematic advantage for models that update the marginal parameters \LZ{\textit{jointly} } with the dependence model, relative to models that fix the marginals at pre-estimated values. This mirrors the findings of \citet{shi2026} and suggests that fully joint hierarchical inference can improve model performance over the more common two-step approach.

\begin{figure}[ht]
    \centering
    \begin{minipage}[t]{0.24\textwidth}
        \centering
        \includegraphics[width=\textwidth, clip=true, trim=5 5 35 0]{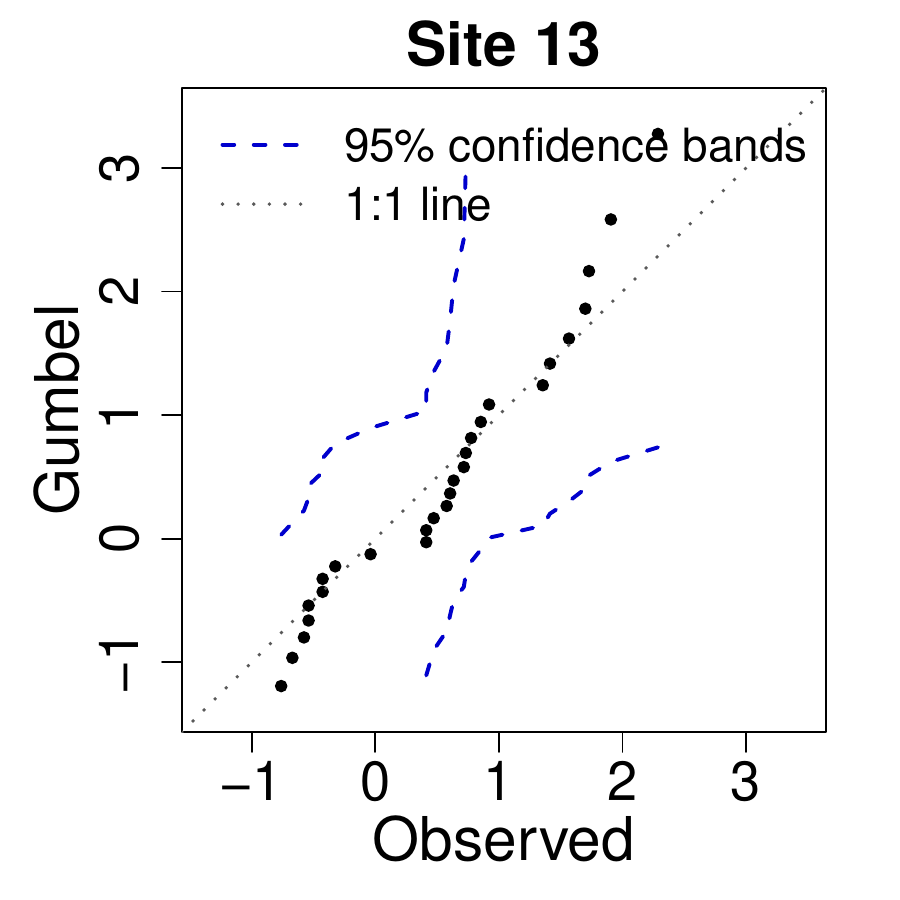}
    \end{minipage}
    \hfill
    \begin{minipage}[t]{0.24\textwidth}
        \centering
        \includegraphics[width=\textwidth, clip=true, trim=5 5 35 0]{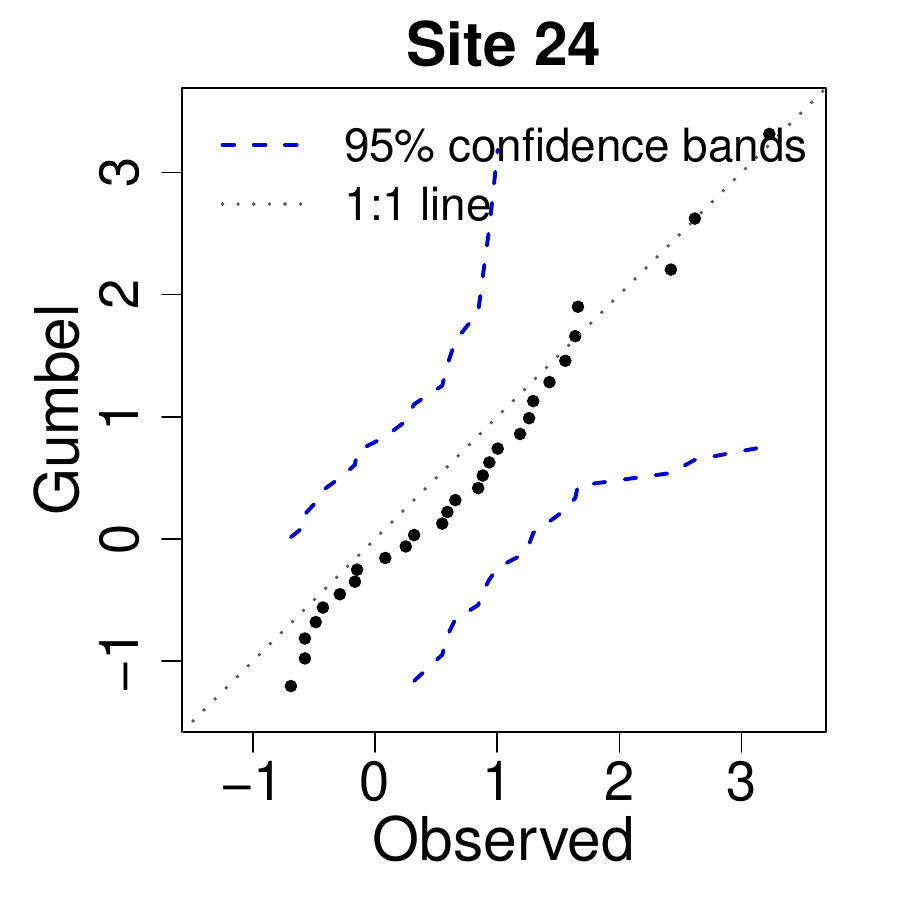}
    \end{minipage}
    \hfill
    \begin{minipage}[t]{0.24\textwidth}
        \centering
        \includegraphics[width=\textwidth, clip=true, trim=5 5 35 0]{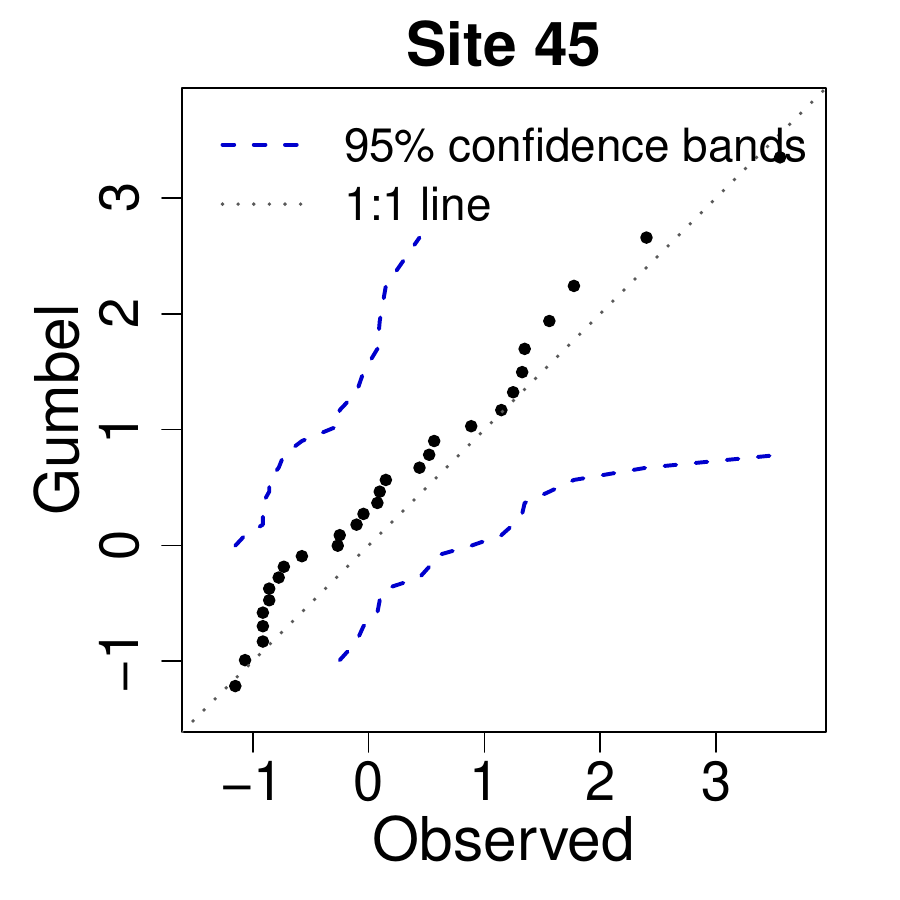}
    \end{minipage}
    \hfill
    \begin{minipage}[t]{0.24\textwidth}
        \centering
        \includegraphics[width=\textwidth, clip=true, trim=5 5 35 0]{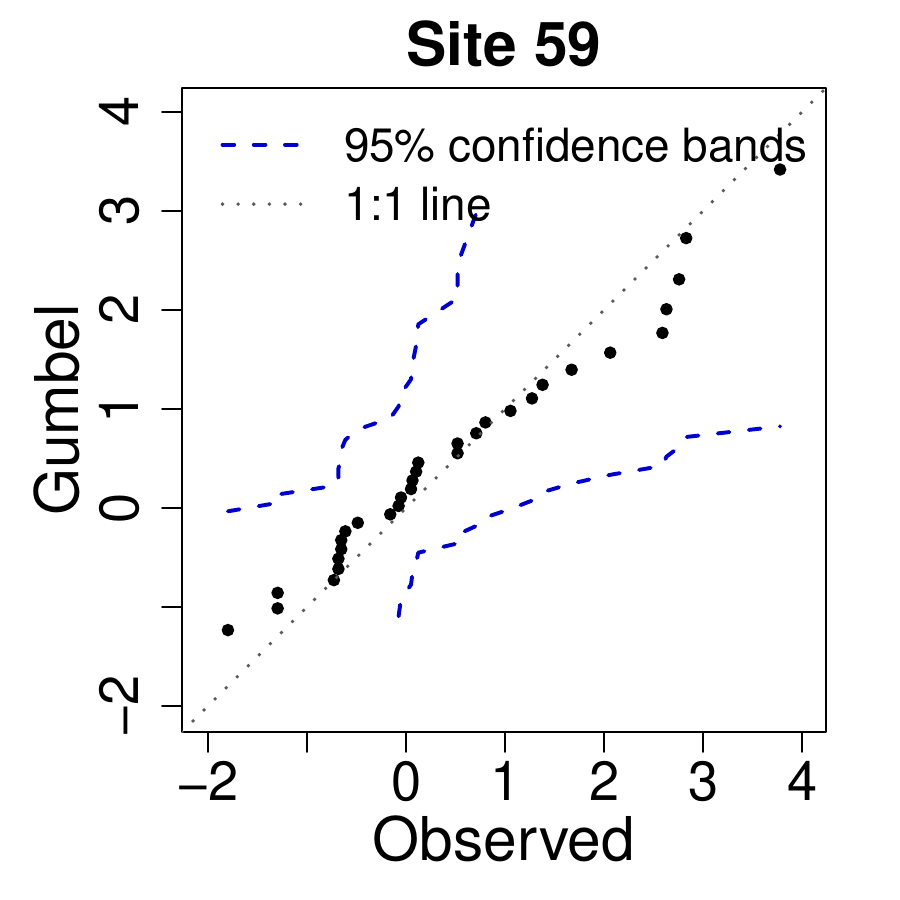}
    \end{minipage}

    \caption{QQ-plots of four randomly selected holdout locations, comparing observed and predicted marginal quantiles for the \texttt{k41b4} model, along with 95\% confidence envelopes. All values are transformed to the Gumbel scale for ease of visualisation.}
    \label{fig:qqplot}
\end{figure}

We also assess marginal fit using empirical quantile plots at the holdout sites compared to threshold exceedance draws from the posterior predictive distribution. \autoref{fig:qqplot} shows QQ-plots for four randomly selected holdout stations under the selected \texttt{k41b4} model. In each case, the 95\% uncertainty band covers the 1:1 line reasonably well, indicating adequate marginal fit.

\subsection{Results}

\begin{figure}[!t]
  \centering
  \subfloat[Kernel configuration.\label{fig:k41b4_grid}]{
    \includegraphics[width=0.42\textwidth, clip=true, trim=0 0 0 36]{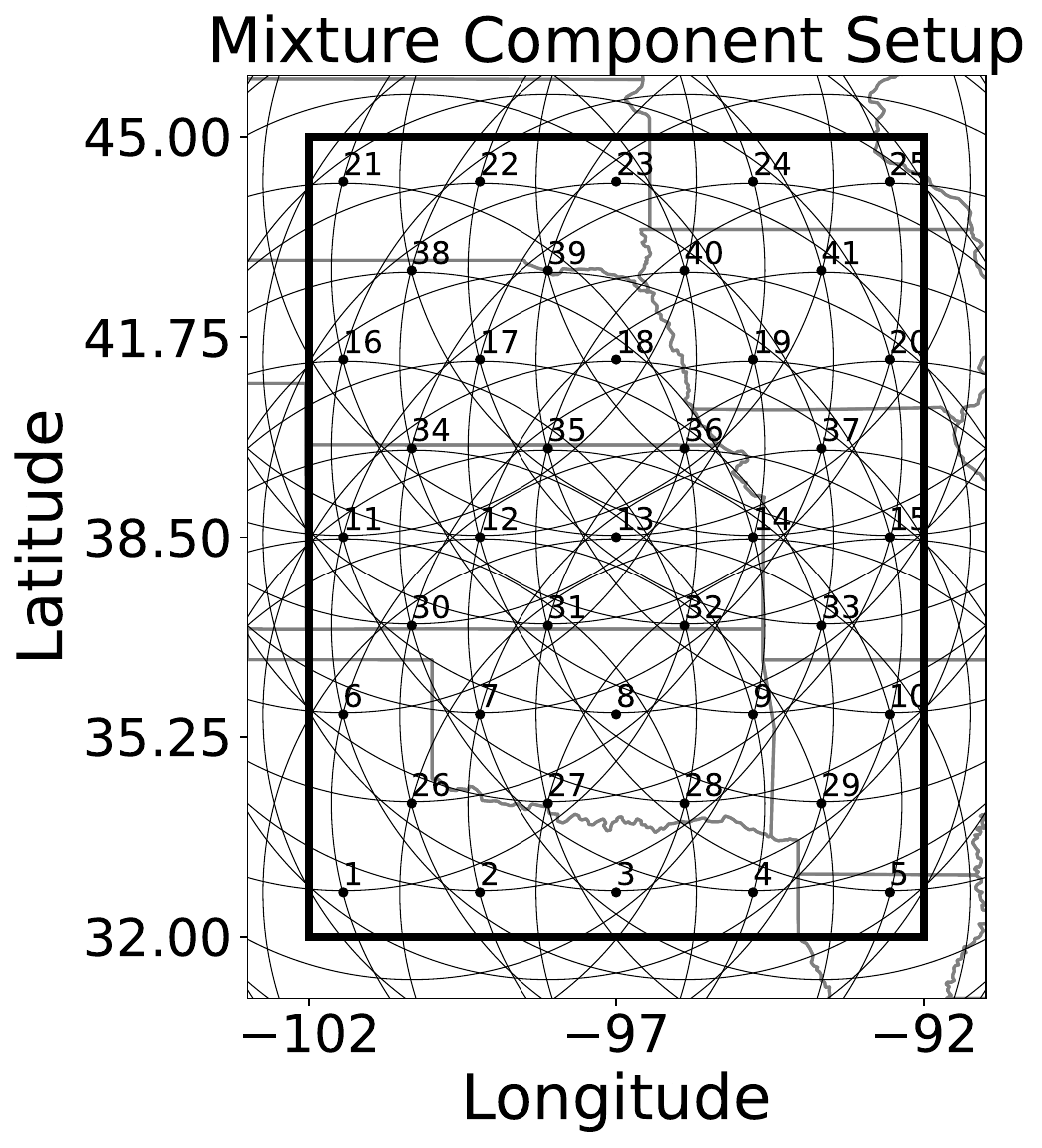}
  }
  \hfill
  \begin{minipage}[b]{0.56\textwidth}
    \centering
    \subfloat[$\phi(\bs)$\label{fig:k41b4_phi_surface}]{
      \includegraphics[width=0.46\linewidth]{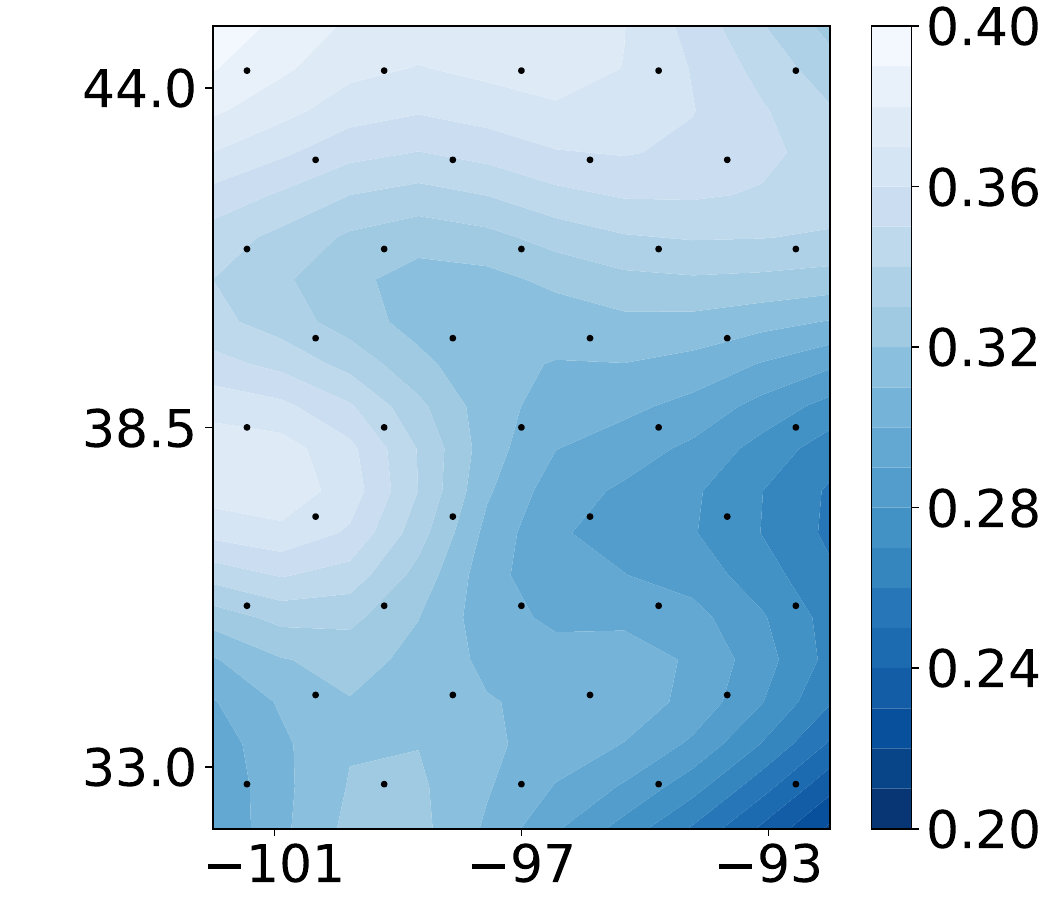}
    }
    \hfill
    \subfloat[$\rho(\bs)$\label{fig:k41b4_rho_surface}]{
      \includegraphics[width=0.46\linewidth]{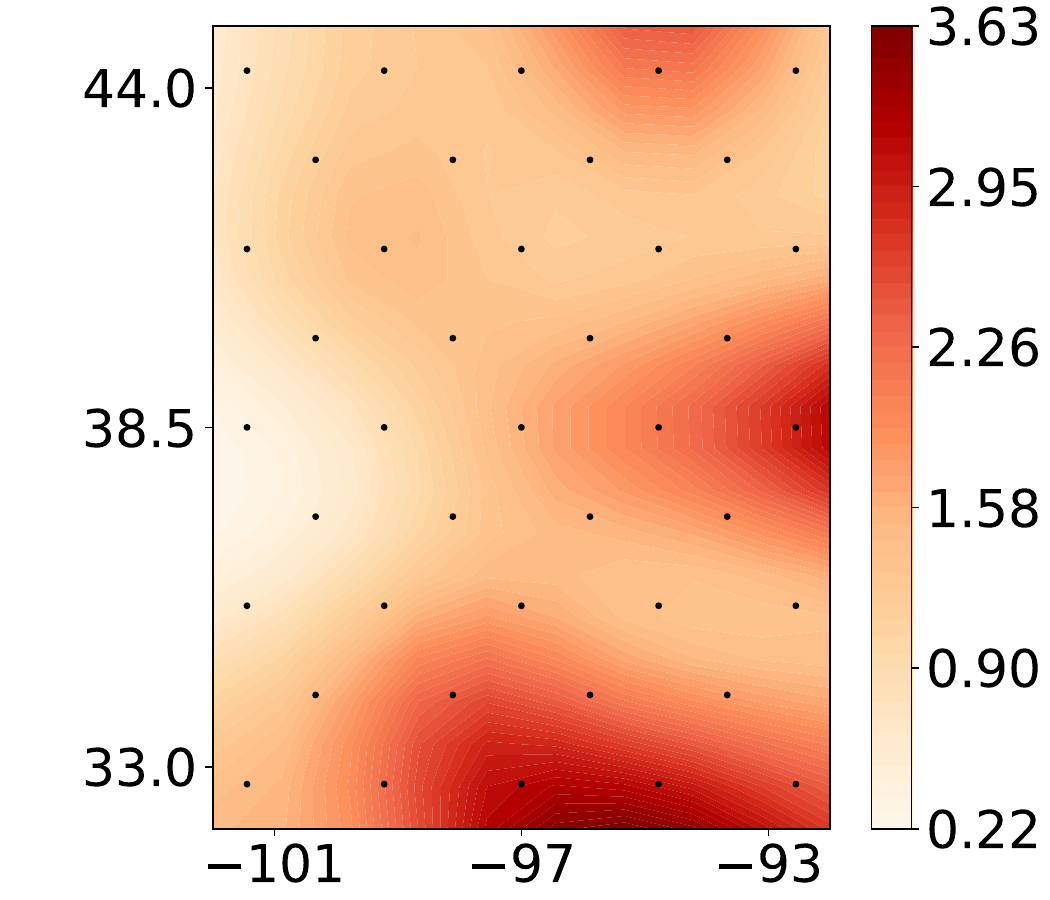}
    }

    \subfloat[$\log \sigma(\bs)$\label{fig:k41b4_sigma_surface}]{
      \includegraphics[width=0.46\linewidth]{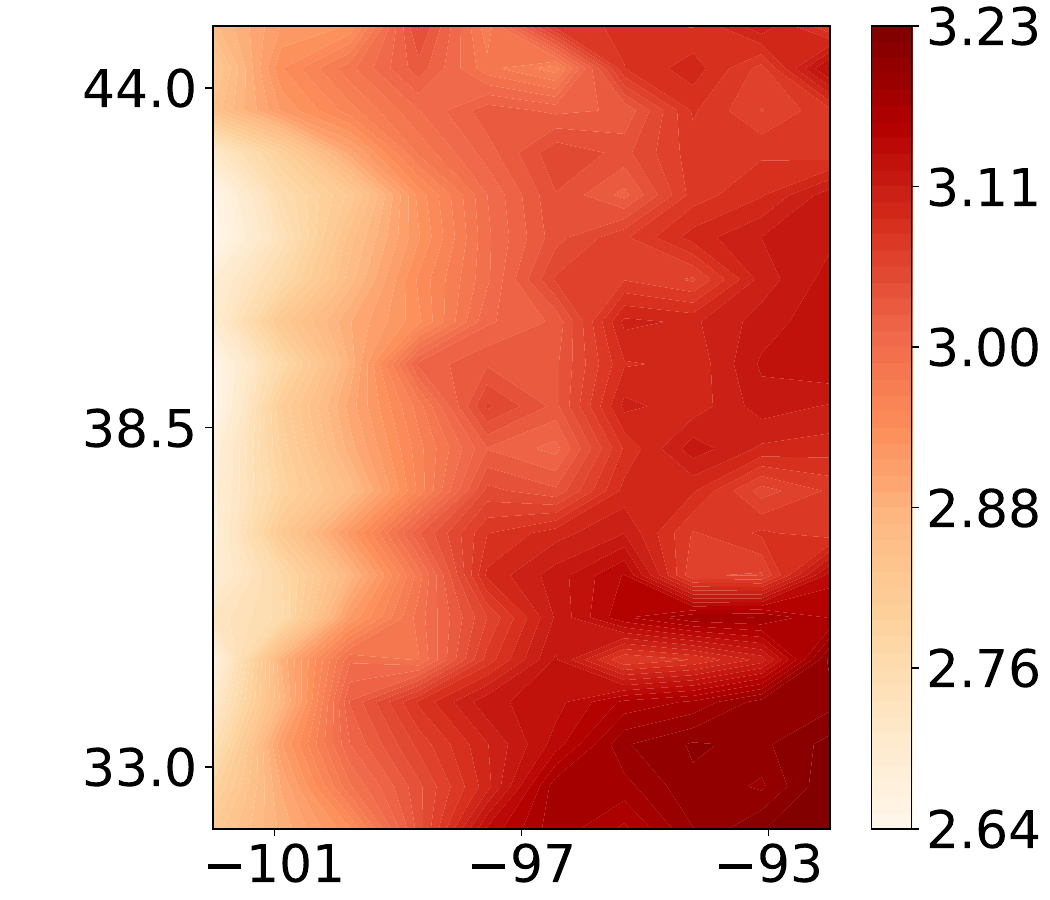}
    }
    \hfill
    \subfloat[$\xi(\bs)$\label{fig:k41b4_xi_surface}]{
      \includegraphics[width=0.46\linewidth]{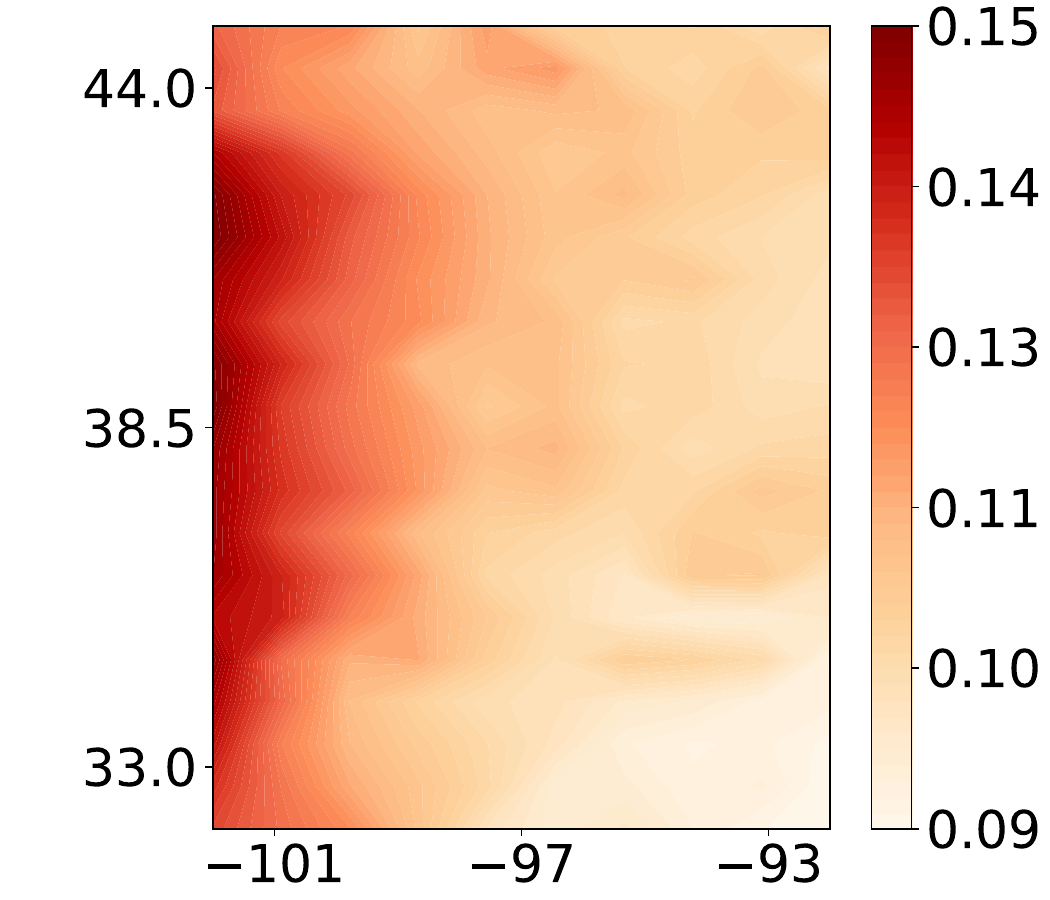}
    }
  \end{minipage}

  \caption{Left: For the \texttt{k41b4} model, the black dots represent the knots and circles represent areas covered by each kernel with posterior mean radius. Right: posterior mean parameter surfaces for the fitted \texttt{k41b4} model.}
  \label{fig:k41b4_results}
\end{figure}

We now present results from the selected \texttt{k41b4} model. \autoref{fig:k41b4_grid} \MS{visualises the knot locations and associated basis, } and \autoref{fig:k41b4_phi_surface}--\autoref{fig:k41b4_xi_surface} display the posterior mean dependence and marginal parameter surfaces. The posterior mean \(\phi(\bs)\) surface lies below \(0.5\) throughout the domain, indicating \BAS{short-range } asymptotic independence across the study region, but with clear spatial variation in the strength of the dependence, with larger values \BAS{occurring } over the western part of the domain and smaller values \BAS{occurring } toward the east and southeast.  This result resembles the spatial structure found in \citet{shi2026}. The \(\rho(\bs)\) surface and the marginal surfaces for \(\log \sigma(\bs)\) and \(\xi(\bs)\) also vary spatially, and these patterns are broadly consistent with the spatial structure found in \citet{shi2026}.

\begin{figure}[htb]
    \centering

    \begin{minipage}[b]{0.495\textwidth}
        \centering
        \BAS{Empirical } $\chi$
        \vspace{5pt}

        \includegraphics[width=\textwidth, clip=true, trim=0 0 0 0]{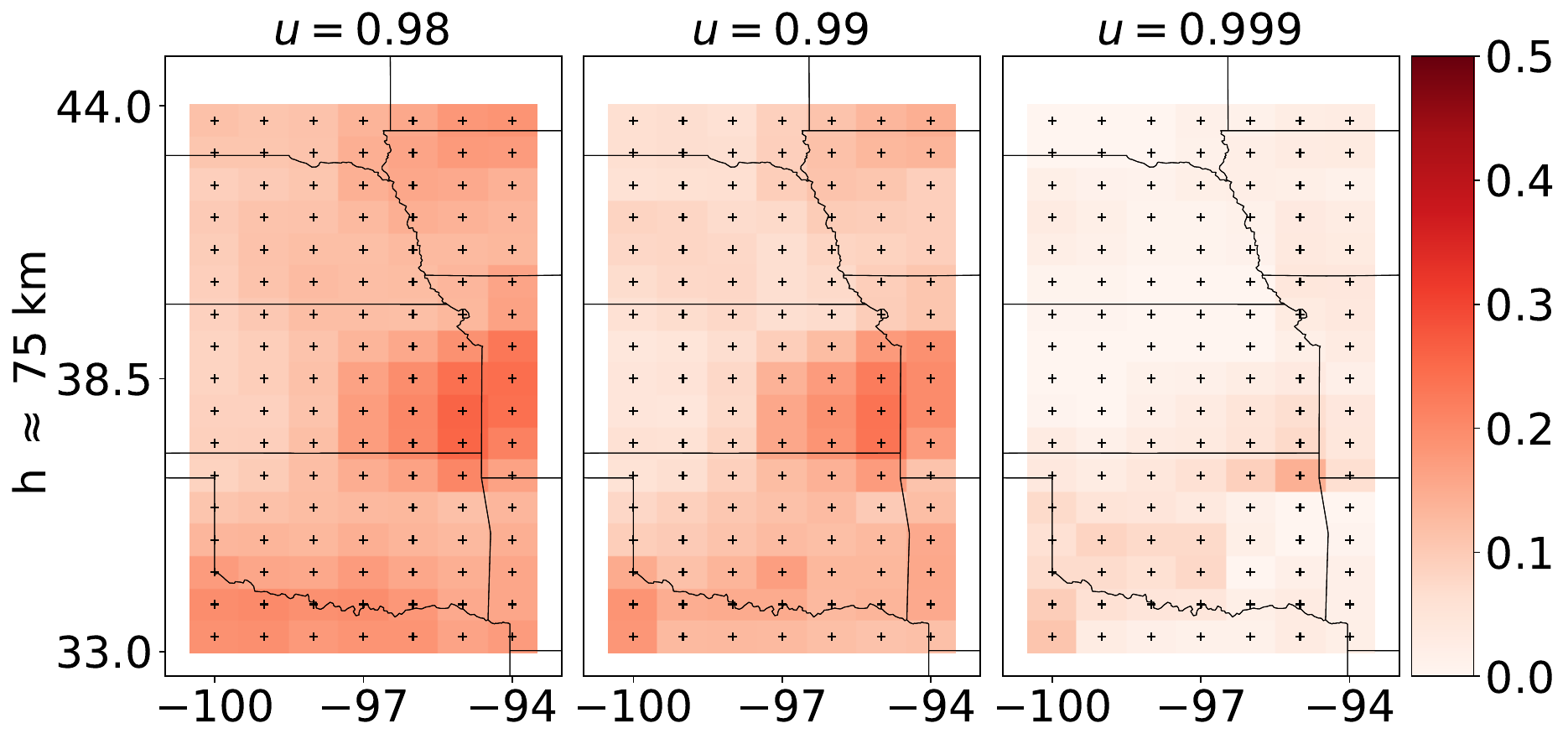}
    \end{minipage}
    \hfill
    \begin{minipage}[b]{0.495\textwidth}
        \centering
        Model-based $\chi$
        \vspace{5pt}

        \includegraphics[width=\textwidth, clip=true, trim=0 0 0 0]{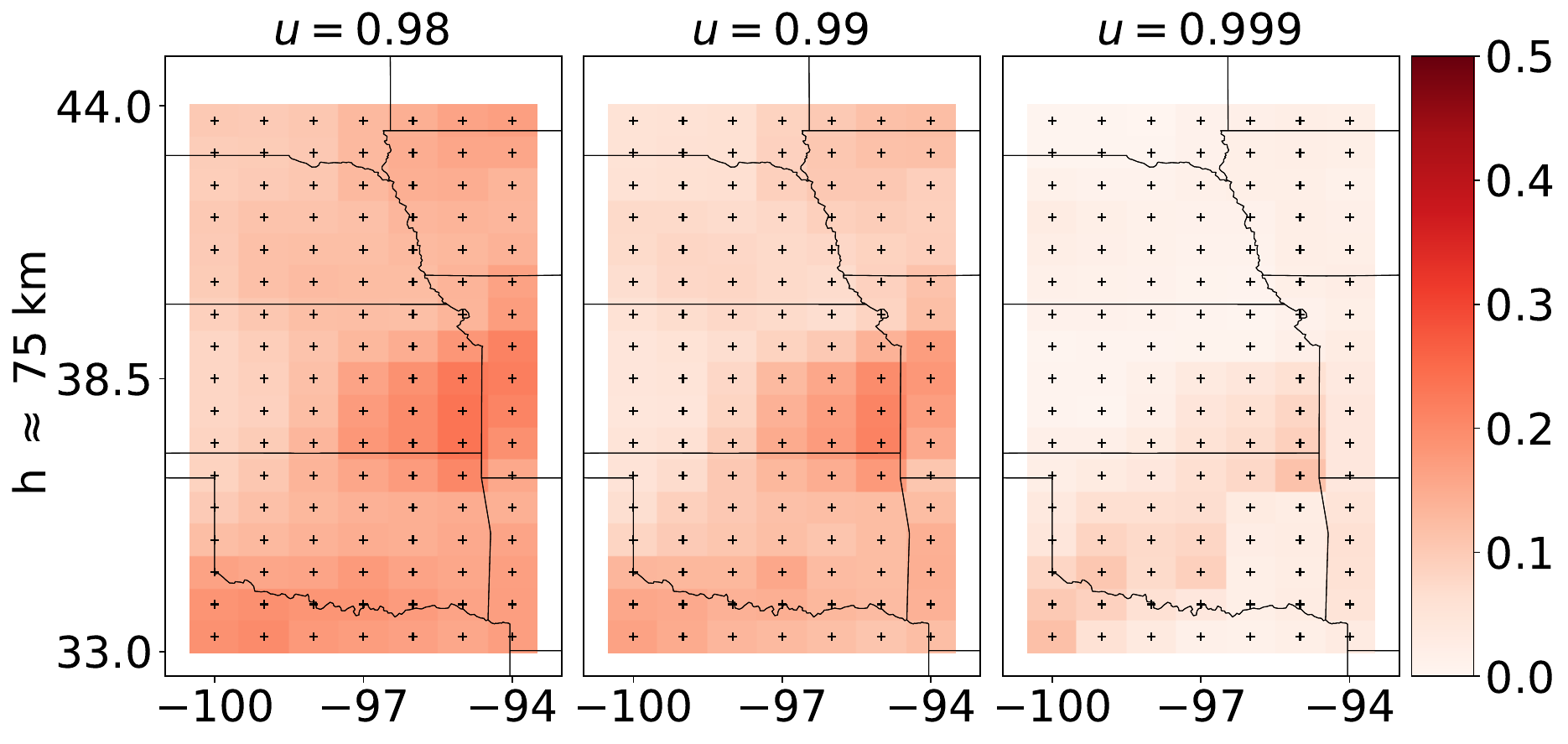}
    \end{minipage}
    \vfill
    \begin{minipage}[b]{0.495\textwidth}
        \centering
        \includegraphics[width=\textwidth, clip=true, trim=0 0 0 0]{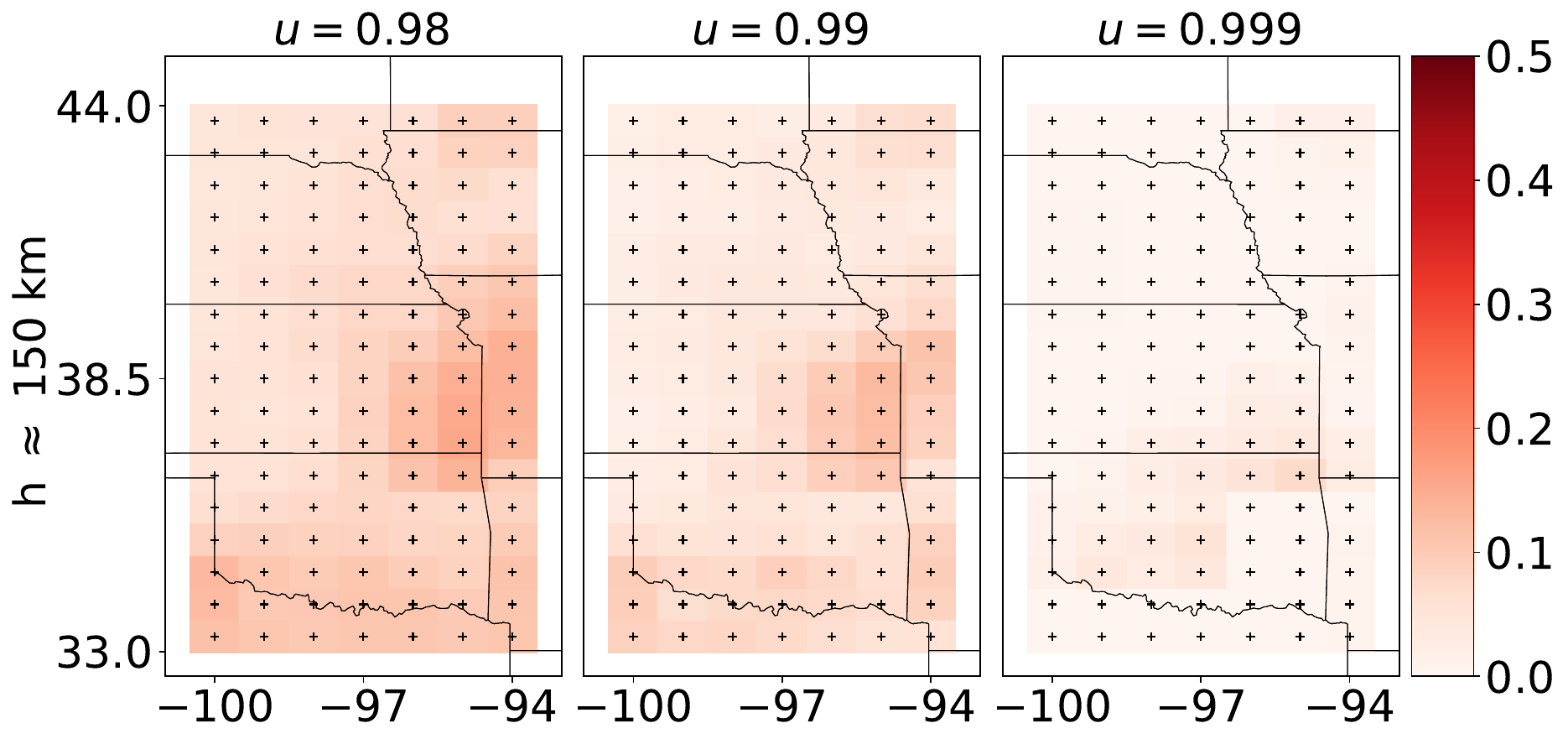}
    \end{minipage}
    \hfill
    \begin{minipage}[b]{0.495\textwidth}
        \centering
        \includegraphics[width=\textwidth, clip=true, trim=0 0 0 0]{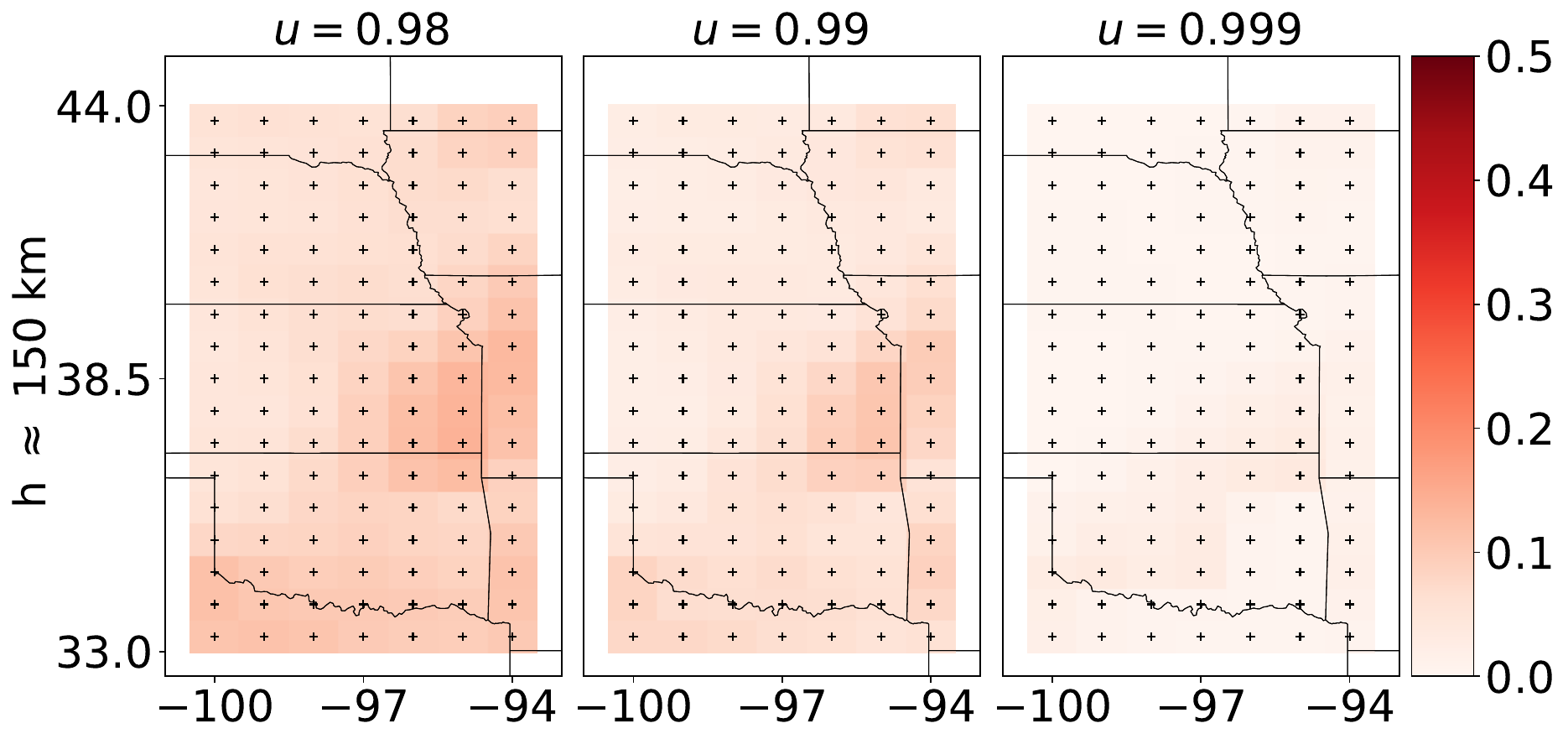}
    \end{minipage}
    \vfill
    \begin{minipage}[b]{0.495\textwidth}
        \centering
        \includegraphics[width=\textwidth, clip=true, trim=0 0 0 0]{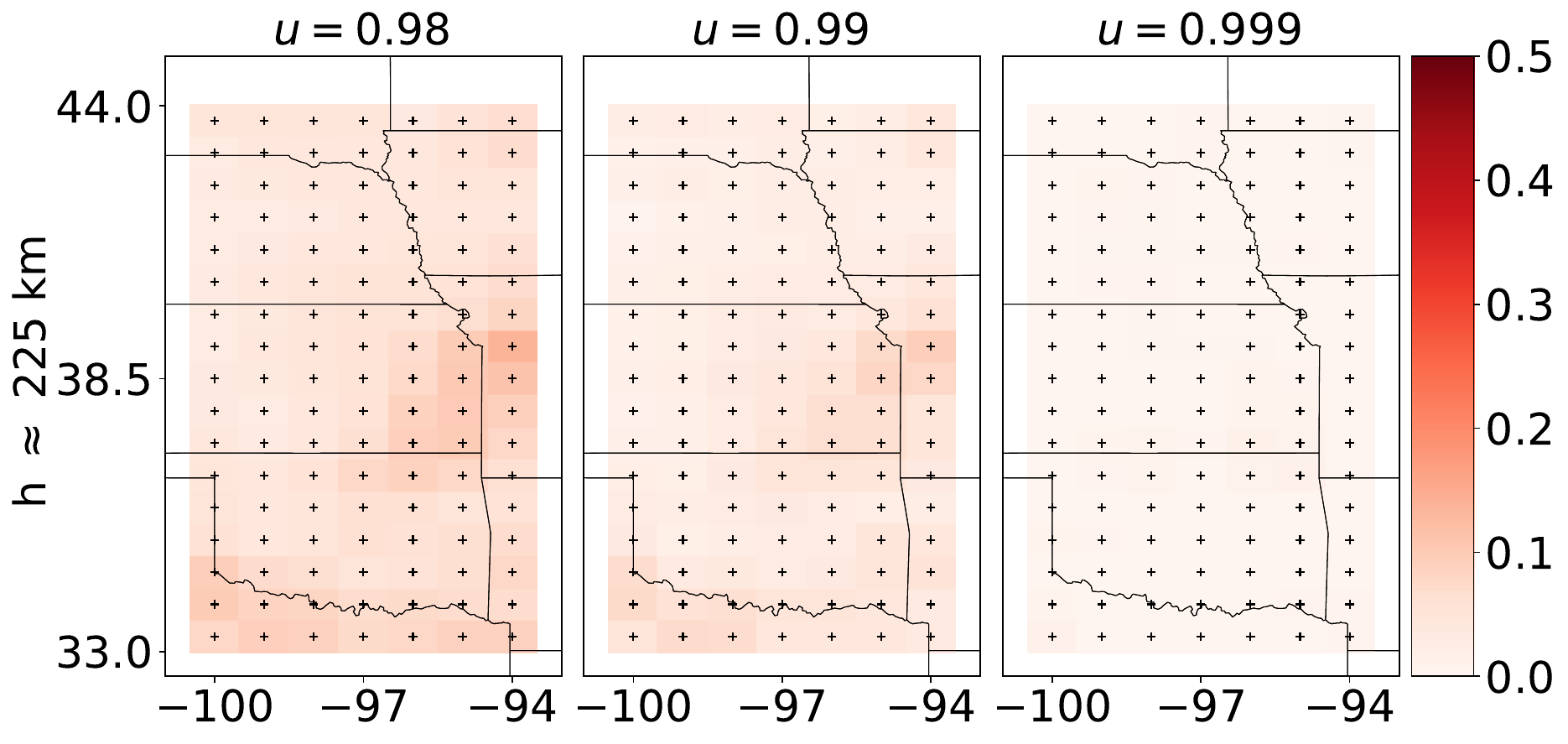}
    \end{minipage}
    \hfill
    \begin{minipage}[b]{0.495\textwidth}
        \centering
        \includegraphics[width=\textwidth, clip=true, trim=0 0 0 0]{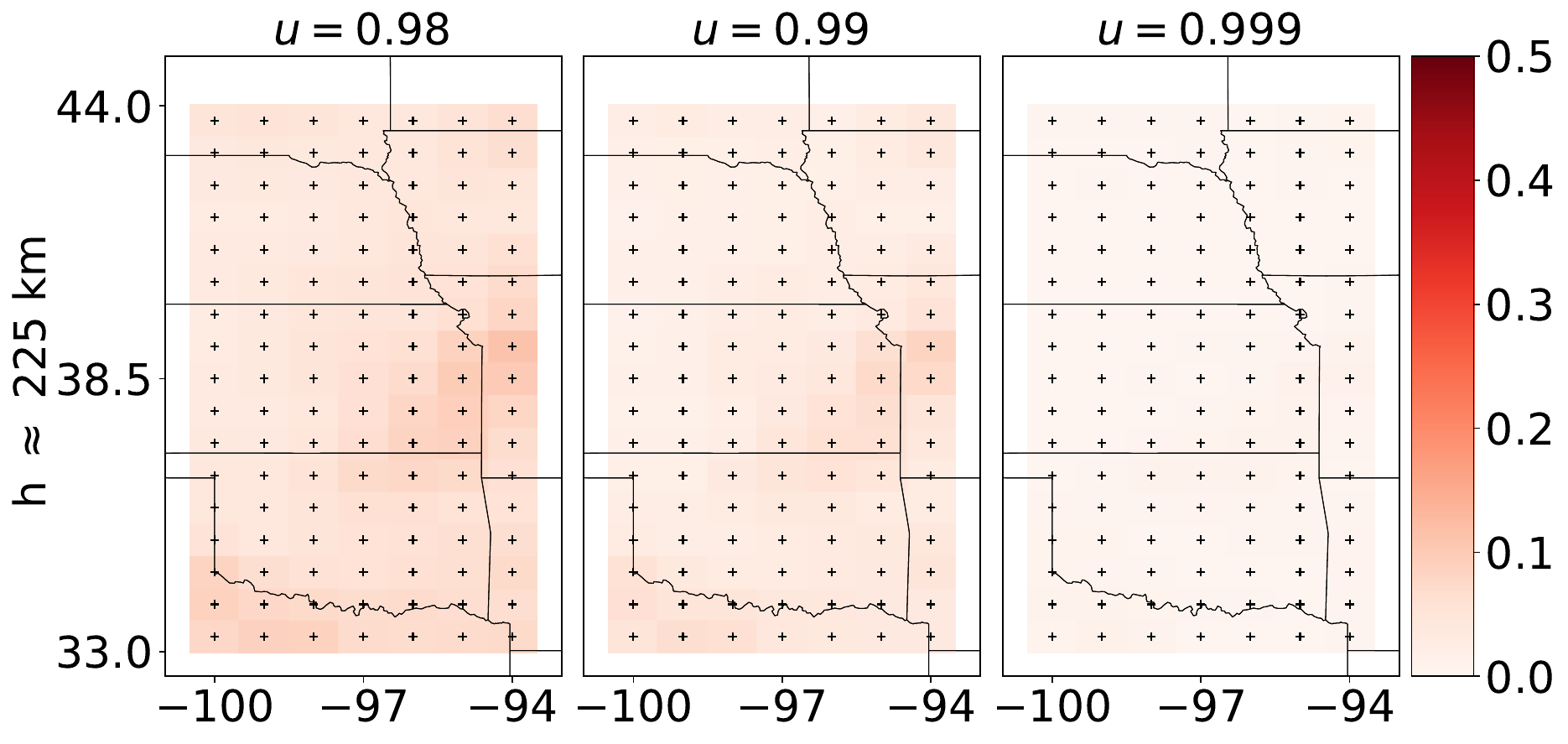}
    \end{minipage}
    \caption{Moving-window  estimates of $\chi_u(h)$  across three quantiles $u$ and three spatial lags $h$. The left-hand panel shows the dataset empirical estimates of $\chi_u(h)$, and the right-hand panel shows the model-based estimates of $\chi_u(h)$, based on the \texttt{k41b4} model.}
    \label{fig:chi_moving_window}
\end{figure}

To assess extremal dependence fit, \autoref{fig:chi_moving_window} compares moving-window local empirical \(\chi_u(h)\) surfaces with their model-based counterparts over several thresholds and spatial lags. The \BAS{empirical } surfaces in the left column are obtained by transforming the observations \(Y_t(\bs)\) to the \(X_t(\bs)\) scale using the fitted marginal GP parameters from the \texttt{k41b4} model and then estimating \(\chi_u(h)\) empirically. The fitted surfaces in the right column are obtained from conditional posterior predictive draws under the fitted model. We see good agreement between the data and the model fit, as the fitted model reproduces the main localised patches of elevated finite-threshold dependence as well as the overall weakening of dependence with increasing lag and threshold. Moreover, the empirical \(\chi_u(h)\) surfaces decrease toward zero as \(u \to 1\), which is consistent with the asymptotic independence implied by \(\phi(\bs) < 0.5\). Overall, these diagnostics indicate that the \texttt{k41b4} model captures both the broad spatial variation in the marginal behaviour and the local subasymptotic tail-dependence structure of the precipitation extremes.

\section{Discussion}

In this paper, we introduced a multiplicative log-Laplace nugget for a broad class of \LZ{random } scale-mixture models. \BAS{For inference using the } censored likelihood, \BAS{augmenting the smooth process with } the nugget removes the need to evaluate \BAS{high}-dimensional Gaussian distribution functions, substantially reducing computational cost.  \BAS{The particular form of the nugget results in closed forms for the marginal density and distribution functions, which eliminates the need for numerical integration and makes computation feasible for models with spatially varying parameters. } At the same time, we show\BAS{ed } analytically that \BAS{including the nugget } preserves the main tail properties of the underlying smooth process.  \BAS{The result is that the total computational cost is dominated by standard spatial statistics operations like factorizing the covariance matrix.  This represents a major shift, as heretofore likelihood-based analysis of spatial extremes has been severely limited by formidable computational obstacles, which have only been ameliorated by using bespoke approximations \citep[e.g.][]{padoan2010likelihood,shaby2014open,wadsworth2014efficient,defondeville2018high}, often entailing considerable compromises.}

We illustrated the proposed modification using the model of \citet{shi2026}\LZ{---the most flexible available model to the best of our knowledge}. Through simulation and data analysis, we showed that the modified model retains the ability to capture qualitatively different forms of asymptotic and subasymptotic tail dependence, while still allowing spatially varying extremal dependence across the domain.

The simulation study and precipitation application also demonstrate that the resulting nuggeted hierarchical model is scalable in practice. Inference can be carried out with standard MCMC, parallelised over time replicates, and numerical routines implemented in \texttt{C++}, the approach is feasible for spatial extreme datasets with both many locations and many time points. In our application, the modified model \BAS{reduced computation time by a factor of around 100 per iteration relative to the GEV implementation in \citet{shi2026}, }while recovering \BAS{broadly similar } spatial dependence patterns.

A further finding, also consistent with \citet{shi2026}, is that models that jointly estimate marginal and dependence parameters outperform analogous two-step approaches in holdout predictive log score. Although joint inference is more demanding to implement because it requires the marginal probability integral transform within the MCMC, the nugget itself does not add an additional layer of numerical integration. As a result, the extra computational cost remains manageable, while the predictive gains suggest that joint estimation is often worthwhile because it propagates uncertainty coherently between marginal and dependence components and improves out-of-sample prediction.

\section{\texorpdfstring{\MS{Supplementary Material}}{Supplementary material}}\label{supplementary-material}
\MS{The supplementary material includes technical proofs and MCMC details.}

\section{Acknowledgments}\label{acknowledgments}
The authors gratefully acknowledge the support of NSF grants DMS-2308680 and DMS-2001433, which made this research possible.

\section{Disclosure statement}\label{disclosure-statement}
No competing interest is declared.


\section{\texorpdfstring{\MS{Author contributions statement}}{Author contributions statement}}
\MS{M.S., L.Z., and B.S. developed the model. M.S. carried out the implementation}, \LZ{while M.S. and L.Z. conducted the simulation study. M.S. also carried out }\MS{analysis and application. All authors wrote and revised the manuscript.}

\bibliographystyle{abbrvnat}
\bibliography{main}

\clearpage
\begin{appendices}
\section*{Appendix}

\section*{Log-Laplace Nugget Parameterisation}\label{sec:appendix_loglaplace}

We record the parameterisation of the log-Laplace nugget used throughout the appendix. Let $\epsilon$ be a $\mathrm{Log\text{-}Laplace}(0,1/\alpha_0)$ variable, parametrised as
\begin{align*}
    \Pr\{\epsilon\leq x\}=     \begin{cases}
    \frac{1}{2}\exp \{\alpha_0 \log x \}, \quad &0 < x \leq 1, \\
    1-\frac{1}{2}\exp \{ -\alpha_0 \log x \}, \quad &x > 1,
    \end{cases}
\end{align*}
where $x>0$ and $\alpha_0 > 0$; the corresponding density function is
\begin{align*}
    f_\epsilon(x) = \begin{cases}
        \dfrac{\alpha_0}{2} x^{\alpha_0 - 1}, \quad &0 < x \leq 1\\
        \dfrac{\alpha_0}{2} x^{-\alpha_0 - 1}, \quad &1 < x
    \end{cases}
\end{align*}
For intuition, $\alpha_0$ controls the tail heaviness of the log-Laplace nugget $\epsilon$, where larger $\alpha_0$ gives us lighter tails.

\clearpage
\section{Technical proofs}

This section provides the proofs of \autoref{thm:marg_tail_equiv}, \autoref{thm:joint_tail_equiv}, and \autoref{prop:pmm_joint_tail}.

\subsection{Marginal and Joint Tail Equivalence}\label{sec:appendix_tail_equiv}
    Recall that
    \[
    X(\bs)=\epsilon(\bs)X^*(\bs),
    \]
    where $X^*(\bs)$ is the latent smooth process with regularly varying tail and $\epsilon(\bs)$ is the log-Laplace nugget. For any pair of sites $\bs_i,\bs_j\in\mathcal S$, let $(\chi_{ij},\eta_{ij})$ denote the upper tail dependence and residual tail dependence coefficients of the nuggeted pair $(X_i,X_j)$, and let $(\chi_{ij}^*,\eta_{ij}^*)$ denote the corresponding coefficients of the smooth pair $(X_i^*,X_j^*)$, with both pairs of coefficients defined through \autoref{eqn:chi_measure} and \autoref{eqn:eta_measure}. We also write $\alpha^*(\bs)$ for the marginal tail index of $X^*(\bs)$.

    Under the condition $\alpha_0>\sup_{\bs\in\mathcal S}\alpha^*(\bs)$, we first show that, for any site $\bs\in\mathcal S$,
    \[
    \Pr\{X(\bs)>x\}\sim \mathbb{E}\{\epsilon(\bs)^{\alpha^*(\bs)}\}\,\Pr\{X^*(\bs)>x\},
    \qquad x\to\infty.
    \]
    We then show that, 
    \MS{for a random scale-mixture model whose underlying $\{Z(\bs)\}$ process is asymptotically independent but positively correlated (i.e., $\eta_{ij}^Z \in (1/2, 1))$},
    for any pair of sites $\bs_i,\bs_j\in\mathcal S$, if \MS{$X_i^*, X_j^*$}, \MS{and } $\min\{X_i^*,X_j^*\}$ all have regularly varying tails and satisfy \MS{$\alpha_0>=2\sup_{\bs\in\mathcal S}\alpha^*(\bs)$}, then
    \[
    \eta_{ij}=\eta_{ij}^*
    \quad \text{and} \quad
    \chi_{ij}\in\big[\LZ{c_{ij}}\chi_{ij}^*,\,\LZ{C_{ij}}\chi_{ij}^*\big].
    \]
    The proof uses Breiman's lemma, which we recall below.
    \begin{lemma}[\citet{breiman1965some}]\label{lemma:breiman}
    \upshape
    Let $X \geq 0$ and $Y \geq 0$ be independent. If $X$ has a regularly varying tail with index $\alpha>0$, that is,
    \[
    \Pr(X>x)=x^{-\alpha}L(x), \qquad x\to\infty,
    \]
    for some slowly varying function $L(\cdot)$, and if
    \[
    \mathbb{E}(Y^{\alpha+\delta})<\infty
    \qquad \text{for some } \delta>0,
    \]
    then
    \[
    \Pr(XY>x)\sim \mathbb{E}(Y^\alpha)\Pr(X>x),
    \qquad x\to\infty.
    \]
    \end{lemma}
	

    \subsubsection{Marginal Tail Equivalence}

    Fix a site $\bs\in\mathcal S$. Write $\epsilon(\bs)=e^{\zeta(\bs)}$, where
    $\zeta(\bs)\iid \text{Laplace}(0,b=1/\alpha_0)$. The moment generating function of
    $\zeta(\bs)$ is
    \[
    M_\zeta(t)=\mathbb{E}\!\left[e^{t\zeta(\bs)}\right]
    =\frac{1}{1-b^2t^2},
    \qquad |t|<\frac{1}{b}=\alpha_0.
    \]
    Hence, for any $p$ such that $|p|<\alpha_0$,
    \[
    \mathbb{E}\!\left\{\epsilon(\bs)^p\right\}
    =\mathbb{E}\!\left[e^{p\zeta(\bs)}\right]
    =M_\zeta(p)
    =\frac{1}{1-(p/\alpha_0)^2}.
    \]
    Now suppose that $\alpha_0>\alpha^*(\bs)$. Then we may choose $\delta>0$ such that
    $\alpha^*(\bs)+\delta<\alpha_0$, and therefore
    \[
    \mathbb{E}\!\left\{\epsilon(\bs)^{\alpha^*(\bs)+\delta}\right\}<\infty.
    \]
    Since $X^*(\bs)$ has a regularly varying upper tail with index $\alpha^*(\bs)$ by assumption,
    Breiman's lemma applies to $X(\bs)=\epsilon(\bs)X^*(\bs)$ and yields
    \[
    \Pr\{X(\bs)>x\}
    =\Pr\{\epsilon(\bs)X^*(\bs)>x\}
    \sim
    \mathbb{E}\!\left\{\epsilon(\bs)^{\alpha^*(\bs)}\right\}\Pr\{X^*(\bs)>x\},
    \qquad x\to\infty.
    \]
    Thus, to satisfy $\alpha^*(\bs) < \alpha_0$ at any site $\bs$, we impose $\alpha_0>\sup_{\bs\in\mathcal S}\alpha^*(\bs)$.

    \subsubsection{Joint Tail Equivalence}
    
    For two sites $\bs_i,\bs_j$, we write
    \[
    \alpha_i^*=\alpha^*(\bs_i),\qquad \alpha_j^*=\alpha^*(\bs_j),
    \qquad
    m_i=\mathbb{E}(\epsilon_i^{\alpha_i^*}),\qquad m_j=\mathbb{E}(\epsilon_j^{\alpha_j^*}).
    \]
    We make use of the following ``sandwich'' inequality
    \begin{equation}\label{eqn:sandwich}
    \begin{split}
    \pr(\min(\epsilon_i,\epsilon_j)\times\min(X_i^*,X_j^*)>x)
    & \leq \pr(\epsilon_i X_i^*>x,\epsilon_j X_j^*>x) \\
    &\leq \pr(\max(\epsilon_i,\epsilon_j)\times \min(X_i^*,X_j^*)>x).
    \end{split}
    \tag{Sandwich}
    \end{equation}
    For $\ell\in\{i,j\}$, by definition,
    \[
    \overline{F}_{X_\ell}\left(F_{X_\ell}^{-1}(u)\right) = 1-u,
    \]
    and by marginal tail equivalence,
    \[
    \overline{F}_{X_\ell}(x) \sim m_\ell \overline{F}_{X_\ell^*}(x).
    \]
    Write the tail of the smooth process as
    \[
    \overline{F}_{X_\ell^*}(x) \sim \tilde{c}_\ell x^{-\alpha_\ell^*}.
    \]
    Hence
    \[
    1-u
    = \overline{F}_{X_\ell}\left(F_{X_\ell}^{-1}(u)\right)
    \sim m_\ell \tilde{c}_\ell \left(F_{X_\ell}^{-1}(u)\right)^{-\alpha_\ell^*},
    \]
    so that
    \[
    F_{X_\ell}^{-1}(u) \sim m_\ell^{1/\alpha_\ell^*}\tilde{c}_\ell^{1/\alpha_\ell^*}(1-u)^{-1/\alpha_\ell^*}.
    \]
    Hence, as $u \rightarrow 1$,
    \begin{align*}
    \Pr&\left[X_i > F_{X_i}^{-1}(u), X_j > F_{X_j}^{-1}(u)\right] \\
    &= \Pr\left[X_i > m_i^{1/\alpha_i^*}\tilde{c}_i^{1/\alpha_i^*}(1-u)^{-1/\alpha_i^*},
    X_j > m_j^{1/\alpha_j^*}\tilde{c}_j^{1/\alpha_j^*}(1-u)^{-1/\alpha_j^*}\right] \\
    &= \Pr\left[\dfrac{\epsilon_i^{\alpha_i^*}}{m_i}\dfrac{(X_i^*)^{\alpha_i^*}}{\tilde{c}_i} > (1-u)^{-1},
    \dfrac{\epsilon_j^{\alpha_j^*}}{m_j}\dfrac{(X_j^*)^{\alpha_j^*}}{\tilde{c}_j} > (1-u)^{-1}\right]
    \tag{$\triangle$}
    \end{align*}
    Applying \autoref{eqn:sandwich}, we have
    \begin{align*}
    \triangle \in
    \Bigg[
    &\Pr\left[\min\left(\dfrac{\epsilon_i^{\alpha_i^*}}{m_i}, \dfrac{\epsilon_j^{\alpha_j^*}}{m_j}\right)\times
    \min\left(\dfrac{(X_i^*)^{\alpha_i^*}}{\tilde{c}_i}, \dfrac{(X_j^*)^{\alpha_j^*}}{\tilde{c}_j}\right) > (1-u)^{-1}\right], \\
    &\Pr\left[\max\left(\dfrac{\epsilon_i^{\alpha_i^*}}{m_i}, \dfrac{\epsilon_j^{\alpha_j^*}}{m_j}\right)\times
    \min\left(\dfrac{(X_i^*)^{\alpha_i^*}}{\tilde{c}_i}, \dfrac{(X_j^*)^{\alpha_j^*}}{\tilde{c}_j}\right) > (1-u)^{-1}\right]
    \Bigg].
    \end{align*}
    We now apply Breiman's lemma to the two bounds. First note that, as $u \rightarrow 1$,
    \begin{align*}
    \Pr&\left[\min\left(\dfrac{(X_i^*)^{\alpha_i^*}}{\tilde{c}_i}, \dfrac{(X_j^*)^{\alpha_j^*}}{\tilde{c}_j}\right) > (1-u)^{-1}\right] \\
    &=
    \Pr\left[X_i^* > \tilde{c}_i^{1/\alpha_i^*}(1-u)^{-1/\alpha_i^*},
    X_j^* > \tilde{c}_j^{1/\alpha_j^*}(1-u)^{-1/\alpha_j^*}\right], \\
    &= \Pr \left[X_i^* > F_{X_i^*}^{-1}(u), X_j^* > F_{X_j^*}^{-1}(u)\right],
    \end{align*}
    which is regularly varying by assumption. Denote its tail exponent by $\kappa_{ij}$, that is,
    \[
    \Pr\left[\min\left(\dfrac{(X_i^*)^{\alpha_i^*}}{\tilde{c}_i}, \dfrac{(X_j^*)^{\alpha_j^*}}{\tilde{c}_j}\right) > t\right]
    \in \mathrm{RV}_{-\kappa_{ij}}.
    \]
    We next show that
    \[
    \min\left(\dfrac{\epsilon_i^{\alpha_i^*}}{m_i}, \dfrac{\epsilon_j^{\alpha_j^*}}{m_j}\right)
    \quad\text{and}\quad
    \max\left(\dfrac{\epsilon_i^{\alpha_i^*}}{m_i}, \dfrac{\epsilon_j^{\alpha_j^*}}{m_j}\right)
    \]
    are lighter tailed.

Indeed,
\begin{align*}
\Pr\left[\min\left(\dfrac{\epsilon_i^{\alpha_i^*}}{m_i}, \dfrac{\epsilon_j^{\alpha_j^*}}{m_j}\right) > t\right]
&= \Pr\left[\dfrac{\epsilon_i^{\alpha_i^*}}{m_i} > t, \dfrac{\epsilon_j^{\alpha_j^*}}{m_j} > t\right] \\
&= \Pr\left[\dfrac{\epsilon_i^{\alpha_i^*}}{m_i} > t\right]\Pr\left[\dfrac{\epsilon_j^{\alpha_j^*}}{m_j} > t\right]
\quad \text{as } \epsilon_i \indep \epsilon_j \\
&\sim L_i(t)L_j(t)t^{-[(\alpha_0/\alpha_i^*)+(\alpha_0/\alpha_j^*)]},
\end{align*}
and
\begin{align*}
\Pr&\left[\max\left(\dfrac{\epsilon_i^{\alpha_i^*}}{m_i}, \dfrac{\epsilon_j^{\alpha_j^*}}{m_j}\right) > t\right]
= 1 - \Pr\left[\dfrac{\epsilon_i^{\alpha_i^*}}{m_i} < t, \dfrac{\epsilon_j^{\alpha_j^*}}{m_j} < t\right] \\
&= 1 - \Pr\left[\dfrac{\epsilon_i^{\alpha_i^*}}{m_i} < t\right]\Pr\left[\dfrac{\epsilon_j^{\alpha_j^*}}{m_j} < t\right]
\quad \text{as } \epsilon_i \indep \epsilon_j \\
&= \Pr\left[\dfrac{\epsilon_i^{\alpha_i^*}}{m_i} > t\right] + \Pr\left[\dfrac{\epsilon_j^{\alpha_j^*}}{m_j} > t\right]  - \Pr\left[\dfrac{\epsilon_i^{\alpha_i^*}}{m_i} > t\right]\Pr\left[\dfrac{\epsilon_j^{\alpha_j^*}}{m_j} > t\right] \\
&\sim L_i(t)t^{-\alpha_0/\alpha_i^*} + L_j(t)t^{-\alpha_0/\alpha_j^*}  - L_i(t)L_j(t)t^{-[(\alpha_0/\alpha_i^*)+(\alpha_0/\alpha_j^*)]} \\
&\sim L(t)t^{-\min(\alpha_0/\alpha_i^*,\,\alpha_0/\alpha_j^*)}.
\end{align*}
For \MS{example}, the transformed Gaussian scale-mixture models with the underlying Gaussian process having \MS{positive } pairwise correlations,
\[
\LZ{\eta_{ij}^Z}=\frac{1+\rho_{ij}}{2}\in\MS{\left(\frac12,1\right)}.
\]
\MS{Since we assumed in general}, the smooth-pair residual tail dependence coefficient satisfies \MS{$\eta_{ij}^* >  1/2$},
\[
\MS{\kappa_{ij}=\frac{1}{\eta_{ij}^*} < 2},
\]
while in the AD case $\kappa_{ij}=1$. Therefore, the assumption
\[
\MS{\alpha_0  \ge  2\sup_{\bs\in\mathcal S}\alpha^*(\bs)}
\]
implies
\[
\min\left(\frac{\alpha_0}{\alpha_i^*},\frac{\alpha_0}{\alpha_j^*}\right)>\kappa_{ij},
\]
so the required moments are finite and Breiman's lemma applies.
Therefore, for some $\delta>0$,
\[
\mathbb{E}\left[\min\left(\dfrac{\epsilon_i^{\alpha_i^*}}{m_i}, \dfrac{\epsilon_j^{\alpha_j^*}}{m_j}\right)^{\kappa_{ij}+\delta}\right] < \infty,
\qquad
\mathbb{E}\left[\max\left(\dfrac{\epsilon_i^{\alpha_i^*}}{m_i}, \dfrac{\epsilon_j^{\alpha_j^*}}{m_j}\right)^{\kappa_{ij}+\delta}\right] < \infty.
\]
Thus Breiman's lemma applies to both bounds.
The lower bound can therefore be written as
\begin{align*}
\Pr &\left[\min\left(\dfrac{\epsilon_i^{\alpha_i^*}}{m_i}, \dfrac{\epsilon_j^{\alpha_j^*}}{m_j}\right)\times
\min\left(\dfrac{(X_i^*)^{\alpha_i^*}}{\tilde{c}_i}, \dfrac{(X_j^*)^{\alpha_j^*}}{\tilde{c}_j}\right) > (1-u)^{-1}\right] \\
&\sim\mathbb{E}\left[\min\left(\dfrac{\epsilon_i^{\alpha_i^*}}{m_i}, \dfrac{\epsilon_j^{\alpha_j^*}}{m_j}\right)\right]
\Pr\left[\min\left(\dfrac{(X_i^*)^{\alpha_i^*}}{\tilde{c}_i}, \dfrac{(X_j^*)^{\alpha_j^*}}{\tilde{c}_j}\right) > (1-u)^{-1}\right].
\end{align*}
Similarly, the upper bound can be written as
\begin{align*}
\Pr &\left[\max\left(\dfrac{\epsilon_i^{\alpha_i^*}}{m_i}, \dfrac{\epsilon_j^{\alpha_j^*}}{m_j}\right)\times
\min\left(\dfrac{(X_i^*)^{\alpha_i^*}}{\tilde{c}_i}, \dfrac{(X_j^*)^{\alpha_j^*}}{\tilde{c}_j}\right) > (1-u)^{-1}\right] \\
&\sim\mathbb{E}\left[\max\left(\dfrac{\epsilon_i^{\alpha_i^*}}{m_i}, \dfrac{\epsilon_j^{\alpha_j^*}}{m_j}\right)\right]
\Pr\left[\min\left(\dfrac{(X_i^*)^{\alpha_i^*}}{\tilde{c}_i}, \dfrac{(X_j^*)^{\alpha_j^*}}{\tilde{c}_j}\right) > (1-u)^{-1}\right].
\end{align*}
Since the constant factor does not change the tail exponent, we obtain
\[
\eta_{ij}=\eta_{ij}^*.
\]
To identify the constants in the bound for $\chi_{ij}$, we calculate
\[
\LZ{c_{ij}}
=
\mathbb{E}\left[\min\left(\dfrac{\epsilon_i^{\alpha_i^*}}{m_i}, \dfrac{\epsilon_j^{\alpha_j^*}}{m_j}\right)\right]
\quad \text{and} \quad
\LZ{C_{ij}}
=
\mathbb{E}\left[\max\left(\dfrac{\epsilon_i^{\alpha_i^*}}{m_i}, \dfrac{\epsilon_j^{\alpha_j^*}}{m_j}\right)\right].
\]
Denote
\[
U:= \dfrac{\epsilon_i^{\alpha_i^*}}{m_i},
\qquad
V:= \dfrac{\epsilon_j^{\alpha_j^*}}{m_j},
\]
and we have
\begin{align*}
\mathbb{E}\left[\min(U,V)\right]
&= \int_0^\infty \Pr(U > t, V > t)\, dt \\
&= \int_0^\infty \Pr\left[\dfrac{\epsilon_i^{\alpha_i^*}}{m_i} > t\right]
\Pr\left[\dfrac{\epsilon_j^{\alpha_j^*}}{m_j} > t\right]dt \\
&= \int_0^\infty \Pr\left[\epsilon_i > (m_i t)^{1/\alpha_i^*}\right]
\Pr\left[\epsilon_j > (m_j t)^{1/\alpha_j^*}\right] dt.
\end{align*}
Without loss of generality, assume $m_i > m_j$, so $1/m_i \le 1/m_j$. Define
\[
t_i = 1/m_i,\qquad t_j = 1/m_j,
\]
and split the integral into
\[
\int_0^\infty \cdots
=
\underbrace{\int_0^{t_i} \cdots}_{I_1}
+
\underbrace{\int_{t_i}^{t_j} \cdots}_{I_2}
+
\underbrace{\int_{t_j}^{\infty} \cdots}_{I_3},
\]
because for $\ell \in \{i,j\}$ we have
\[
\Pr\left[\epsilon_\ell > (m_\ell t)^{1/\alpha_\ell^*}\right]
=
\begin{cases}
1-\frac{1}{2}(m_\ell t)^{p_\ell} & t < 1/m_\ell, \\
\frac{1}{2}(m_\ell t)^{-p_\ell} & t \ge 1/m_\ell,
\end{cases}
\]
where $p_\ell = \alpha_0 / \alpha_\ell^*$.
Then,
\begin{align*}
I_1 &= \int_0^{t_i} \left[1 - \frac{1}{2}(m_it)^{p_i}\right]\left[1 - \frac{1}{2}(m_jt)^{p_j}\right]dt \\
&= \dfrac{1}{m_i} - \frac{1}{2}\frac{1}{(p_i + 1)m_i}-\frac{1}{2}\frac{m_j^{p_j}}{(p_j+1)m_i^{p_j+1}} + \frac{1}{4}\frac{m_j^{p_j}}{(p_i+p_j+1)m_i^{p_j+1}}, \\
I_2 &= \int_{t_i}^{t_j}\left[\frac{1}{2}(m_it)^{-p_i}\right]\left[1 - \frac{1}{2}(m_jt)^{p_j}\right]dt \\
&= \frac{1}{2}\frac{1}{p_i-1}\left(\dfrac{1}{m_i} - \dfrac{m_j^{p_i-1}}{m_i^{p_i}}\right) - \frac{1}{4}\frac{1}{1+p_j-p_i}\left(\frac{m_j^{p_i-1}}{m_i^{p_i}} - \frac{m_j^{p_j}}{m_i^{p_j+1}}\right), \\
I_3 &= \int_{t_j}^\infty \left[\frac{1}{2}(m_it)^{-p_i}\right]\left[\frac{1}{2}(m_jt)^{-p_j}\right]dt \\
&= \frac{1}{4} \dfrac{m_i^{-p_i}m_j^{p_i-1}}{p_i+p_j-1}.
\end{align*}
Thus,
\[
\LZ{c_{ij}}=\mathbb{E}[\min(U, V)] = I_1 + I_2 + I_3.
\]
Since $\mathbb{E}[\max(U, V)] + \mathbb{E}[\min(U, V)] = \mathbb{E}[U+V] = 2$, we have
\[
\LZ{C_{ij}}=\mathbb{E}[\max(U, V)] = 2 - (I_1 + I_2 + I_3).
\]
Therefore,
\[
\chi_{ij}\in\big[\LZ{c_{ij}}\chi_{ij}^*,\,\LZ{C_{ij}}\chi_{ij}^*\big].
\]

\subsection{\texorpdfstring{Proof of \autoref{prop:pmm_joint_tail}}{Proof of Proposition 1}}

The model of \citet{majumder2024modeling} is
\begin{equation*}\label{eqn:reetam-PMM}
    \tilde{X}^*(\bs) = \delta \tilde{R}(\bs) + (1-\delta)\tilde{W}(\bs)
\end{equation*}
where $\tilde{R}(\bs)$ and $\tilde{W}(\bs)$ have $\mathrm{Exp}(1)$ margins. Consequently, $\tilde{X}^*(\bs)$ has marginal distribution
\begin{equation*}\label{eqn:reetam-PMM-CDF}
    F_{\tilde{X}^*}(x) = 1 - \dfrac{1-\delta}{1-2\delta}\exp \left\{-\dfrac{x}{1-\delta} \right\} + \dfrac{\delta}{1-2\delta}\exp\left\{-\dfrac{x}{\delta}\right\}.
\end{equation*}
Because the copula is invariant under strictly increasing marginal transformations, it is more convenient to work with
\begin{equation*}
    X^*(\bs) = \exp(\MS{{\tilde{X}^*(\bs)}}) = R(\bs)^\delta W(\bs)^{1-\delta}
\end{equation*}
where $R(\bs)$ and $W(\bs)$ are marginally standard Pareto.

We now derive the joint tail decay rate of this model. Let $x := F_{X^*}^{-1}(u)$. Then
\begin{align*}
    \lim_{u \rightarrow 1}\Pr[X_1^* > F_{X_1^*}^{-1}(u)&, X_2^* > F_{X_2^*}^{-1}(u)] \\
    &= \lim_{u \rightarrow 1} \Pr[R_1^\delta W_1^{1-\delta} > x, R_2^\delta W_2^{1-\delta} > x],
\end{align*}
so we study the asymptotic \MS{behaviour } of the right-hand side as $u \to 1$.

To bound this probability, we use the sandwich inequality \autoref{eqn:sandwich}:
\begin{align*}
    \Pr[\min(R_1^\delta, R_2^\delta) &\cdot \min(W_1^{1-\delta}, W_2^{1-\delta}) > x] \\
    & \leq \Pr[X_1^* > F_{X_1^*}^{-1}(u), X_2^* > F_{X_2^*}^{-1}(u)]\\
    & \le \Pr[\max(R_1^\delta, R_2^\delta) \cdot \min(W_1^{1-\delta}, W_2^{1-\delta}) > x]
\end{align*}
Hence, it suffices to determine the tail \MS{behaviour } of $\Pr\{\min(R_1^\delta,R_2^\delta)>x\}$, $\Pr\{\min(W_1^{1-\delta},W_2^{1-\delta})>x\}$, and $\Pr\{\max(R_1^\delta,R_2^\delta)>x\}$.

We first \MS{analyse } $\Pr\{\min(R_1^\delta,R_2^\delta)>x\}$. In \citet{majumder2024modeling}, $R(\bs)$ is induced by a Brown--Resnick process satisfying
\begin{equation*}
    \MS{\Pr[\tilde{R}_1 < r_1, \tilde{R}_2 < r_2] } = \exp[-\Lambda(r_1, r_2)]
\end{equation*}
where
\begin{equation*}
    \Lambda(r_1, r_2) = \dfrac{1}{r_1}\Phi\left\{\dfrac{a}{2} - \dfrac{1}{a}\log\left(\dfrac{r_1}{r_2}\right)\right\} + \dfrac{1}{r_2}\Phi\left\{\dfrac{a}{2} - \dfrac{1}{a}\log\left(\dfrac{r_2}{r_1}\right)\right\}
\end{equation*}
in which
\begin{equation*}
    a = [2\gamma(h)]^{1/2}, \quad \gamma(h) = (h/\rho_R)^{\alpha_R}, \quad \Phi\{\cdot\} \text{ is standard Gaussian distribution}.
\end{equation*}
After transforming to standard Pareto margins, we obtain
\begin{align*}
    \Pr[R_1 < r_1 &, R_2 < r_2] \\
    &= \Pr\left[-\dfrac{1}{\log(1-\tfrac{1}{R_1})} < -\dfrac{1}{\log(1-\tfrac{1}{r_1})}, -\dfrac{1}{\log(1-\tfrac{1}{R_2})} < -\dfrac{1}{\log(1-\tfrac{1}{r_2})}\right] \\
    &= \exp\left[-\Lambda\left(-\dfrac{1}{\log(1-\tfrac{1}{r_1})}, -\dfrac{1}{\log(1-\tfrac{1}{r_2})}\right)\right]
\end{align*}
Therefore, as $x \to \infty$,
\begin{align*}
    \Pr[\min(R_1^\delta&, R_2^\delta) > x] \\
    &= \Pr[R_1 > x^{1/\delta}, R_2 > x^{1/\delta}] \\
    &= 1 - \Pr[R_1 \leq x^{1/\delta}] - \Pr[R_2 \leq x^{1/\delta}] + \Pr[R_1 \leq x^{1/\delta}, R_2 \leq x^{1/\delta}] \\
    &= \dfrac{2}{x^{1/\delta}} - \left\{1 - \exp\left[-\Lambda\left(-\dfrac{1}{\log(1-\tfrac{1}{x^{1/\delta}})}, -\dfrac{1}{\log(1-\tfrac{1}{x^{1/\delta}})}\right)\right]\right\} \\
    &\sim \dfrac{2}{x^{1/\delta}} - \Lambda\left(-\dfrac{1}{\log(1-\tfrac{1}{x^{1/\delta}})}, -\dfrac{1}{\log(1-\tfrac{1}{x^{1/\delta}})}\right) \\ 
    &= \dfrac{2}{x^{1/\delta}} + 2\Phi\left(\dfrac{a}{2}\right)\left[\MS{\log \left(1 - \dfrac{1}{x^{1/\delta}}\right)}\right] \\
    &\sim L_1(x)x^{-1/\delta} 
\end{align*}
By the same argument, we also obtain
\begin{align*}
    \Pr[\max(R_1^\delta, R_2^\delta) > x] &= 1 - \Pr[\max(R_1^\delta, R_2^\delta) < x] \\
    &\sim L_2(x)x^{-1/\delta}
\end{align*}
For the Gaussian-copula component, \citet{ledford1996statistics} show that the joint survival function with unit Pareto margins satisfies
\begin{align*}
    \Pr[\min(W_1^{1-\delta}, W_2^{1-\delta}) > x] \sim L_3(x)x^{-1/\{\eta_W(1-\delta)\}}
\end{align*}
where $\eta_W=(1+\rho)/2$ and $\rho=\mathrm{Cor}(Z_1,Z_2)$.

Combining these with Breiman's lemma and the sandwich inequality yields
{\small
\setlength{\abovedisplayskip}{1pt}
\setlength{\belowdisplayskip}{1pt}
\setlength{\abovedisplayshortskip}{1pt}
\setlength{\belowdisplayshortskip}{1pt}
\begin{align*}
    \MS{\Pr\left[X_1^* > F_{X_1^*}^{-1}(u), X_2^* > F_{X_2^*}^{-1}(u)\right]} &\sim \begin{cases}
        L(x)x^{-1/\{\eta_W(1-\delta)\}} & \eta_W \geq \delta / (1-\delta), \\
        \mathbb{E}\left[\min(W_1, W_2)^{(1-\delta)/\delta}\right]x^{-1/\delta} & \eta_W < \delta / (1-\delta).
    \end{cases}
\end{align*}
}
Thus the joint exceedance probability is regularly varying.

\clearpage
\section{MCMC Details}\label{sec:Appendix_MCMC}

This section gives the full hierarchical specification used for posterior sampling. Here $t=1,\ldots,T$ indexes replicates, $j=1,\ldots,D$ indexes sites, and $k=1,\ldots,K$ indexes knots. The observed series $Y_{tj}$ is treated as an anomaly process. We assume temporal independence across replicates over time.
We restate the censored likelihood from \autoref{sec:hier_model}. Let
$\mathcal{C}_t=\{j:Y_{tj}\le u_{j}\}$ and $\mathcal{E}_t=\{j:Y_{tj}>u_{j}\}$, with $j=1,\ldots,D$ and $x_{0j}=F_{X_{j}}^{-1}(p)$. Then
\[
L_t(\bY_t\mid \bS_t,\bZ_t,\bphi,\bgamma,\brho,l,\alpha_0,\bsigma,\bxi)
=
\prod_{j\in\mathcal{C}_t}
F_\epsilon\!\left(\frac{x_{0j}}{X^*_{tj}}\right)
\;\times\;
\prod_{j\in\mathcal{E}_t}
f_{X\mid X^*}(x_{tj}\mid X^*_{tj})
\left|\frac{dX_{tj}}{dY_{tj}}\right|.
\]
For $j\in\mathcal{E}_t$,
\[
\left|\frac{dX_{tj}}{dY_{tj}}\right|
=
\frac{g_{Y_{tj}}(Y_{tj})}{f_{X_{tj}}(x_{tj})},
\qquad
g_{Y_{tj}}(y)=(1-p)\,h_{u_{j}}(y\mid \sigma_{j},\xi_{j}),
\]
where $h_{u_{j}}$ is the GP density associated with $H_{u_{j}}$. Under temporal independence of the anomaly replicates, the full likelihood is the product of $L_t$ at each replicate.

The latent process is
\[
X_t(\bs)=\epsilon_t(\bs)\,X_t^*(\bs),
\qquad
X_t^*(\bs)=R_t(\bs)^{\phi(\bs)}\,g\{Z_t(\bs)\},
\]
with
\[
R_t(\bs) = \sum_{k=1}^K B_k(\bs;l)\,S_{tk},
\]
where \(B_k(\cdot;l)\) are \citet{Wendland1995} basis functions with radius \(l\), and
\begin{align*}
S_{tk}\mid \gamma_k &\sim \mathrm{Stable}(\alpha=1/2,\beta=1,\gamma_k,\delta=0) \equiv \text{L\'evy}(0,\gamma_k), \\
l &\sim \mathrm{Half\mbox{-}}t_{\nu=1}(0,3).
\end{align*}
Similar to \citet{shi2026}, we fix the scale parameter $\bgamma$ at 1, as varying it does not play any role in modulating the tail dependence characteristics of the model. 

The nugget terms are i.i.d. across sites, and we enforce $\alpha_0 > 1$ by
\[
\epsilon_t(\bs_j)\iid \log\text{-Laplace}(0,1/\alpha_0),
\qquad
\alpha_0=1+\exp(\vartheta),\ \vartheta\sim N(3,0.5^2).
\]

The latent Gaussian field satisfies
\begin{equation*}
\bZ_t=(Z_t(\bs_1),\ldots,Z_t(\bs_D))^\top\mid\brho \sim \mathcal{N}(\mathbf 0,\bSigma_{\brho}),
\end{equation*}
where $\bSigma_{\brho}$ is a locally isotropic nonstationary Matérn covariance \citep{paciorek2006spatial,risser2015regression}, with entries
\[
C(\bs_i,\bs_j)
=
\zeta(\bs_i)\zeta(\bs_j)\,
\frac{\sqrt{\rho(\bs_i)\rho(\bs_j)}}{\{\rho(\bs_i)+\rho(\bs_j)\}/2}\;
\mathcal{M}_\nu\!\left(
\frac{\|\bs_i-\bs_j\|}{\sqrt{\{\rho(\bs_i)+\rho(\bs_j)\}/2}}
\right),
\]
where $\zeta(\bs) \equiv 1$ and $\nu$ is fixed in our implementation.

The surfaces $\phi(\bs)$ and $\rho(\bs)$ are represented using Gaussian kernel basis functions centred at the $K$ knots. Similar to \citet{shi2026}, the prior for $\phi$ is centred at the AI - AD transition boundary and assigns relatively little mass near the edges of its support, which correspond to extremely strong or extremely weak tail dependence. Knot-level priors are
\[
\phi_k\sim\mathrm{Beta}(2,2),\qquad
\rho_k\sim\mathrm{Half\mbox{-}Normal}(0,10^2),\qquad k=1,\ldots,K.
\]
Marginal GP parameters are modelled as
\[
\log \sigma_{j}=\bc_{j}^{\top}\bbeta_{\sigma},
\qquad
\xi_{j}=\bd_{j}^{\top}\bbeta_{\xi},
\]
with
\[
\bbeta_{\sigma}\mid \sigma_{\bbeta_{\sigma}}\sim \mathrm{MVN}(\mathbf 0,\sigma_{\bbeta_{\sigma}}^2\bI),\quad
\bbeta_{\xi}\mid \sigma_{\bbeta_{\xi}}\sim \mathrm{MVN}(\mathbf 0,\sigma_{\bbeta_{\xi}}^2\bI),
\]
\[
\sigma_{\bbeta_{\sigma}}\sim \mathrm{Half\mbox{-}}t_{\nu=2}(0,1),\qquad
\sigma_{\bbeta_{\xi}}\sim \mathrm{Half\mbox{-}}t_{\nu=2}(0,1).
\]

Let
\[
\Theta=\{\bphi,\brho,l,\alpha_0,\bbeta_{\sigma},\bbeta_{\xi},\sigma_{\bbeta_{\sigma}},\sigma_{\bbeta_{\xi}}\}.
\]
The posterior target is
\[
p\!\left(\Theta,\{\bS_t,\bZ_t\}_{t=1}^T\mid \bY_{1:T}\right)
\propto
\prod_{t=1}^T
L_t(\bY_t\mid \bS_t,\bZ_t,\Theta,\bgamma)\,
p(\bS_t\mid \bgamma)\,
p(\bZ_t\mid \brho)\,
p(\Theta).
\]

We sample from the posterior using an adaptive random-walk Metropolis-within-Gibbs algorithm \citep{shaby2010exploring}. Parameters shared across all replicates are updated sequentially, and replicate-specific latent blocks \((\bS_t,\bZ_t)\), \(t=1,\ldots,T\), are updated in parallel. 

The sampler is implemented in \texttt{Python}, with parallelisation via \texttt{mpi4py} and \texttt{OpenMPI} \citep{mpi4py}. For Model~\ref{Model-Shi}, evaluations of \(F_X\) and \(F_X^{-1}\) are implemented in \texttt{C++} using \texttt{GSL} \citep{gough2009gnu}, and likelihood evaluations are JIT-compiled with \texttt{Numba} \citep{numba}.

\end{appendices}

\end{document}